\documentclass[10pt]{article}
\usepackage{graphicx}
\usepackage{amsmath}
\usepackage{amssymb}
\usepackage{caption2}
\begin{document}
\begin{center}
{\large {\bf \sc{  Analysis of  the ${\frac{1}{2}}^{\pm}$  pentaquark states in the diquark-diquark-antiquark model with QCD sum rules }}} \\[2mm]
Zhi-Gang Wang \footnote{E-mail: zgwang@aliyun.com.  }     \\
 Department of Physics, North China Electric Power University, Baoding 071003, P. R. China
\end{center}

\begin{abstract}
In this article, we  construct  both the axialvector-diquark-axialvector-diquark-antiquark type and axialvector-diquark-scalar-diquark-antiquark type interpolating currents, then calculate the contributions of the vacuum condensates up to dimension-10 in the operator product expansion, and  study the masses and pole residues of the  $J^P={\frac{1}{2}}^\pm$   hidden-charm pentaquark states    with the QCD sum rules in a systematic way. In calculations, we use the  formula $\mu=\sqrt{M^2_{P}-(2{\mathbb{M}}_c)^2}$  to determine  the energy scales of the QCD spectral densities.  We take into account the $SU(3)$ breaking effects of the light quarks, and obtain the masses of the hidden-charm pentaquark states with the strangeness $S=0,\,-1,\,-2,\,-3$, which can be confronted with the experimental data in the future.
\end{abstract}



\section{Introduction}

Recently,  the  LHCb collaboration studied the $\Lambda_b^0\to J/\psi K^- p$ decays, and observed  two pentaquark candidates $P_c(4380)$ and $P_c(4450)$ in the $J/\psi p$ mass spectrum   with the significances of more than $9\,\sigma$  \cite{LHCb-4380}. The  measured masses and widths are  $M_{P_c(4380)}=4380\pm 8\pm 29\,\rm{MeV}$, $M_{P_c(4450)}=4449.8\pm 1.7\pm 2.5\,\rm{MeV}$, $\Gamma_{P_c(4380)}=205\pm 18\pm 86\,\rm{MeV}$, and  $\Gamma_{P_c(4450)}=39\pm 5\pm 19\,\rm{MeV}$, respectively.
The decays $P_c(4380)\to J/\psi p$ take place through relative S-wave while the decays $P_c(4450)\to J/\psi p$ take place through relative P-wave, the decays  $P_c(4380)\to J/\psi p$ are kinematically favored and  the $P_c(4380)$ has larger width.
The  preferred spin-parity of the  $P_c(4380)$ and $P_c(4450)$ are   $J^P={\frac{3}{2}}^-$ and ${\frac{5}{2}}^+$, respectively.
 There have been  several  attempted assignments, such as the $\Sigma_c \bar{D}^*$, $\Sigma_c^* \bar{D}^*$, $\chi_{c1}p$, $J/\psi N(1440)$, $J/\psi N(1520)$ molecule-like  pentaquark  states \cite{mole-penta} (or not the molecular pentaquark states \cite{mole-penta-No}),  the diquark-diquark-antiquark type pentaquark states \cite{Maiani1507,di-di-anti-penta,XGHe1507,Wang1508,WangHuang1508,Cheng1509,mass-Anisovich}, the diquark-triquark  type  pentaquark states \cite{di-tri-penta}, re-scattering effects \cite{rescattering-penta}, etc.

 The quarks have color $SU(3)$ symmetry,  we can construct the pentaquark configurations according to the routine  ${\rm quark}\to {\rm diquark}\to {\rm pentaquark}$,
\begin{eqnarray}
(3\otimes 3)\otimes(3\otimes 3) \otimes\overline{3}&=&(\overline{3}\oplus 6)\otimes(\overline{3}\oplus 6)\otimes\overline{3} =\overline{3}\otimes\overline{3}\otimes\overline{3}\oplus\cdots =1\oplus  \cdots \, ,
\end{eqnarray}
where the $1$, $3$ ($\overline{3}$) and $6$   denote the color singlet, triplet (antitriplet) and sextet, respectively.
The diquarks $q^{T}_j C\Gamma q^{\prime}_k$ have  five  structures  in Dirac spinor space, where $C\Gamma=C\gamma_5$, $C$, $C\gamma_\mu \gamma_5$,  $C\gamma_\mu $ and $C\sigma_{\mu\nu}$ for the scalar, pseudoscalar, vector, axialvector  and  tensor diquarks, respectively, and the $j$ and $k$ are color indexes. The
attractive interactions of one-gluon exchange  favor  formation of
the diquarks in  color antitriplet $\overline{3}_{ c}$, flavor
antitriplet $\overline{3}_{ f}$ and spin singlet $1_s$ \cite{One-gluon},
 while the favored configurations are the scalar ($C\gamma_5$) and axialvector ($C\gamma_\mu$) diquark states \cite{WangDiquark,WangLDiquark}.  The calculations based on the QCD sum rules indicate that the  heavy-light scalar and axialvector  diquark states have almost  degenerate masses \cite{WangDiquark},  while  the masses of the
 light  axialvector  diquark states lie   $(150-200)\,\rm{MeV}$ above that of the  light scalar diquark states \cite{WangLDiquark}, if they have the same quark constituents.

 In Refs.\cite{Wang1508,WangHuang1508}, we choose the light scalar diquark and heavy axialvector diquark (or heavy scalar diquark) as the basic constituents,  construct both  the scalar-diquark-axialvector-diquark-antiquark type and scalar-diquark-scalar-diquark-antiquark type interpolating currents, which are supposed to  couple potentially to the lowest pentaquark  states according to the light scalar diquark constituent \cite{WangLDiquark}, then calculate the contributions of the vacuum condensates up to dimension-10 in the operator product expansion and study the masses and pole residues  of the $J^P={\frac{3}{2}}^-$, ${\frac{5}{2}}^+$ and ${\frac{1}{2}}^\pm$ hidden-charm pentaquark states    with the QCD sum rules. The numerical results favor    assigning  the $P_c(4380)$ and $P_c(4450)$ to be the ${\frac{3}{2}}^-$ and ${\frac{5}{2}}^+$ diquark-diquark-antiquark type pentaquark states, respectively \cite{Wang1508}. In Ref.\cite{Wang1508}, we  take the energy scale formula $ \mu=\sqrt{M_{P}^2-(2{\mathbb{M}}_c)^2}$ to determine the energy scales of the QCD spectral densities, the resulting pole contributions are about $(40-60)\%$, and the  contributions of the vacuum condensates of dimension 10 are less than $5\%$, the two criteria
(pole dominance at the phenomenological side and convergence of the operator product expansion) of the conventional QCD sum rules can be satisfied. Now we extend
our previous work to study the ${\frac{1}{2}}^\pm$  hidden-charm pentaquark states in a systematic way, where the energy scale formula $ \mu=\sqrt{M_{P}^2-(2{\mathbb{M}}_c)^2}$ serves as an additional constraint on the predicted masses.  All the predictions can be confronted with  the experimental data in the future,  and the assignments of the $P_c(4380)$ and $P_c(4450)$ in the scenario of the diquark-diquark-antiquark type  pentaquark states  can be testified.

  In this article, we take the light axialvector  diquarks and heavy axialvector diquarks (and heavy scalar diquarks) as the basic constituents, and study the axialvector-diquark-axialvector-diquark-antiquark type and axialvector-diquark-scalar-diquark-antiquark type pentaquark configurations.
 Now we illustrate how to construct the pentaquark configurations in the diquark-diquark-antiquark  model  according to the spin-parity $J^P$,
\begin{eqnarray}
1^+_{q_1q_2}\otimes 0^+_{q_3c} \otimes {\frac{1}{2}}^-_{\bar{c}} &=&  \underline{{\frac{1}{2}}^-_{q_1q_2q_3c\bar{c}}} \oplus {\frac{3}{2}}^-_{q_1q_2q_3c\bar{c}} \, , \\
1^+_{q_1q_2}\otimes 1^+_{q_3c} \otimes {\frac{1}{2}}^-_{\bar{c}} &=& \left[0^+_{q_1q_2q_3c}\oplus 1^+_{q_1q_2q_3c}\oplus  2^+_{q_1q_2q_3c}\right]\otimes {\frac{1}{2}}^-_{\bar{c}} \nonumber \\
&=&\underline{{\frac{1}{2}}^-_{q_1q_2q_3c\bar{c}}} \oplus \left[{\frac{1}{2}}^-_{q_1q_2q_3c\bar{c}} \oplus {\frac{3}{2}}^-_{q_1q_2q_3c\bar{c}}\right]\oplus \left[ {\frac{3}{2}}^-_{q_1q_2q_3c\bar{c}}\oplus {\frac{5}{2}}^-_{q_1q_2q_3c\bar{c}}\right] \, , \\
1^+_{q_1q_2}\otimes 0^+_{q_3c} \otimes \left[1^-\otimes {\frac{1}{2}}^-_{\bar{c}}\right] &=& 1^+_{q_1q_2}\otimes 0^+_{q_3c} \otimes  \left[{\frac{1}{2}}^+_{\bar{c}}\oplus {\frac{3}{2}}^+_{\bar{c}}\right]  \nonumber \\
&=& \left[\underline{{\frac{1}{2}}^+_{q_1q_2q_3c\bar{c}}}\oplus {\frac{3}{2}}^+_{q_1q_2q_3c\bar{c}}\right]\oplus \left[ {\frac{1}{2}}^+_{q_1q_2q_3c\bar{c}}\oplus {\frac{3}{2}}^+_{q_1q_2q_3c\bar{c}} \oplus {\frac{5}{2}}^+_{q_1q_2q_3c\bar{c}} \right] \, ,  \\
1^+_{q_1q_2}\otimes 1^+_{q_3c} \otimes \left[1^-\otimes {\frac{1}{2}}^-_{\bar{c}} \right]&=& \left[0^+_{q_1q_2q_3c}\oplus 1^+_{q_1q_2q_3c}\oplus  2^+_{q_1q_2q_3c}\right]\otimes \left[{\frac{1}{2}}^+_{\bar{c}}\oplus {\frac{3}{2}}^+_{\bar{c}}\right]\nonumber \\
&=&\underline{{\frac{1}{2}}^+_{q_1q_2q_3c\bar{c}}} \oplus \left[{\frac{1}{2}}^+_{q_1q_2q_3c\bar{c}} \oplus {\frac{3}{2}}^+_{q_1q_2q_3c\bar{c}}\right]\oplus \left[ {\frac{3}{2}}^+_{q_1q_2q_3c\bar{c}}\oplus {\frac{5}{2}}^+_{q_1q_2q_3c\bar{c}}\right]  \nonumber\\
&&  \oplus {\frac{3}{2}}^+_{q_1q_2q_3c\bar{c}} \oplus\left[{\frac{1}{2}}^+_{q_1q_2q_3c\bar{c}} \oplus {\frac{3}{2}}^+_{q_1q_2q_3c\bar{c}}\oplus {\frac{5}{2}}^+_{q_1q_2q_3c\bar{c}}\right] \nonumber \\
&&\oplus\left[{\frac{1}{2}}^+_{q_1q_2q_3c\bar{c}} \oplus{\frac{3}{2}}^+_{q_1q_2q_3c\bar{c}}\oplus {\frac{5}{2}}^+_{q_1q_2q_3c\bar{c}}\oplus {\frac{7}{2}}^+_{q_1q_2q_3c\bar{c}}\right]\, ,
\end{eqnarray}
where the $1^-$ denotes the contribution of the additional P-wave to the spin-parity,  the subscripts $q_1q_2$, $q_3c$, $\cdots$ denote the quark   constituents. The quark and antiquark have opposite parity,   we usually take it for granted that the quarks have positive parity while the antiquarks have negative parity, so the $\bar{c}$-quark has $J^P={\frac{1}{2}}^-$.

 In this article, we study the  pentaquark  states with the  underlined spin-parity ${\frac{1}{2}}^{\pm}_{q_1q_2q_3c\bar{c}}$\, , which  are supposed to be the lowest pentaquark states with the light  axialvector diquarks.  We  construct   both the axialvector-diquark-axialvector-diquark-antiquark type and axialvector-diquark-scalar-diquark-antiquark type    currents $J(x)=J^{j_L j_H }_{q_1q_2 q_3}(x)$ according to Eqs.(2-3), where the superscripts $j_L$ and $j_H$ denote the spins of the light diquark and heavy diquark, respectively, the subscripts $q_1$, $q_2$, $q_3$ denote the light quark constituents. We calculate the vacuum condensate up to dimension-10 in the operator product expansion, and study the masses and pole residues of the lowest pentaquark states in a systematic way.

 The article is arranged as follows:
  we derive the QCD sum rules for the masses and pole residues of  the
$ {\frac{1}{2}}^{\pm}$ pentaquark states in Sect.2;  in Sect.3, we present the numerical results and discussions; and Sect.4 is reserved for our
conclusion.

\section{QCD sum rules for  the $ {\frac{1}{2}}^{\pm}$ pentaquark states}

We write down the two-point correlation functions $\Pi_{q_1q_2q_3}^{j_Lj_H}(p)$ in the QCD sum rules,
\begin{eqnarray}
\Pi_{q_1q_2q_3}^{j_Lj_H}(p)&=&i\int d^4x e^{ip \cdot x} \langle0|T\left\{J_{q_1q_2q_3}^{j_Lj_H}(x)\bar{J}_{q_1q_2q_3}^{j_Lj_H}(0)\right\}|0\rangle \, ,
\end{eqnarray}
where
\begin{eqnarray}
J^{11}_{uuu}(x)&=&\varepsilon^{ila} \varepsilon^{ijk}\varepsilon^{lmn}  u^T_j(x) C\gamma_\mu u_k(x)u^T_m(x) C\gamma^\mu c_n(x)  C\bar{c}^{T}_{a}(x) \, , \nonumber\\
J^{11}_{uud}(x)&=&\frac{\varepsilon^{ila} \varepsilon^{ijk}\varepsilon^{lmn}}{\sqrt{3}} \left[ u^T_j(x) C\gamma_\mu u_k(x)d^T_m(x) C\gamma^\mu c_n(x)+2u^T_j(x) C\gamma_\mu d_k(x)u^T_m(x) C\gamma^\mu c_n(x) \right] C\bar{c}^{T}_{a}(x) \, , \nonumber\\
J^{11}_{udd}(x)&=&\frac{\varepsilon^{ila} \varepsilon^{ijk}\varepsilon^{lmn}}{\sqrt{3}} \left[ d^T_j(x) C\gamma_\mu d_k(x)u^T_m(x) C\gamma^\mu c_n(x)+2d^T_j(x) C\gamma_\mu u_k(x)d^T_m(x) C\gamma^\mu c_n(x) \right] C\bar{c}^{T}_{a}(x) \, , \nonumber\\
J^{11}_{ddd}(x)&=&\varepsilon^{ila} \varepsilon^{ijk}\varepsilon^{lmn}  d^T_j(x) C\gamma_\mu d_k(x)d^T_m(x) C\gamma^\mu c_n(x)  C\bar{c}^{T}_{a}(x) \, ,
\end{eqnarray}

\begin{eqnarray}
 J^{11}_{uus}(x)&=&\frac{\varepsilon^{ila} \varepsilon^{ijk}\varepsilon^{lmn}}{\sqrt{3}} \left[ u^T_j(x) C\gamma_\mu u_k(x)s^T_m(x) C\gamma^\mu c_n(x)+2u^T_j(x) C\gamma_\mu s_k(x)u^T_m(x) C\gamma^\mu c_n(x) \right] C\bar{c}^{T}_{a}(x) \, , \nonumber\\
  J^{11}_{uds}(x)&=&\frac{\varepsilon^{ila} \varepsilon^{ijk}\varepsilon^{lmn}}{\sqrt{3}} \left[ u^T_j(x) C\gamma_\mu d_k(x)s^T_m(x) C\gamma^\mu c_n(x)+u^T_j(x) C\gamma_\mu s_k(x)d^T_m(x) C\gamma^\mu c_n(x) \right. \nonumber\\
 &&\left.+d^T_j(x) C\gamma_\mu s_k(x)u^T_m(x) C\gamma^\mu c_n(x) \right] C\bar{c}^{T}_{a}(x) \, , \nonumber\\
  J^{11}_{dds}(x)&=&\frac{\varepsilon^{ila} \varepsilon^{ijk}\varepsilon^{lmn}}{\sqrt{3}} \left[ d^T_j(x) C\gamma_\mu d_k(x)s^T_m(x) C\gamma^\mu c_n(x)+2d^T_j(x) C\gamma_\mu s_k(x)d^T_m(x) C\gamma^\mu c_n(x) \right] C\bar{c}^{T}_{a}(x) \, , \nonumber\\
 \end{eqnarray}

\begin{eqnarray}
  J^{11}_{uss}(x)&=&\frac{\varepsilon^{ila} \varepsilon^{ijk}\varepsilon^{lmn}}{\sqrt{3}} \left[ s^T_j(x) C\gamma_\mu s_k(x)u^T_m(x) C\gamma^\mu c_n(x)+2s^T_j(x) C\gamma_\mu u_k(x)s^T_m(x) C\gamma^\mu c_n(x) \right] C\bar{c}^{T}_{a}(x) \, , \nonumber\\
   J^{11}_{dss}(x)&=&\frac{\varepsilon^{ila} \varepsilon^{ijk}\varepsilon^{lmn}}{\sqrt{3}} \left[ s^T_j(x) C\gamma_\mu s_k(x)d^T_m(x) C\gamma^\mu c_n(x)+2s^T_j(x) C\gamma_\mu d_k(x)s^T_m(x) C\gamma^\mu c_n(x) \right] C\bar{c}^{T}_{a}(x) \, , \nonumber\\
   \end{eqnarray}

\begin{eqnarray}
  J^{11}_{sss}(x)&=&\varepsilon^{ila} \varepsilon^{ijk}\varepsilon^{lmn}  s^T_j(x) C\gamma_\mu s_k(x)s^T_m(x) C\gamma^\mu c_n(x)  C\bar{c}^{T}_{a}(x) \, ,
\end{eqnarray}

\begin{eqnarray}
J_{uuu}^{10}(x)&=&\frac{\varepsilon^{ila} \varepsilon^{ijk}\varepsilon^{lmn}}{\sqrt{3}}   u^T_j(x) C\gamma_\mu u_k(x) u^T_m(x) C\gamma_5 c_n(x) \gamma_5  \gamma^\mu C\bar{c}^{T}_{a}(x) \, , \nonumber\\
J^{10}_{uud}(x)&=&\frac{\varepsilon^{ila} \varepsilon^{ijk}\varepsilon^{lmn}}{\sqrt{3}} \left[ u^T_j(x) C\gamma_\mu u_k(x) d^T_m(x) C\gamma_5 c_n(x)+2u^T_j(x) C\gamma_\mu d_k(x) u^T_m(x) C\gamma_5 c_n(x)\right] \gamma_5 \gamma^\mu  C\bar{c}^{T}_{a}(x) \, ,  \nonumber\\
J^{10}_{udd}(x)&=&\frac{\varepsilon^{ila} \varepsilon^{ijk}\varepsilon^{lmn}}{\sqrt{3}} \left[ d^T_j(x) C\gamma_\mu d_k(x) u^T_m(x) C\gamma_5 c_n(x)+2d^T_j(x) C\gamma_\mu u_k(x) d^T_m(x) C\gamma_5 c_n(x)\right]  \gamma_5 \gamma^\mu C\bar{c}^{T}_{a}(x) \, , \nonumber \\
J_{ddd}^{10}(x)&=&\frac{\varepsilon^{ila} \varepsilon^{ijk}\varepsilon^{lmn}}{\sqrt{3}}   d^T_j(x) C\gamma_\mu d_k(x) d^T_m(x) C\gamma_5 c_n(x) \gamma_5  \gamma^\mu C\bar{c}^{T}_{a}(x) \, ,
\end{eqnarray}

\begin{eqnarray}
J^{10}_{uus}(x)&=&\frac{\varepsilon^{ila} \varepsilon^{ijk}\varepsilon^{lmn}}{\sqrt{3}} \left[ u^T_j(x) C\gamma_\mu u_k(x) s^T_m(x) C\gamma_5 c_n(x)+2u^T_j(x) C\gamma_\mu s_k(x) u^T_m(x) C\gamma_5 c_n(x)\right] \gamma_5 \gamma^\mu  C\bar{c}^{T}_{a}(x) \, ,  \nonumber\\
J^{10}_{uds}(x)&=&\frac{\varepsilon^{ila} \varepsilon^{ijk}\varepsilon^{lmn}}{\sqrt{3}} \left[ u^T_j(x) C\gamma_\mu d_k(x) s^T_m(x) C\gamma_5 c_n(x)+u^T_j(x) C\gamma_\mu s_k(x) d^T_m(x) C\gamma_5 c_n(x)\right.  \nonumber\\
&&\left.+ d^T_j(x) C\gamma_\mu s_k(x) u^T_m(x) C\gamma_5 c_n(x)\right] \gamma_5 \gamma^\mu  C\bar{c}^{T}_{a}(x) \, ,\nonumber\\
J^{10}_{dds}(x)&=&\frac{\varepsilon^{ila} \varepsilon^{ijk}\varepsilon^{lmn}}{\sqrt{3}} \left[ d^T_j(x) C\gamma_\mu d_k(x) s^T_m(x) C\gamma_5 c_n(x)+2d^T_j(x) C\gamma_\mu s_k(x) d^T_m(x) C\gamma_5 c_n(x)\right] \gamma_5 \gamma^\mu  C\bar{c}^{T}_{a}(x) \, ,  \nonumber\\
\end{eqnarray}

\begin{eqnarray}
J^{10}_{uss}(x)&=&\frac{\varepsilon^{ila} \varepsilon^{ijk}\varepsilon^{lmn}}{\sqrt{3}} \left[ s^T_j(x) C\gamma_\mu s_k(x) u^T_m(x) C\gamma_5 c_n(x)+2s^T_j(x) C\gamma_\mu u_k(x) s^T_m(x) C\gamma_5 c_n(x)\right]  \gamma_5 \gamma^\mu C\bar{c}^{T}_{a}(x) \, , \nonumber \\
J^{10}_{dss}(x)&=&\frac{\varepsilon^{ila} \varepsilon^{ijk}\varepsilon^{lmn}}{\sqrt{3}} \left[ s^T_j(x) C\gamma_\mu s_k(x) d^T_m(x) C\gamma_5 c_n(x)+2s^T_j(x) C\gamma_\mu d_k(x) s^T_m(x) C\gamma_5 c_n(x)\right]  \gamma_5 \gamma^\mu C\bar{c}^{T}_{a}(x) \, , \nonumber \\
\end{eqnarray}

\begin{eqnarray}
J^{10}_{sss}(x)&=&\frac{\varepsilon^{ila} \varepsilon^{ijk}\varepsilon^{lmn}}{\sqrt{3}}   s^T_j(x) C\gamma_\mu s_k(x) s^T_m(x) C\gamma_5 c_n(x) \gamma_5  \gamma^\mu C\bar{c}^{T}_{a}(x) \, ,
 \end{eqnarray}
the $i$, $j$, $k$, $l$, $m$, $n$ and $a$ are color indices, the $C$ is the charge conjugation matrix. In the currents $J^{10}_{q_1q_2q_3}(x)$, the light axialvector diquark combines with the heavy scalar diquark to form a tetraquark  with $J^P=1^+$ in color triplet according to Eq.(2), while in the currents $J^{11}_{q_1q_2q_3}(x)$, the light axialvector diquark combines with the heavy axialvector diquark to form a tetraquark  with $J^P=0^+$ in color triplet according to Eq.(3). Then they couple with the antiquark to form pentaquark states with $J^P={\frac{1}{2}}^-$ in color singlet.    In this article, we take the isospin limit,
and classify the currents  couple to the pentaquark states with degenerate masses  into the following 8 types,
\begin{eqnarray}
&&J^{11}_{uuu,\mu}(x)\, , \, \, \, J^{11}_{uud,\mu}(x)\, , \, \, \, J^{11}_{udd,\mu}(x)\, , \, \, \,J^{11}_{ddd,\mu}(x) \, ; \nonumber\\
&&J^{11}_{uus,\mu}(x)\, , \, \, \,  J^{11}_{uds,\mu}(x)\, , \, \, \,  J^{11}_{dds,\mu}(x) \, ; \nonumber\\
&&  J^{11}_{uss,\mu}(x)\, , \, \, \,   J^{11}_{dss,\mu}(x)\, ; \nonumber\\
&&  J^{11}_{sss,\mu}(x) \, ; \nonumber\\
&& J_{uuu,\mu}^{10}(x)\, , \, \, \, J^{10}_{uud,\mu}(x)\, , \, \, \, J^{10}_{udd,\mu}(x)\, , \, \, \,J_{ddd,\mu}^{10}(x)\, ; \nonumber\\
&&J^{10}_{uus,\mu}(x)\, , \, \, \, J^{10}_{uds,\mu}(x)\, , \, \, \, J^{10}_{dds,\mu}(x) \, ; \nonumber\\
&&J^{10}_{uss,\mu}(x)\, , \, \, \, J^{10}_{dss,\mu}(x) \, ; \nonumber \\
&&J^{10}_{sss,\mu}(x) \, .
 \end{eqnarray}
In calculations, we choose the first current in each type.

The currents $J_{q_1q_2q_3}^{j_Lj_H}(0)$ have negative parity, and    couple potentially to the ${\frac{1}{2}}^-$   hidden-charm  pentaquark  states $P_{q_1q_2q_3}^{j_Lj_H \frac{1}{2}}( {\frac{1}{2}}^-)$, where the superscript $\frac{1}{2}$ denotes the total angular momentum  of  the anti-$c$-quark $\bar{c}$,
\begin{eqnarray}
\langle 0| J_{q_1q_2q_3}^{j_Lj_H}(0)|P_{q_1q_2q_3}^{j_Lj_H \frac{1}{2}}( {\frac{1}{2}}^-)(p)\rangle &=&\lambda^{-}_{P} U^{-}(p,s) \, ,
\end{eqnarray}
the $\lambda^{-}_P$ are the pole residues, the spinors $U^{-}(p,s)$   satisfy the Dirac equations $(\not\!\!p-M_{P,-})U^{-}(p,s)=0$, the $s$ are the polarization  vectors of the pentaquark states.   On the other hand, the currents $J_{q_1q_2q_3}^{j_Lj_H}(0)$  also couple potentially to the ${\frac{1}{2}}^+$   hidden-charm  pentaquark states $P_{q_1q_2q_3}^{j_Lj_H \frac{1}{2}}( {\frac{1}{2}}^+)$ as multiplying $i \gamma_{5}$ to the currents $J_{q_1q_2q_3}^{j_Lj_H}(x)$ change their parity,
 \begin{eqnarray}
\langle 0| J_{q_1q_2q_3}^{j_Lj_H}(0)|P_{q_1q_2q_3}^{j_Lj_H \frac{1}{2}}( {\frac{1}{2}}^+)(p)\rangle &=&\lambda^{+}_{P}i\gamma_5 U^{+}(p,s) \, ,
\end{eqnarray}
the spinors $U^{\pm}(p,s)$ (pole residues $\lambda_i^\pm$)   have analogous  properties \cite{Chung82,Bagan93,Oka96,WangHbaryon}. We can study the $J^P={\frac{1}{2}}^+$   hidden-charm  pentaquark  states without introducing the additional P-wave explicitly, see Eqs.(4-5).

 We  insert  a complete set  of intermediate pentaquark states with the
same quantum numbers as the current operators $J(x)$   and $i\gamma_5 J(x)$ into the correlation functions
$\Pi_{q_1q_2q_3}^{j_Lj_H}(p)$   to obtain the hadronic representation
\cite{SVZ79,PRT85}. After isolating the pole terms of the lowest
states of the hidden-charm  pentaquark states, we obtain the
following results:
\begin{eqnarray}
  \Pi_{q_1q_2q_3}^{j_Lj_H}(p) & = &  {\lambda^{-}_{P}}^2  {\!\not\!{p}+ M_{P,-} \over M_{P,-}^{2}-p^{2}  } +{\lambda^{+}_{P}}^2  {\!\not\!{p}- M_{P,+} \over M_{P,+}^{2}-p^{2}  }      +\cdots \, ,
    \end{eqnarray}
where the $M_{\pm}$ are the masses of the lowest pentaquark states with the
 parity $\pm$,  respectively.

Now we obtain the hadronic spectral densities  through the dispersion relation,
\begin{eqnarray}
\frac{{\rm Im}\Pi_{q_1q_2q_3}^{j_Lj_H}(s)}{\pi}&=&\!\not\!{p} \left[{\lambda^{-}_{P}}^2 \delta\left(s-M_{P,-}^2\right)+{\lambda^{+}_{P}}^2 \delta\left(s-M_{P,+}^2\right)\right] \nonumber\\
&& +\left[M_{P,-}{\lambda^{-}_{P}}^2 \delta\left(s-M_{P,-}^2\right)-M_{P,+}{\lambda^{+}_{P}}^2 \delta\left(s-M_{P,+}^2\right)\right]\, , \nonumber\\
&=&\!\not\!{p} \,\rho_{q_1q_2q_3,H}^{j_Lj_H,1}(s)+\rho_{q_1q_2q_3,H}^{j_Lj_H,0}(s) \, ,
\end{eqnarray}
where the subscript index $H$ denotes the hadron side,  then we introduce the weight function $\exp\left(-\frac{s}{T^2}\right)$ to obtain the QCD sum rules at the hadron  side,
\begin{eqnarray}
\int_{4m_c^2}^{s_0}ds \left[\sqrt{s}\rho_{q_1q_2q_3,H}^{j_Lj_H,1}(s)+\rho_{q_1q_2q_3,H}^{j_Lj_H,0}(s)\right]\exp\left( -\frac{s}{T^2}\right)
&=&2M_{P,-}{\lambda^{-}_{P}}^2\exp\left( -\frac{M_{P,-}^2}{T^2}\right) \, , \\
\int_{4m_c^2}^{s_0}ds \left[\sqrt{s}\rho_{q_1q_2q_3,H}^{j_Lj_H,1}(s)-\rho_{q_1q_2q_3,H}^{j_Lj_H,0}(s)\right]\exp\left( -\frac{s}{T^2}\right)
&=&2M_{P,+}{\lambda^{+}_{P}}^2\exp\left( -\frac{M_{P,+}^2}{T^2}\right) \, ,
\end{eqnarray}
where the $s_0$ are the continuum threshold parameters and the $T^2$ are the Borel parameters. We separate the contributions of   the
negative-parity (positive-parity) pentaquark states from the positive-parity (negative-parity) pentaquark states explicitly.

In the following, we briefly outline  the operator product expansion for the correlation functions $\Pi_{q_1q_2q_3}^{j_Lj_H}(p)$   in perturbative QCD. Firstly,  we contract the $u$, $s$ and $c$ quark fields in the correlation functions
$\Pi_{q_1q_2q_3}^{j_Lj_H}(p)$    with Wick theorem, and obtain the results:
\begin{eqnarray}
\Pi_{uuu}^{11}(p)&=&-i\,\varepsilon_{ila}\varepsilon_{ijk}\varepsilon_{lmn}\varepsilon_{i^{\prime}l^{\prime}a^{\prime}}\varepsilon_{i^{\prime}j^{\prime}k^{\prime}}
\varepsilon_{l^{\prime}m^{\prime}n^{\prime}}\int d^4x e^{ip\cdot x}C C_{a^{\prime}a}^T(-x)C \nonumber\\
&&\left\{ 2  Tr\left[\gamma_\mu U_{kk^\prime}(x) \gamma_\nu C U^{T}_{jj^\prime}(x)C\right] \,Tr\left[\gamma^\mu C_{nn^\prime}(x) \gamma^\nu C U^{T}_{mm^\prime}(x)C\right] \right. \nonumber\\
&&\left.-4  Tr \left[\gamma_\mu U_{kk^\prime}(x) \gamma_\nu C U^{T}_{mj^\prime}(x)C \gamma^\mu C_{nn^\prime}(x) \gamma^\nu C U^{T}_{jm^\prime}(x)C\right]  \right\} \, ,
\end{eqnarray}

\begin{eqnarray}
\Pi_{uus}^{11}(p)&=&-\frac{i}{3}\,\varepsilon_{ila}\varepsilon_{ijk}\varepsilon_{lmn}\varepsilon_{i^{\prime}l^{\prime}a^{\prime}}\varepsilon_{i^{\prime}j^{\prime}k^{\prime}}
\varepsilon_{l^{\prime}m^{\prime}n^{\prime}}\int d^4x e^{ip\cdot x} C C_{a^{\prime}a}^T(-x)C\nonumber\\
&&\left\{ 2  Tr\left[\gamma_\mu U_{kk^\prime}(x) \gamma_\nu C U^{T}_{jj^\prime}(x)C\right] \,Tr\left[\gamma^\mu C_{nn^\prime}(x) \gamma^\nu C S^{T}_{mm^\prime}(x)C\right] \right. \nonumber\\
&& +4 Tr\left[\gamma_\mu S_{kk^\prime}(x) \gamma_\nu C U^{T}_{jj^\prime}(x)C\right] \,Tr\left[\gamma^\mu C_{nn^\prime}(x) \gamma^\nu C U^{T}_{mm^\prime}(x)C\right]  \nonumber\\
&&-4 Tr \left[\gamma_\mu U_{kk^\prime}(x) \gamma_\nu C U^{T}_{mj^\prime}(x)C \gamma^\mu C_{nn^\prime}(x) \gamma^\nu C S^{T}_{jm^\prime}(x)C\right]  \nonumber\\
 &&-4 Tr \left[\gamma_\mu U_{kk^\prime}(x) \gamma_\nu C S^{T}_{mj^\prime}(x)C \gamma^\mu C_{nn^\prime}(x) \gamma^\nu C U^{T}_{jm^\prime}(x)C\right]  \nonumber\\
 &&\left.-4 Tr \left[\gamma_\mu S_{kk^\prime}(x) \gamma_\nu C U^{T}_{mj^\prime}(x)C \gamma^\mu C_{nn^\prime}(x) \gamma^\nu C U^{T}_{jm^\prime}(x)C\right] \right\} \, ,
 \end{eqnarray}

 \begin{eqnarray}
\Pi_{uss}^{11}(p)&=&-\frac{i}{3}\,\varepsilon_{ila}\varepsilon_{ijk}\varepsilon_{lmn}\varepsilon_{i^{\prime}l^{\prime}a^{\prime}}\varepsilon_{i^{\prime}j^{\prime}k^{\prime}}
\varepsilon_{l^{\prime}m^{\prime}n^{\prime}}\int d^4x e^{ip\cdot x}C C_{a^{\prime}a}^T(-x)C \nonumber\\
&&\left\{ 2  Tr\left[\gamma_\mu S_{kk^\prime}(x) \gamma_\nu C S^{T}_{jj^\prime}(x)C\right] \,Tr\left[\gamma^\mu C_{nn^\prime}(x) \gamma^\nu C U^{T}_{mm^\prime}(x)C\right] \right. \nonumber\\
&& +4 Tr\left[\gamma_\mu U_{kk^\prime}(x) \gamma_\nu C S^{T}_{jj^\prime}(x)C\right] \,Tr\left[\gamma^\mu C_{nn^\prime}(x) \gamma^\nu C S^{T}_{mm^\prime}(x)C\right]
   \nonumber\\
&&-4 Tr \left[\gamma_\mu S_{kk^\prime}(x) \gamma_\nu C S^{T}_{mj^\prime}(x)C \gamma^\mu C_{nn^\prime}(x) \gamma^\nu C U^{T}_{jm^\prime}(x)C\right] \nonumber\\
 &&-4  Tr \left[\gamma_\mu S_{kk^\prime}(x) \gamma_\nu C U^{T}_{mj^\prime}(x)C \gamma^\mu C_{nn^\prime}(x) \gamma^\nu C S^{T}_{jm^\prime}(x)C\right]  \nonumber\\
 &&\left.-4  Tr \left[\gamma_\mu U_{kk^\prime}(x) \gamma_\nu C S^{T}_{mj^\prime}(x)C \gamma^\mu C_{nn^\prime}(x) \gamma^\nu C S^{T}_{jm^\prime}(x)C\right] \right\}\, ,
 \end{eqnarray}

 \begin{eqnarray}
\Pi_{sss}^{11}(p)&=&-i\,\varepsilon_{ila}\varepsilon_{ijk}\varepsilon_{lmn}\varepsilon_{i^{\prime}l^{\prime}a^{\prime}}\varepsilon_{i^{\prime}j^{\prime}k^{\prime}}
\varepsilon_{l^{\prime}m^{\prime}n^{\prime}}\int d^4x e^{ip\cdot x}C C_{a^{\prime}a}^T(-x)C \nonumber\\
&&\left\{ 2  Tr\left[\gamma_\mu S_{kk^\prime}(x) \gamma_\nu C S^{T}_{jj^\prime}(x)C\right] \,Tr\left[\gamma^\mu C_{nn^\prime}(x) \gamma^\nu C S^{T}_{mm^\prime}(x)C\right] \right. \nonumber\\
&&\left.-4  Tr \left[\gamma_\mu S_{kk^\prime}(x) \gamma_\nu C S^{T}_{mj^\prime}(x)C \gamma^\mu C_{nn^\prime}(x) \gamma^\nu C S^{T}_{jm^\prime}(x)C\right] \right\} \, ,
\end{eqnarray}

\begin{eqnarray}
\Pi_{uuu}^{10}(p)&=&-i\,\varepsilon_{ila}\varepsilon_{ijk}\varepsilon_{lmn}\varepsilon_{i^{\prime}l^{\prime}a^{\prime}}\varepsilon_{i^{\prime}j^{\prime}k^{\prime}}
\varepsilon_{l^{\prime}m^{\prime}n^{\prime}}\int d^4x e^{ip\cdot x}\gamma_5\gamma^\mu C C_{a^{\prime}a}^T(-x)C\gamma^\nu\gamma_5\nonumber\\
&&\left\{ 2  Tr\left[\gamma_\mu U_{kk^\prime}(x) \gamma_\nu C U^{T}_{jj^\prime}(x)C\right] \,Tr\left[\gamma_5 C_{nn^\prime}(x) \gamma_5 C U^{T}_{mm^\prime}(x)C\right]\gamma_5\gamma^\mu  \right. \nonumber\\
&&\left.-4  Tr \left[\gamma_\mu U_{kk^\prime}(x) \gamma_\nu C U^{T}_{mj^\prime}(x)C \gamma_5 C_{nn^\prime}(x) \gamma_5 C U^{T}_{jm^\prime}(x)C\right]
  \right\} \, ,
\end{eqnarray}

\begin{eqnarray}
\Pi_{uus}^{10}(p)&=&-\frac{i}{3}\,\varepsilon_{ila}\varepsilon_{ijk}\varepsilon_{lmn}\varepsilon_{i^{\prime}l^{\prime}a^{\prime}}\varepsilon_{i^{\prime}j^{\prime}k^{\prime}}
\varepsilon_{l^{\prime}m^{\prime}n^{\prime}}\int d^4x e^{ip\cdot x}\gamma_5\gamma^\mu C C_{a^{\prime}a}^T(-x)C\gamma^\nu\gamma_5 \nonumber\\
&&\left\{ 2  Tr\left[\gamma_\mu U_{kk^\prime}(x) \gamma_\nu C U^{T}_{jj^\prime}(x)C\right] \,Tr\left[\gamma_5 C_{nn^\prime}(x) \gamma_5 C S^{T}_{mm^\prime}(x)C\right] \right. \nonumber\\
&& +4 Tr\left[\gamma_\mu S_{kk^\prime}(x) \gamma_\nu C U^{T}_{jj^\prime}(x)C\right] \,Tr\left[\gamma_5 C_{nn^\prime}(x) \gamma_5 C U^{T}_{mm^\prime}(x)C\right]
    \nonumber\\
&&-4 Tr\left[\gamma_\mu U_{kk^\prime}(x) \gamma_\nu C U^{T}_{mj^\prime}(x)C \gamma_5 C_{nn^\prime}(x) \gamma_5 C S^{T}_{jm^\prime}(x)C\right] \nonumber\\
 &&-4  Tr \left[\gamma_\mu U_{kk^\prime}(x) \gamma_\nu C S^{T}_{mj^\prime}(x)C \gamma_5 C_{nn^\prime}(x) \gamma_5 C U^{T}_{jm^\prime}(x)C\right]  \nonumber\\
 &&\left.-4  Tr \left[\gamma_\mu S_{kk^\prime}(x) \gamma_\nu C U^{T}_{mj^\prime}(x)C \gamma_5 C_{nn^\prime}(x) \gamma_5 C U^{T}_{jm^\prime}(x)C\right] \right\}\, ,
\end{eqnarray}

\begin{eqnarray}
\Pi_{uss}^{10}(p)&=&-\frac{i}{3}\,\varepsilon_{ila}\varepsilon_{ijk}\varepsilon_{lmn}\varepsilon_{i^{\prime}l^{\prime}a^{\prime}}\varepsilon_{i^{\prime}j^{\prime}k^{\prime}}
\varepsilon_{l^{\prime}m^{\prime}n^{\prime}}\int d^4x e^{ip\cdot x} \gamma_5\gamma^\mu C C_{a^{\prime}a}^T(-x)C\gamma^\nu\gamma_5\nonumber\\
&&\left\{ 2  Tr\left[\gamma_\mu S_{kk^\prime}(x) \gamma_\nu C S^{T}_{jj^\prime}(x)C\right] \,Tr\left[\gamma_5 C_{nn^\prime}(x) \gamma_5 C U^{T}_{mm^\prime}(x)C\right] \right. \nonumber\\
&& +4  Tr\left[\gamma_\mu U_{kk^\prime}(x) \gamma_\nu C S^{T}_{jj^\prime}(x)C\right] \,Tr\left[\gamma_5 C_{nn^\prime}(x) \gamma_5 C S^{T}_{mm^\prime}(x)C\right]   \nonumber\\
&&-4  Tr \left[\gamma_\mu S_{kk^\prime}(x) \gamma_\nu C S^{T}_{mj^\prime}(x)C \gamma_5 C_{nn^\prime}(x) \gamma_5 C U^{T}_{jm^\prime}(x)C\right]\nonumber\\
 &&-4  Tr \left[\gamma_\mu S_{kk^\prime}(x) \gamma_\nu C U^{T}_{mj^\prime}(x)C \gamma_5 C_{nn^\prime}(x) \gamma_5 C S^{T}_{jm^\prime}(x)C\right] \nonumber\\
 &&\left.-4  Tr \left[\gamma_\mu U_{kk^\prime}(x) \gamma_\nu C S^{T}_{mj^\prime}(x)C \gamma_5 C_{nn^\prime}(x) \gamma_5 C S^{T}_{jm^\prime}(x)C\right]  \right\}\, ,
\end{eqnarray}

\begin{eqnarray}
\Pi_{sss}^{10}(p)&=&-i\,\varepsilon_{ila}\varepsilon_{ijk}\varepsilon_{lmn}\varepsilon_{i^{\prime}l^{\prime}a^{\prime}}\varepsilon_{i^{\prime}j^{\prime}k^{\prime}}
\varepsilon_{l^{\prime}m^{\prime}n^{\prime}}\int d^4x e^{ip\cdot x}\gamma_5\gamma^\mu C C_{a^{\prime}a}^T(-x)C\gamma^\nu\gamma_5 \nonumber\\
&&\left\{ 2  Tr\left[\gamma_\mu S_{kk^\prime}(x) \gamma_\nu C S^{T}_{jj^\prime}(x)C\right] \,Tr\left[\gamma_5 C_{nn^\prime}(x) \gamma_5 C S^{T}_{mm^\prime}(x)C\right] \right. \nonumber\\
&&\left.-4  Tr \left[\gamma_\mu S_{kk^\prime}(x) \gamma_\nu C S^{T}_{mj^\prime}(x)C \gamma_5 C_{nn^\prime}(x) \gamma_5 C S^{T}_{jm^\prime}(x)C\right] \right\} \, ,
\end{eqnarray}
where
the $U_{ij}(x)$, $S_{ij}(x)$ and $C_{ij}(x)$ are the full $u$, $s$ and $c$ quark propagators respectively,
 \begin{eqnarray}
U_{ij}(x)&=& \frac{i\delta_{ij}\!\not\!{x}}{ 2\pi^2x^4}-\frac{\delta_{ij}\langle
\bar{q}q\rangle}{12} -\frac{\delta_{ij}x^2\langle \bar{q}g_s\sigma Gq\rangle}{192} -\frac{ig_sG^{a}_{\alpha\beta}t^a_{ij}(\!\not\!{x}
\sigma^{\alpha\beta}+\sigma^{\alpha\beta} \!\not\!{x})}{32\pi^2x^2}    -\frac{1}{8}\langle\bar{q}_j\sigma^{\mu\nu}q_i \rangle \sigma_{\mu\nu}+\cdots \, , \nonumber\\
S_{ij}(x)&=& \frac{i\delta_{ij}\!\not\!{x}}{ 2\pi^2x^4}
-\frac{\delta_{ij}m_s}{4\pi^2x^2}-\frac{\delta_{ij}\langle
\bar{s}s\rangle}{12} +\frac{i\delta_{ij}\!\not\!{x}m_s
\langle\bar{s}s\rangle}{48}-\frac{\delta_{ij}x^2\langle \bar{s}g_s\sigma Gs\rangle}{192}+\frac{i\delta_{ij}x^2\!\not\!{x} m_s\langle \bar{s}g_s\sigma
 Gs\rangle }{1152}\nonumber\\
&& -\frac{ig_s G^{a}_{\alpha\beta}t^a_{ij}(\!\not\!{x}
\sigma^{\alpha\beta}+\sigma^{\alpha\beta} \!\not\!{x})}{32\pi^2x^2}  -\frac{1}{8}\langle\bar{s}_j\sigma^{\mu\nu}s_i \rangle \sigma_{\mu\nu}+\cdots \, ,
\end{eqnarray}

\begin{eqnarray}
C_{ij}(x)&=&\frac{i}{(2\pi)^4}\int d^4k e^{-ik \cdot x} \left\{
\frac{\delta_{ij}}{\!\not\!{k}-m_c}
-\frac{g_sG^n_{\alpha\beta}t^n_{ij}}{4}\frac{\sigma^{\alpha\beta}(\!\not\!{k}+m_c)+(\!\not\!{k}+m_c)
\sigma^{\alpha\beta}}{(k^2-m_c^2)^2}\right.\nonumber\\
&&\left. -\frac{g_s^2 (t^at^b)_{ij} G^a_{\alpha\beta}G^b_{\mu\nu}(f^{\alpha\beta\mu\nu}+f^{\alpha\mu\beta\nu}+f^{\alpha\mu\nu\beta}) }{4(k^2-m_c^2)^5}+\cdots\right\} \, ,\nonumber\\
f^{\alpha\beta\mu\nu}&=&(\!\not\!{k}+m_c)\gamma^\alpha(\!\not\!{k}+m_c)\gamma^\beta(\!\not\!{k}+m_c)\gamma^\mu(\!\not\!{k}+m_c)\gamma^\nu(\!\not\!{k}+m_c)\, ,
\end{eqnarray}
and  $t^n=\frac{\lambda^n}{2}$, the $\lambda^n$ is the Gell-Mann matrix   \cite{PRT85}, then compute  the integrals both in the coordinate and momentum spaces to obtain the correlation functions $\Pi_{q_1q_2q_3}^{j_Lj_H}(p)$,   therefore the QCD spectral densities $\rho_{q_1q_2q_3}^{j_Lj_H,1}(s)$ and $\widetilde{\rho}_{q_1q_2q_3}^{j_Lj_H,0}(s)$ at the quark level through the dispersion  relation,
\begin{eqnarray}
\frac{{\rm Im}\Pi_{q_1q_2q_3}^{j_Lj_H}(s)}{\pi}&=&\!\not\!{p} \,\rho_{q_1q_2q_3}^{j_Lj_H,1}(s)+m_c\widetilde{\rho}_{q_1q_2q_3}^{j_Lj_H,0}(s) \, .
\end{eqnarray}
The explicit expressions of the $\rho_{q_1q_2q_3}^{j_Lj_H,1}(s)$ and $\widetilde{\rho}_{q_1q_2q_3}^{j_Lj_H,0}(s)$   are given in the appendix.
In Eq.(30), we retain the term $\langle\bar{q}_j\sigma_{\mu\nu}q_i \rangle$ ($\langle\bar{s}_j\sigma_{\mu\nu}s_i \rangle$)  comes from the Fierz re-arrangement of the $\langle q_i \bar{q}_j\rangle$ ($\langle s_i \bar{s}_j\rangle$) to  absorb the gluons  emitted from other  quark lines to form $\langle\bar{q}_j g_s G^a_{\alpha\beta} t^a_{mn}\sigma_{\mu\nu} q_i \rangle$ ($\langle\bar{s}_j g_s G^a_{\alpha\beta} t^a_{mn}\sigma_{\mu\nu} s_i \rangle$)  to extract the mixed condensate  $\langle\bar{q}g_s\sigma G q\rangle$ ($\langle\bar{s}g_s\sigma G s\rangle$). A number of terms involving the mixed condensates  $\langle\bar{q}g_s\sigma G q\rangle$ and $\langle\bar{s}g_s\sigma G s\rangle$ appear and play an important role in the QCD sum rules.

 Once the analytical QCD spectral densities $\rho_{q_1q_2q_3}^{j_Lj_H,1}(s)$ and $\widetilde{\rho}_{q_1q_2q_3}^{j_Lj_H,0}(s)$ are obtained,  we can take the
quark-hadron duality below the continuum thresholds  $s_0$ and introduce the weight function $\exp\left(-\frac{s}{T^2}\right)$ to obtain  the following QCD sum rules:
\begin{eqnarray}
2M_{P,-}{\lambda^{-}_{P}}^2\exp\left( -\frac{M_{P,-}^2}{T^2}\right)
&=& \int_{4m_c^2}^{s_0}ds \left[\sqrt{s}\rho_{q_1q_2q_3}^{j_Lj_H,1}(s)+m_c\widetilde{\rho}_{q_1q_2q_3}^{j_Lj_H,0}(s)\right]\exp\left( -\frac{s}{T^2}\right)\, ,\\
2M_{P,+}{\lambda^{+}_{P}}^2\exp\left( -\frac{M_{P,+}^2}{T^2}\right)
&=& \int_{4m_c^2}^{s_0}ds \left[\sqrt{s}\rho_{q_1q_2q_3}^{j_Lj_H,1}(s)-m_c\widetilde{\rho}_{q_1q_2q_3}^{j_Lj_H,0}(s)\right]\exp\left( -\frac{s}{T^2}\right)\, ,
\end{eqnarray}
where we take into account the contributions of the terms $D_0$, $D_3$, $D_5$, $D_6$, $D_8$, $D_9$ and $D_{10}$,
\begin{eqnarray}
D_0&=& {\rm perturbative \,\,\,\, terms}\, , \nonumber\\
D_3&\propto& \langle \bar{q}q\rangle\, , \,\langle \bar{s}s\rangle\, ,  \nonumber\\
D_5&\propto& \langle \bar{q}g_s\sigma Gq\rangle\, , \,\langle \bar{s}g_s\sigma Gs\rangle \, , \nonumber\\
D_6&\propto& \langle \bar{q}q\rangle^2\, , \, \langle \bar{q}q\rangle \langle \bar{s}s\rangle\, , \,\langle \bar{s}s\rangle^2  \, , \nonumber\\
D_8&\propto& \langle\bar{q}q\rangle\langle \bar{q}g_s\sigma Gq\rangle\, , \,\langle\bar{s}s\rangle\langle \bar{q}g_s\sigma Gq\rangle\, , \,\langle\bar{q}q\rangle\langle \bar{s}g_s\sigma Gs\rangle\, , \, \langle\bar{s}s\rangle\langle \bar{s}g_s\sigma Gs\rangle\, ,   \nonumber\\
D_9&\propto& \langle \bar{q}q\rangle^3\, , \,\langle \bar{q}q\rangle\langle \bar{s}s\rangle^2\, , \, \langle \bar{q}q\rangle^2 \langle \bar{s}s\rangle\, , \, \langle \bar{s}s\rangle^3\, ,  \nonumber\\
D_{10}&\propto& \langle \bar{q}g_s\sigma Gq\rangle^2\, , \, \langle \bar{q}g_s\sigma Gq\rangle \langle \bar{s}g_s\sigma Gs\rangle\, , \,\langle \bar{s}g_s\sigma Gs\rangle^2  \, .
\end{eqnarray}
In this article, we carry out the
operator product expansion to the vacuum condensates  up to dimension-10, and
assume vacuum saturation for the  higher dimension vacuum condensates.

We differentiate   Eqs.(33-34) with respect to  $\frac{1}{T^2}$, then eliminate the
 pole residues $\lambda^{\pm}_{P}$ and obtain the QCD sum rules for
 the masses of the pentaquark states,
 \begin{eqnarray}
 M^2_{P,-} &=& \frac{\int_{4m_c^2}^{s_0}ds s\left[\sqrt{s}\rho_{q_1q_2q_3}^{j_Lj_H,1}(s)+m_c\widetilde{\rho}_{q_1q_2q_3}^{j_Lj_H,0}(s)\right]\exp\left( -\frac{s}{T^2}\right)}{\int_{4m_c^2}^{s_0}ds \left[\sqrt{s}\rho_{q_1q_2q_3}^{j_Lj_H,1}(s)+m_c\widetilde{\rho}_{q_1q_2q_3}^{j_Lj_H,0}(s)\right]\exp\left( -\frac{s}{T^2}\right)}\, ,\\
 M^2_{P,+} &=& \frac{\int_{4m_c^2}^{s_0}ds s\left[\sqrt{s}\rho_{q_1q_2q_3}^{j_Lj_H,1}(s)-m_c\widetilde{\rho}_{q_1q_2q_3}^{j_Lj_H,0}(s)\right]\exp\left( -\frac{s}{T^2}\right)}{\int_{4m_c^2}^{s_0}ds \left[\sqrt{s}\rho_{q_1q_2q_3}^{j_Lj_H,1}(s)-m_c\widetilde{\rho}_{q_1q_2q_3}^{j_Lj_H,0}(s)\right]\exp\left( -\frac{s}{T^2}\right)}\, .
\end{eqnarray}
Once the masses $M_{P,\pm}$ are obtained, we can take them as input parameters and obtain the pole residues from the QCD sum rules in Eqs.(33-34).
The   gluon condensates  are associated with  large numerical denominators, their contributions to total QCD spectral densities are  less (or much less) than the contributions of the dimension 10 vacuum condensates  $D_{10}$ for the pentaquark currents with the axialvector diquark constituents \cite{Wang1508,WangHuang1508}. We obtain the masses through  fractions, see Eqs.(36-37), the effects of the gluon condensates can be safely absorbed into the pole residues $\lambda^{\pm}_{P}$ and tiny effects on the masses can be  safely neglected.

\section{Numerical results and discussions}
The vacuum condensates are taken to be the standard values
$\langle\bar{q}q \rangle=-(0.24\pm 0.01\, \rm{GeV})^3$,  $\langle\bar{s}s \rangle=(0.8\pm0.1)\langle\bar{q}q \rangle$,
$\langle\bar{q}g_s\sigma G q \rangle=m_0^2\langle \bar{q}q \rangle$, $\langle\bar{s}g_s\sigma G s \rangle=m_0^2\langle \bar{s}s \rangle$,
$m_0^2=(0.8 \pm 0.1)\,\rm{GeV}^2$    at the energy scale  $\mu=1\, \rm{GeV}$
\cite{SVZ79,PRT85}.
The quark condensates and mixed quark condensates  evolve with the   renormalization group equation,
$\langle\bar{q}q \rangle(\mu)=\langle\bar{q}q \rangle(Q)\left[\frac{\alpha_{s}(Q)}{\alpha_{s}(\mu)}\right]^{\frac{4}{9}}$, $\langle\bar{s}s \rangle(\mu)=\langle\bar{s}s \rangle(Q)\left[\frac{\alpha_{s}(Q)}{\alpha_{s}(\mu)}\right]^{\frac{4}{9}}$,
 $\langle\bar{q}g_s \sigma Gq \rangle(\mu)=\langle\bar{q}g_s \sigma Gq \rangle(Q)\left[\frac{\alpha_{s}(Q)}{\alpha_{s}(\mu)}\right]^{\frac{2}{27}}$ and $\langle\bar{s}g_s \sigma Gs \rangle(\mu)=\langle\bar{s}g_s \sigma Gs \rangle(Q)\left[\frac{\alpha_{s}(Q)}{\alpha_{s}(\mu)}\right]^{\frac{2}{27}}$.

In the article, we take the $\overline{MS}$ masses  $m_{c}(m_c)=(1.275\pm0.025)\,\rm{GeV}$ and $m_s(\mu=2\,\rm{GeV})=(0.095\pm0.005)\,\rm{GeV}$
 from the Particle Data Group \cite{PDG}, and take into account
the energy-scale dependence of  the $\overline{MS}$ masses from the renormalization group equation,
\begin{eqnarray}
m_c(\mu)&=&m_c(m_c)\left[\frac{\alpha_{s}(\mu)}{\alpha_{s}(m_c)}\right]^{\frac{12}{25}} \, ,\nonumber\\
m_s(\mu)&=&m_s({\rm 2GeV} )\left[\frac{\alpha_{s}(\mu)}{\alpha_{s}({\rm 2GeV})}\right]^{\frac{4}{9}} \, ,\nonumber\\
\alpha_s(\mu)&=&\frac{1}{b_0t}\left[1-\frac{b_1}{b_0^2}\frac{\log t}{t} +\frac{b_1^2(\log^2{t}-\log{t}-1)+b_0b_2}{b_0^4t^2}\right]\, ,
\end{eqnarray}
  where $t=\log \frac{\mu^2}{\Lambda^2}$, $b_0=\frac{33-2n_f}{12\pi}$, $b_1=\frac{153-19n_f}{24\pi^2}$, $b_2=\frac{2857-\frac{5033}{9}n_f+\frac{325}{27}n_f^2}{128\pi^3}$,  $\Lambda=213\,\rm{MeV}$, $296\,\rm{MeV}$  and  $339\,\rm{MeV}$ for the flavors  $n_f=5$, $4$ and $3$, respectively  \cite{PDG}.
 Furthermore, we set the small masses of the $u$ and $d$ quarks equal zero,  $m_u=m_d=0$.

In Ref.\cite{ WangHbaryon},  we  study the  $J^P={1\over 2}^{\pm}$ and ${3\over 2}^{\pm}$ heavy, doubly-heavy and triply-heavy baryon states  with the QCD sum rules in a systematic way by subtracting the contributions from the corresponding $J^P={1\over 2}^{\mp}$ and ${3\over 2}^{\mp}$  heavy, doubly-heavy and triply-heavy baryon states, the continuum threshold parameters $\sqrt{s_0}=M_{\rm{gr}}+ (0.6-0.8)\,\rm{GeV}$ work well, where subscript $\rm{gr}$ denotes the ground state baryons.
The pentaquark states are another type baryon states according to the fractional spins, in Ref.\cite{Wang1508},  we take the continuum threshold parameters as
$\sqrt{s_0}= M_{P_c(4380/4450)}+(0.6-0.8)\,\rm{GeV}$, which also works well.
In this article, we take the continuum threshold parameters
$\sqrt{s_0}= M_{P}+(0.6-0.8)\,\rm{GeV}$ as an additional  constraint.

The hidden charm (or bottom) five-quark systems  $q_1q_2q_3Q\bar{Q}$ could be described
by a double-well potential, just like the double heavy four-quark systems $q_1\bar{q}_2Q\bar{Q}$ \cite{Wang-tetraquark}.   We introduce the color indexes $i$, $j$ and $k$ firstly.  In the five-quark system $q_1q_2q_3Q\bar{Q}$,
the light quarks $q_1$ and $q_2$ combine together to form a light diquark $\mathcal{D}^i_{q_1q_2}$ in   color antitriplet,
\begin{eqnarray}
q_1+q_2 &\to& \mathcal{D}^{i}_{q_1q_2}\, ,
\end{eqnarray}
 the $Q$-quark serves as a static well potential and  combines with the light quark $q_3$  to form a heavy diquark $\mathcal{D}^j_{q_3Q}$ in  color antitriplet,
\begin{eqnarray}
q_3+Q &\to & \mathcal{D}^j_{q_3Q} \, ,
\end{eqnarray}
 the $\bar{Q}$-quark serves  as another static well potential and combines with the light diquark $\mathcal{D}^i_{q_1q_2}$  to
form a heavy triquark  in  color triplet,
\begin{eqnarray}
\bar{Q}^k+ \mathcal{D}^i_{q_1q_2} &\to & \varepsilon^{jki}\bar{Q}^k\mathcal{D}^i_{q_1q_2} \, .
\end{eqnarray}
 Then the heavy diquark  $\mathcal{D}^j_{q_3Q}$   combines with the heavy triquark  $\varepsilon^{jki}\bar{Q}^k\mathcal{D}^i_{q_1q_2}$  to form a pentaquark state in  color singlet,
\begin{eqnarray}
\varepsilon^{jki}\bar{Q}^k\mathcal{D}^i_{q_1q_2}+\mathcal{D}^j_{q_3Q} &\to & \varepsilon^{ijk}\mathcal{D}^i_{q_1q_2}\mathcal{D}^j_{q_3Q}\bar{Q}^k\, .
\end{eqnarray}
The interpolating currents in Eqs.(7-14) can also be understood in this way.

Now we can see that the double  heavy five-quark system   is characterized by the effective heavy quark masses ${\mathbb{M}}_Q$ and
the virtuality $V=\sqrt{M^2_{P}-(2{\mathbb{M}}_Q)^2}$, just like   the double heavy four-quark systems $q_1\bar{q}_2Q\bar{Q}$ \cite{Wang-tetraquark}.   The  QCD sum rules have three typical energy scales $\mu^2$, $T^2$, $V^2$, we
 set the energy  scales to be  $ \mu^2=V^2={\mathcal{O}}(T^2)$, and obtain energy scale formula,
 \begin{eqnarray}
 \mu&=&\sqrt{M_{P}^2-(2{\mathbb{M}}_Q)^2}\, ,
   \end{eqnarray}
   to determine the energy scales of the QCD spectral densities \cite{Wang1508,WangHuang1508}. In previous work \cite{Wang1508}, we take the  value ${\mathbb{M}}_c=1.8\,\rm{GeV}$ determined in the double heavy four-quark systems \cite{Wang-tetraquark} and obtain the values $\mu=2.5\,\rm{GeV}$ and $\mu=2.6\,\rm{GeV}$ for the hidden-charm pentaquark states $P_c(4380)$  and $P_c(4450)$, respectively. The energy scale formula works well.

 We can rewrite Eq.(43) into the following form,
\begin{eqnarray}
 M_{P}^2=4{\mathbb{M}}_c^2+\mu^2\, ,
   \end{eqnarray}
which indicates that the masses increase (or decrease) with  increase (or decrease) of the energy scales,
\begin{eqnarray}
\mu\uparrow  \, \, \, \, \,  M_{P} \uparrow \, ,\nonumber\\
\mu\downarrow  \, \, \, \, \, M_{P} \downarrow \, .
\end{eqnarray}
On the other hand, the calculations based on the QCD sum rules in Eqs.(36-37) indicate that the masses decrease  (or increase) with  increase (or decrease) of the energy scales,
\begin{eqnarray}
\mu\uparrow  \, \, \, \, \,  M_{P} \downarrow \, ,\nonumber\\
\mu\downarrow  \, \, \, \, \,  M_{P} \uparrow \, .
\end{eqnarray}
We can search for   a compromise and obtain  the optimal energy scales $\mu$ and masses $M_{P}$.

In the present QCD sum rules, we choose the  Borel parameters $T^2$ and continuum threshold
parameters $s_0$  to satisfy the  following four criteria:

$\bullet$ Pole dominance at the phenomenological side;

$\bullet$ Convergence of the operator product expansion;

$\bullet$ Appearance of the Borel platforms;

$\bullet$ Satisfying the energy scale formula.

Now we  search for the optimal    Borel parameters $T^2$ and continuum threshold parameters $s_0$ according to  the four criteria.
 The resulting Borel parameters $T^2$, continuum threshold parameters $s_0$, pole contributions, and contributions of the vacuum condensates of dimension 9 and 10 are shown   explicitly in Table 1. From the table, we can see that the first two criteria of the QCD sum rules are satisfied, and we expect to make reasonable predictions.
In Table 2, we present the contributions of different terms in the operator product expansion with the central values of the input parameters. In calculations,
 we observe  that the main contributions come from the terms $D_0$, $D_3$ and $D_6$, see Table 2, the operator product expansion is well convergent.   We should  admit that the convergent behavior of the operator product expansion is  not so good  as that in the conventional case, where the $D_0$ dominates the QCD sum rules, but we can find a comparatively reasonable
work window to extract the hadronic information \cite{ZhangJR}.

We take into account  all uncertainties  of the input   parameters,
and obtain the values of the masses and pole residues of
 the ${1\over 2}^{\pm}$   hidden-charm pentaquark states, which are shown in Figs.1-2 and Table 3. From Figs.1-2 and Table 3, we can see that the last two criteria are also satisfied.
 In Table 3, we also present the corresponding thresholds of the $J/\psi B_{10}$ and $J/\psi B_8$, where the $B_8$ and $B_{10}$ denote the octet and decuplet baryons with the constituents $q_1q_2q_3$, respectively.   From the table, we can see that the decays to the  $J/\psi B_{10}$,
\begin{eqnarray}
P_{q_1q_2q_3}^{11\frac{1}{2}}\left({\frac{1}{2}^-}\right),\,P_{q_1q_2q_3}^{10\frac{1}{2}}\left({\frac{1}{2}^-}\right) &\to& J/\psi B_{10}\, ,
\end{eqnarray}
for example,
\begin{eqnarray}
P_{uuu}^{11\frac{1}{2}}\left({\frac{1}{2}^-}\right),\,P_{uuu}^{10\frac{1}{2}}\left({\frac{1}{2}^-}\right) &\to& J/\psi \Delta^{++} \, , \nonumber\\
P_{uus}^{11\frac{1}{2}}\left({\frac{1}{2}^-}\right),\,P_{uus}^{10\frac{1}{2}}\left({\frac{1}{2}^-}\right) &\to& J/\psi \Sigma^{*+} \, , \nonumber\\
P_{uss}^{11\frac{1}{2}}\left({\frac{1}{2}^-}\right),\,P_{uss}^{10\frac{1}{2}}\left({\frac{1}{2}^-}\right) &\to& J/\psi \Xi^{*0} \, , \nonumber\\
P_{sss}^{11\frac{1}{2}}\left({\frac{1}{2}^-}\right),\,P_{sss}^{10\frac{1}{2}}\left({\frac{1}{2}^-}\right) &\to& J/\psi \Omega^{-} \, ,
\end{eqnarray}
can (or maybe) take place, but the decay  widths are rather small due to the small available phase-spaces; on the other hand, the decays,
\begin{eqnarray}
P_{q_1q_2q_3}^{11\frac{1}{2}}\left({\frac{1}{2}^-}\right),\,P_{q_1q_2q_3}^{10\frac{1}{2}}\left({\frac{1}{2}^-}\right) &\to& J/\psi B_{8}\, ,\\
P_{q_1q_2q_3}^{11\frac{1}{2}}\left({\frac{1}{2}^+}\right) &\to& J/\psi B_{10}\, , J/\psi B_{8}\, ,
\end{eqnarray}
for example,
\begin{eqnarray}
P_{uud}^{11\frac{1}{2}}\left({\frac{1}{2}^\pm}\right),\,P_{uud}^{10\frac{1}{2}}\left({\frac{1}{2}^-}\right) &\to& J/\psi p \, , \nonumber\\
P_{uus}^{11\frac{1}{2}}\left({\frac{1}{2}^\pm}\right),\,P_{uus}^{10\frac{1}{2}}\left({\frac{1}{2}^-}\right) &\to& J/\psi \Sigma^{+} \, , \nonumber\\
P_{uss}^{11\frac{1}{2}}\left({\frac{1}{2}^\pm}\right),\,P_{uss}^{10\frac{1}{2}}\left({\frac{1}{2}^-}\right) &\to& J/\psi \Xi^{0} \,   , \nonumber\\
P_{uuu}^{11\frac{1}{2}}\left({\frac{1}{2}^+}\right) &\to& J/\psi \Delta^{++} \, , \nonumber\\
P_{uus}^{11\frac{1}{2}}\left({\frac{1}{2}^+}\right) &\to& J/\psi \Sigma^{*+} \, , \nonumber\\
P_{uss}^{11\frac{1}{2}}\left({\frac{1}{2}^+}\right) &\to& J/\psi \Xi^{*0} \, , \nonumber\\
P_{sss}^{11\frac{1}{2}}\left({\frac{1}{2}^+}\right) &\to& J/\psi \Omega^{-} \, ,
\end{eqnarray}
can take place more easily,   the decay widths are larger due to the larger available phase-spaces; furthermore, the decays
\begin{eqnarray}
P_{q_1q_2q_3}^{10\frac{1}{2}}\left({\frac{1}{2}^+}\right) &\to&J/\psi B_{8}\, , J/\psi B_{10}\, ,
\end{eqnarray}
for example,
\begin{eqnarray}
P_{uud}^{10\frac{1}{2}}\left({\frac{1}{2}^+}\right) &\to& J/\psi p \, , \nonumber\\
P_{uus}^{10\frac{1}{2}}\left({\frac{1}{2}^+}\right) &\to& J/\psi \Sigma^{+} \, , \nonumber\\
P_{uss}^{10\frac{1}{2}}\left({\frac{1}{2}^+}\right) &\to& J/\psi \Xi^{0} \,   , \nonumber\\
P_{uuu}^{10\frac{1}{2}}\left({\frac{1}{2}^+}\right) &\to& J/\psi \Delta^{++} \, , \nonumber\\
P_{uus}^{10\frac{1}{2}}\left({\frac{1}{2}^+}\right) &\to& J/\psi \Sigma^{*+} \, , \nonumber\\
P_{uss}^{10\frac{1}{2}}\left({\frac{1}{2}^+}\right) &\to& J/\psi \Xi^{*0} \, , \nonumber\\
P_{sss}^{10\frac{1}{2}}\left({\frac{1}{2}^+}\right) &\to& J/\psi \Omega^{-} \, ,
\end{eqnarray}
can take place fluently,   the decay  widths are rather large due to the large available phase-spaces.
We can search for the pentaquark states in the $J/\psi B_8$ and $J/\psi B_{10}$ mass spectrum in the decays of the bottom baryons to the final states $J/\psi B_8$ and $J/\psi B_{10}$ associated with the light vector mesons or pseudoscalar mesons \cite{Maiani1507,XGHe1507,Cheng1509},
for example,
\begin{eqnarray}
\Omega_b^- &\to& P_{uss}^{11\frac{1}{2}}\left({\frac{1}{2}}^{\pm}\right)K^- \to J/\psi \Xi^{*0} K^-\, , \nonumber\\
\Omega_b^- &\to& P_{sss}^{11\frac{1}{2}}\left({\frac{1}{2}}^{\pm}\right)\phi \to J/\psi \Omega^- \phi\, .
\end{eqnarray}

In Ref.\cite{mass-Anisovich}, the authors study the  non-strange and strange pentaquarks with hidden-charm  in the diquark-diquark-antiquark model by considering the simple spin-spin interactions, and evaluate the masses, where the scalar and axialvector diquarks  are chosen. The predicted masses are different from ours, see Refs.\cite{Wang1508,WangHuang1508} and present work, the differences originate partly from the fact that in Ref.\cite{mass-Anisovich} the $P_c(4450)$ is taken as the $P_{uud}^{11\frac{1}{2}}({\frac{5}{2}}^-)$ pentaquark state, while in Refs.\cite{Wang1508,WangHuang1508} the $P_c(4450)$ is taken as  the $P_{uud}^{11\frac{3}{2}}({\frac{5}{2}}^+)$ pentaquark state.

\begin{table}
\begin{center}
\begin{tabular}{|c|c|c|c|c|c|c|c|}\hline\hline
                                                        &$T^2 (\rm{GeV}^2)$  &$\sqrt{s_0} (\rm{GeV})$   &pole         &$D_9$         &$D_{10}$ \\ \hline
$P_{uuu}^{11\frac{1}{2}}\left({\frac{1}{2}^-}\right)$   & $3.2-3.6$          &$5.1\pm0.1$               &$(38-60)\%$  &$(16-22)\%$   &$\sim1\%$   \\ \hline
$P_{uus}^{11\frac{1}{2}}\left({\frac{1}{2}^-}\right)$   & $3.3-3.7$          &$5.2\pm0.1$               &$(39-60)\%$  &$(10-15)\%$   &$\leq1\%$   \\ \hline
$P_{uss}^{11\frac{1}{2}}\left({\frac{1}{2}^-}\right)$   & $3.4-3.8$          &$5.3\pm0.1$               &$(40-61)\%$  &$(7-10)\%$    &$<1\%$   \\ \hline
$P_{sss}^{11\frac{1}{2}}\left({\frac{1}{2}^-}\right)$   & $3.5-3.9$          &$5.4\pm0.1$               &$(42-62)\%$  &$(5-7)\%$     &$<1\%$   \\ \hline

$P_{uuu}^{10\frac{1}{2}}\left({\frac{1}{2}^-}\right)$   & $3.3-3.7$          &$5.1\pm0.1$               &$(39-60)\%$  &$(10-13)\%$   &$(3-4)\%$   \\ \hline
$P_{uus}^{10\frac{1}{2}}\left({\frac{1}{2}^-}\right)$   & $3.4-3.8$          &$5.2\pm0.1$               &$(41-61)\%$  &$(6-9)\%$     &$(2-3)\%$   \\ \hline
$P_{uss}^{10\frac{1}{2}}\left({\frac{1}{2}^-}\right)$   & $3.5-3.9$          &$5.3\pm0.1$               &$(42-62)\%$  &$(4-6)\%$     &$(1-2)\%$   \\ \hline
$P_{sss}^{10\frac{1}{2}}\left({\frac{1}{2}^-}\right)$   & $3.6-4.0$          &$5.4\pm0.1$               &$(43-62)\%$  &$(3-4)\%$     &$\sim1\%$   \\ \hline

$P_{uuu}^{11\frac{1}{2}}\left({\frac{1}{2}^+}\right)$   & $3.3-3.7$          &$5.3\pm0.1$               &$(37-59)\%$  &$(15-21)\%$   &$\leq1\%$   \\ \hline
$P_{uus}^{11\frac{1}{2}}\left({\frac{1}{2}^+}\right)$   & $3.4-3.8$          &$5.4\pm0.1$               &$(38-59)\%$  &$(10-14)\%$   &$<1\%$   \\ \hline
$P_{uss}^{11\frac{1}{2}}\left({\frac{1}{2}^+}\right)$   & $3.5-3.9$          &$5.5\pm0.1$               &$(39-60)\%$  &$(7-10)\%$    &$<1\%$   \\ \hline
$P_{sss}^{11\frac{1}{2}}\left({\frac{1}{2}^+}\right)$   & $3.6-4.0$          &$5.6\pm0.1$               &$(40-60)\%$  &$(5-7)\%$     &$<1\%$   \\ \hline

$P_{uuu}^{10\frac{1}{2}}\left({\frac{1}{2}^+}\right)$   & $3.3-3.7$          &$5.8\pm0.1$               &$(60-78)\%$  &$-(3-5)\%$    &$<1\%$   \\ \hline
$P_{uus}^{10\frac{1}{2}}\left({\frac{1}{2}^+}\right)$   & $3.4-3.8$          &$5.9\pm0.1$               &$(61-79)\%$  &$-(2-4)\%$    &$<1\%$   \\ \hline
$P_{uss}^{10\frac{1}{2}}\left({\frac{1}{2}^+}\right)$   & $3.5-3.9$          &$6.0\pm0.1$               &$(62-79)\%$  &$-(1-2)\%$    &$<1\%$   \\ \hline
$P_{sss}^{10\frac{1}{2}}\left({\frac{1}{2}^+}\right)$   & $3.6-4.0$          &$6.1\pm0.1$               &$(63-80)\%$  &$\sim-1\%$    &$<1\%$   \\ \hline
 \hline
\end{tabular}
\end{center}
\caption{ The Borel parameters, continuum threshold parameters, pole contributions, contributions of the vacuum condensates of dimension 9 and dimension 10. }
\end{table}

\begin{table}
\begin{center}
\begin{tabular}{|c|c|c|c|c|c|c|c|}\hline\hline
    Pentaquark states                                      & Contributions     \\ \hline
$P_{uuu}^{11\frac{1}{2}}\left({\frac{1}{2}^-}\right)$      &$D_0 \gg D_6 > D_3 \gg |D_5|\approx |D_8|\approx D_9\gg D_{10}$        \\ \hline
$P_{uus}^{11\frac{1}{2}}\left({\frac{1}{2}^-}\right)$      &$D_0 \gg D_6 > D_3 \gg |D_5|\approx |D_8|> D_9\gg D_{10}$    \\ \hline
$P_{uss}^{11\frac{1}{2}}\left({\frac{1}{2}^-}\right)$      &$D_0 \gg D_6 > D_3 \gg |D_5|\approx |D_8|> D_9\gg D_{10}$ \\ \hline
$P_{sss}^{11\frac{1}{2}}\left({\frac{1}{2}^-}\right)$      &$D_0 \gg D_6 > D_3 > |D_5|> |D_8|> D_9\gg D_{10}$   \\ \hline

$P_{uuu}^{10\frac{1}{2}}\left({\frac{1}{2}^-}\right)$      &$D_6 > D_3 > D_0 > |D_8|\gg |D_5|> D_9\gg D_{10}$   \\ \hline
$P_{uus}^{10\frac{1}{2}}\left({\frac{1}{2}^-}\right)$      &$D_0 \approx D_3 > D_6 \gg |D_8|>|D_5|\gg D_9\gg D_{10}$   \\ \hline
$P_{uss}^{10\frac{1}{2}}\left({\frac{1}{2}^-}\right)$      &$D_0 \gg D_3 > D_6 \gg |D_8|>|D_5|\gg D_9\gg D_{10}$   \\ \hline
$P_{sss}^{10\frac{1}{2}}\left({\frac{1}{2}^-}\right)$      &$D_0 \gg D_3 \gg D_6 \gg |D_8|>|D_5|\gg D_9\gg D_{10}$   \\ \hline

$P_{uuu}^{11\frac{1}{2}}\left({\frac{1}{2}^+}\right)$      &$D_0 \gg D_6 > D_3 \gg |D_5|\approx|D_8|\approx D_9\gg D_{10}$   \\ \hline
$P_{uus}^{11\frac{1}{2}}\left({\frac{1}{2}^+}\right)$      &$D_0 \gg D_6 > D_3 \gg |D_5|\approx|D_8|> D_9\gg D_{10}$   \\ \hline
$P_{uss}^{11\frac{1}{2}}\left({\frac{1}{2}^+}\right)$      &$D_0 \gg D_6 > D_3 \gg |D_5|\approx|D_8|> D_9\gg D_{10}$   \\ \hline
$P_{sss}^{11\frac{1}{2}}\left({\frac{1}{2}^+}\right)$      &$D_0 \gg D_6 > D_3 \gg |D_5|>|D_8|> D_9\gg D_{10}$   \\ \hline

$P_{uuu}^{10\frac{1}{2}}\left({\frac{1}{2}^+}\right)$      &$D_0 \gg D_3 \gg |D_6| > |D_5|\gg D_8> |D_9|\gg |D_{10}|$   \\ \hline
$P_{uus}^{10\frac{1}{2}}\left({\frac{1}{2}^+}\right)$      &$D_0 \gg D_3 \gg |D_6| > |D_5|\gg D_8> |D_9|\gg |D_{10}|$   \\ \hline
$P_{uss}^{10\frac{1}{2}}\left({\frac{1}{2}^+}\right)$      &$D_0 \gg D_3 \gg |D_6| > |D_5|\gg D_8\gg |D_9|\gg |D_{10}|$   \\ \hline
$P_{sss}^{10\frac{1}{2}}\left({\frac{1}{2}^+}\right)$      &$D_0 \gg D_3 \gg |D_6| > |D_5|\gg|D_9|> D_8\gg |D_{10}|$   \\ \hline
 \hline
\end{tabular}
\end{center}
\caption{ The  contributions of different terms  in the operator product expansion with the central values of the input parameters. }
\end{table}

\begin{table}
\begin{center}
\begin{tabular}{|c|c|c|c|c|c|c|c|}\hline\hline
                                                      &$\mu(\rm{GeV})$ &$M_{P}(\rm{GeV})$ &$\lambda_{P}(\rm{GeV}^6)$    &$M_{B_{10}J/\psi(B_8J/\psi)}(\rm{GeV})$\\ \hline
$P_{uuu}^{11\frac{1}{2}}\left({\frac{1}{2}^-}\right)$ &$2.5$           &$4.35\pm0.15$     &$(3.72\pm0.76)\times10^{-3}$  & 4.33 (4.04)\\ \hline
$P_{uus}^{11\frac{1}{2}}\left({\frac{1}{2}^-}\right)$ &$2.6$           &$4.47\pm0.15$     &$(4.50\pm0.85)\times10^{-3}$  & 4.48 (4.29)\\ \hline
$P_{uss}^{11\frac{1}{2}}\left({\frac{1}{2}^-}\right)$ &$2.8$           &$4.58\pm0.14$     &$(5.43\pm0.96)\times10^{-3}$  & 4.63 (4.41)\\ \hline
$P_{sss}^{11\frac{1}{2}}\left({\frac{1}{2}^-}\right)$ &$3.0$           &$4.68\pm0.13$     &$(6.47\pm1.10)\times10^{-3}$  & 4.77\\ \hline

$P_{uuu}^{10\frac{1}{2}}\left({\frac{1}{2}^-}\right)$ &$2.5$           &$4.42\pm0.12$     &$(4.14\pm0.70)\times10^{-3}$  & 4.33 (4.04)\\ \hline
$P_{uus}^{10\frac{1}{2}}\left({\frac{1}{2}^-}\right)$ &$2.7$           &$4.51\pm0.11$     &$(4.97\pm0.79)\times10^{-3}$  & 4.48 (4.29)\\ \hline
$P_{uss}^{10\frac{1}{2}}\left({\frac{1}{2}^-}\right)$ &$2.9$           &$4.60\pm0.11$     &$(5.87\pm0.89)\times10^{-3}$  & 4.63 (4.41) \\ \hline
$P_{sss}^{10\frac{1}{2}}\left({\frac{1}{2}^-}\right)$ &$3.0$           &$4.71\pm0.11$     &$(6.84\pm1.00)\times10^{-3}$  & 4.77\\ \hline

$P_{uuu}^{11\frac{1}{2}}\left({\frac{1}{2}^+}\right)$ &$2.8$           &$4.56\pm0.15$     &$(1.97\pm0.40)\times10^{-3}$  & 4.33 (4.04)\\ \hline
$P_{uus}^{11\frac{1}{2}}\left({\frac{1}{2}^+}\right)$ &$3.0$           &$4.67\pm0.14$     &$(2.42\pm0.47)\times10^{-3}$  & 4.48 (4.29)\\ \hline
$P_{uss}^{11\frac{1}{2}}\left({\frac{1}{2}^+}\right)$ &$3.1$           &$4.78\pm0.13$     &$(2.88\pm0.53)\times10^{-3}$  & 4.63 (4.41)\\ \hline
$P_{sss}^{11\frac{1}{2}}\left({\frac{1}{2}^+}\right)$ &$3.3$           &$4.89\pm0.13$     &$(3.44\pm0.61)\times10^{-3}$  & 4.77\\ \hline

$P_{uuu}^{10\frac{1}{2}}\left({\frac{1}{2}^+}\right)$ &$3.6$           &$5.12\pm0.08$     &$(8.01\pm0.92)\times10^{-3}$  & 4.33 (4.04)\\ \hline
$P_{uus}^{10\frac{1}{2}}\left({\frac{1}{2}^+}\right)$ &$3.7$           &$5.19\pm0.08$     &$(8.92\pm1.05)\times10^{-3}$  & 4.48 (4.29)\\ \hline
$P_{uss}^{10\frac{1}{2}}\left({\frac{1}{2}^+}\right)$ &$3.8$           &$5.26\pm0.08$     &$(9.93\pm1.21)\times10^{-3}$  & 4.63 (4.41)\\ \hline
$P_{sss}^{10\frac{1}{2}}\left({\frac{1}{2}^+}\right)$ &$4.0$           &$5.40\pm0.08$     &$(12.17\pm1.28)\times10^{-3}$ & 4.77\\ \hline
   \hline
\end{tabular}
\end{center}
\caption{ The   energy scales,  masses and pole residues of the pentaquark states, where the $B_{10}$ and $B_8$ denote the decuplet  and octet baryons with the quark constituents $q_1q_2q_3$ respectively. We take the isospin limit, the pentaquark states $P_{uuu}^{11\frac{1}{2}}\left({\frac{1}{2}^-}\right)$, $P_{uud}^{11\frac{1}{2}}\left({\frac{1}{2}^-}\right)$,
$P_{udd}^{11\frac{1}{2}}\left({\frac{1}{2}^-}\right)$ an $P_{ddd}^{11\frac{1}{2}}\left({\frac{1}{2}^-}\right)$ have   degenerate   masses. Other states are implied.   }
\end{table}

\begin{figure}
 \centering
 \includegraphics[totalheight=5cm,width=6cm]{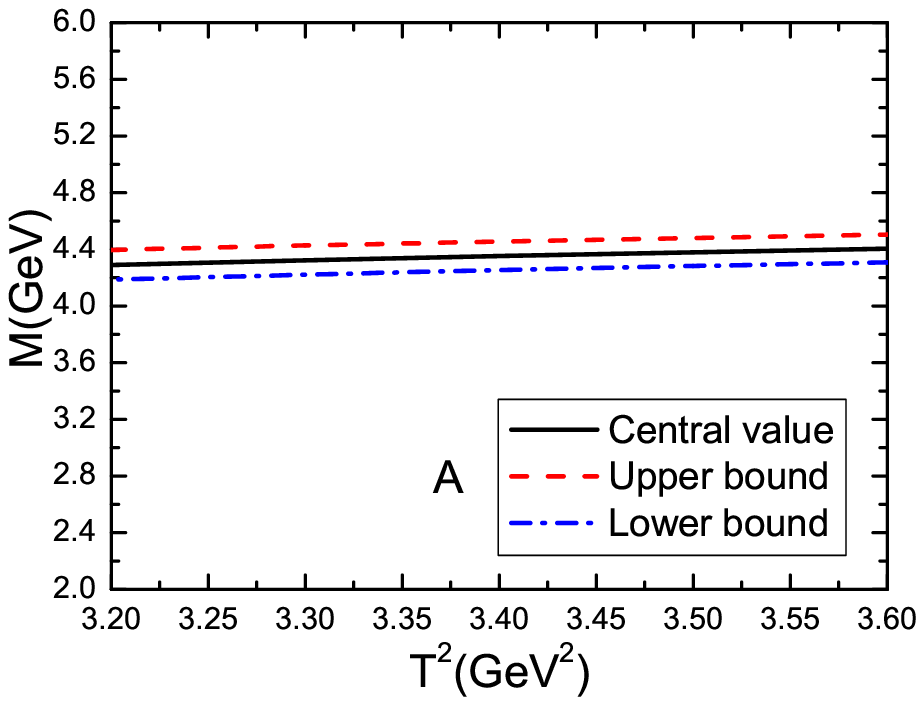}
 \includegraphics[totalheight=5cm,width=6cm]{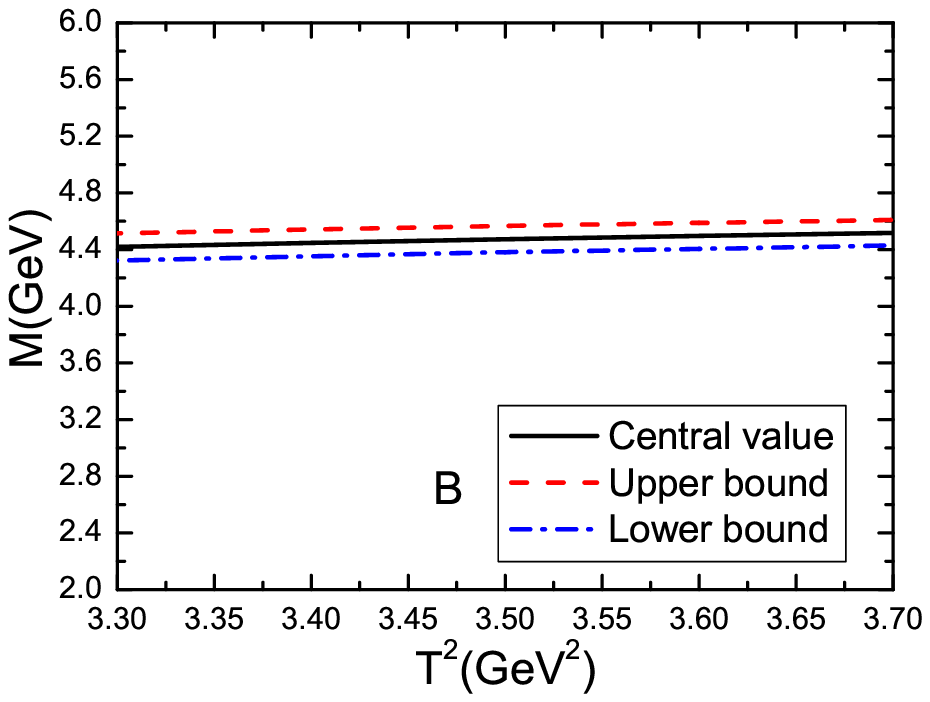}
 \includegraphics[totalheight=5cm,width=6cm]{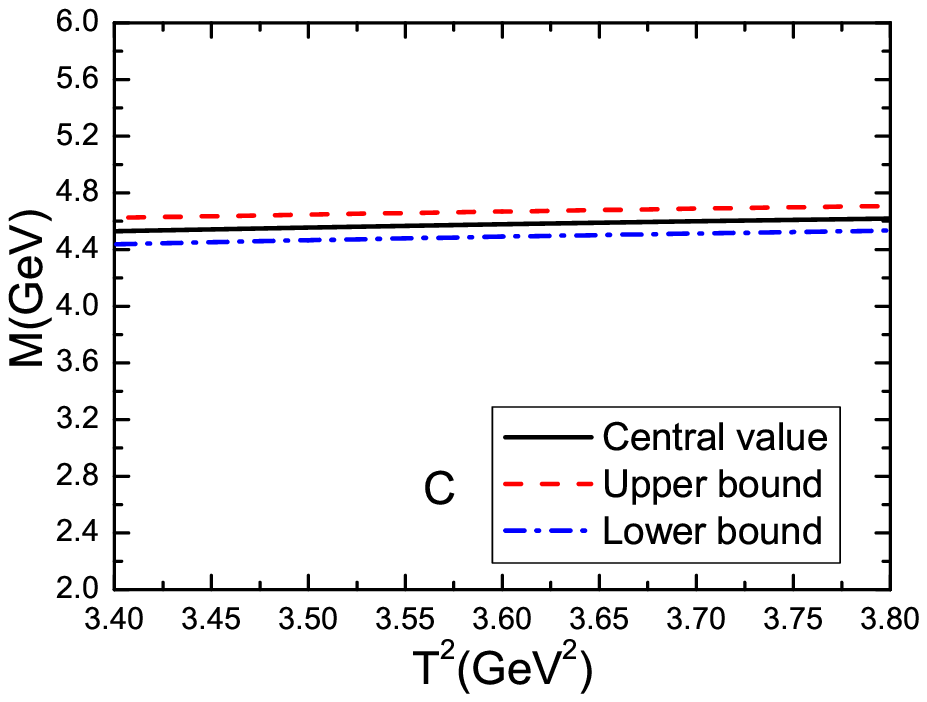}
 \includegraphics[totalheight=5cm,width=6cm]{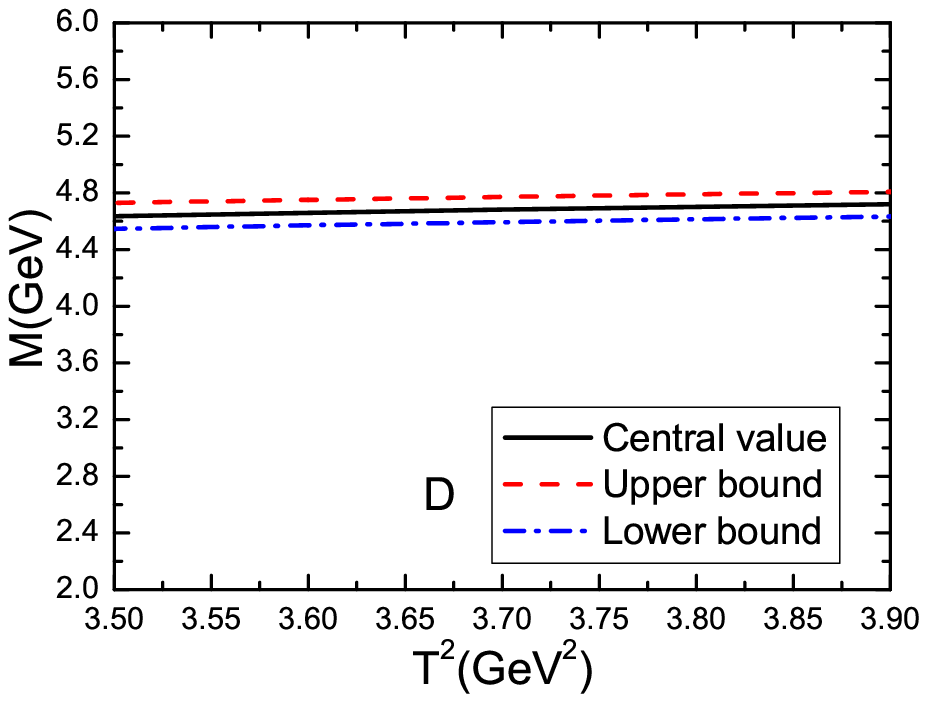}
 \includegraphics[totalheight=5cm,width=6cm]{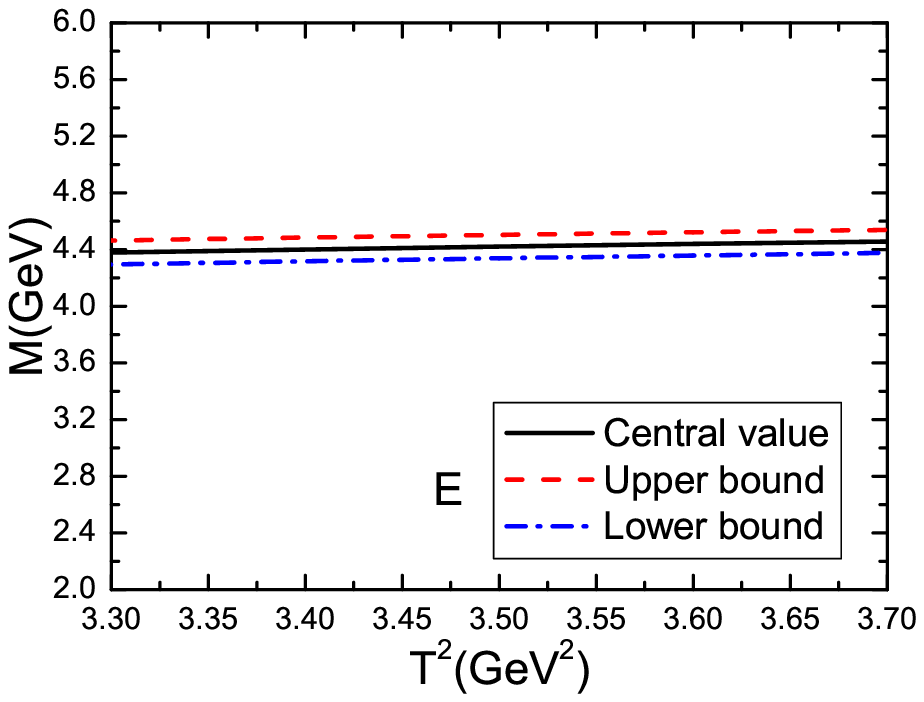}
 \includegraphics[totalheight=5cm,width=6cm]{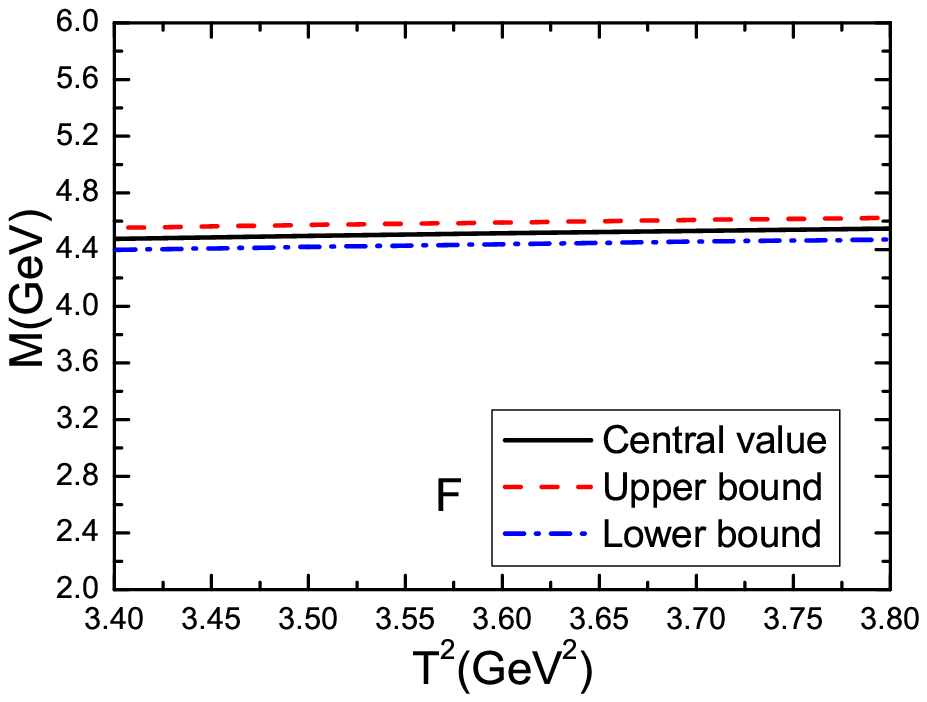}
 \includegraphics[totalheight=5cm,width=6cm]{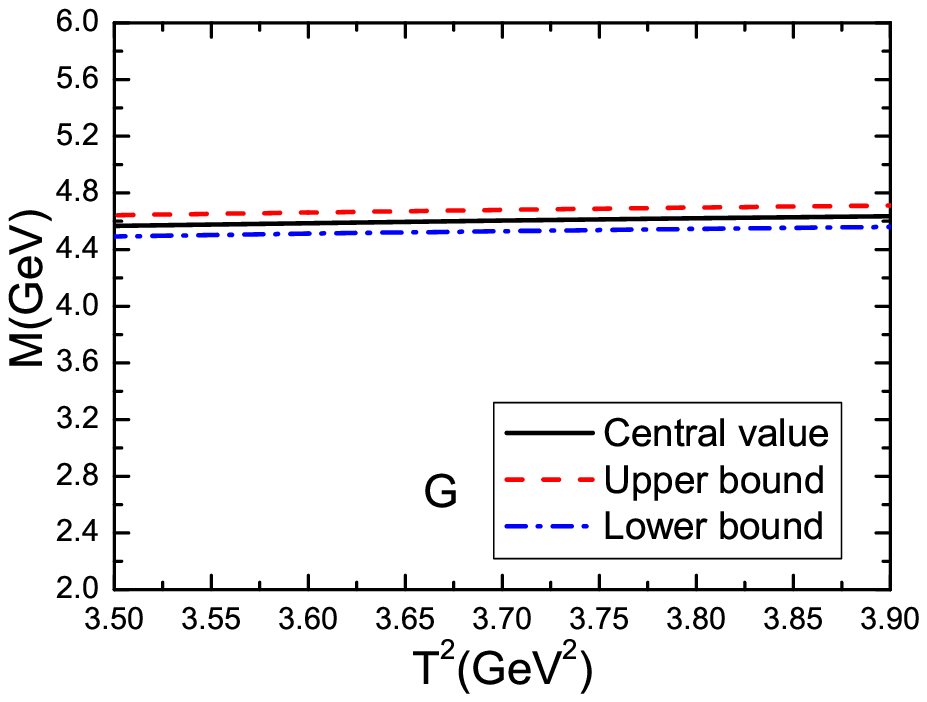}
 \includegraphics[totalheight=5cm,width=6cm]{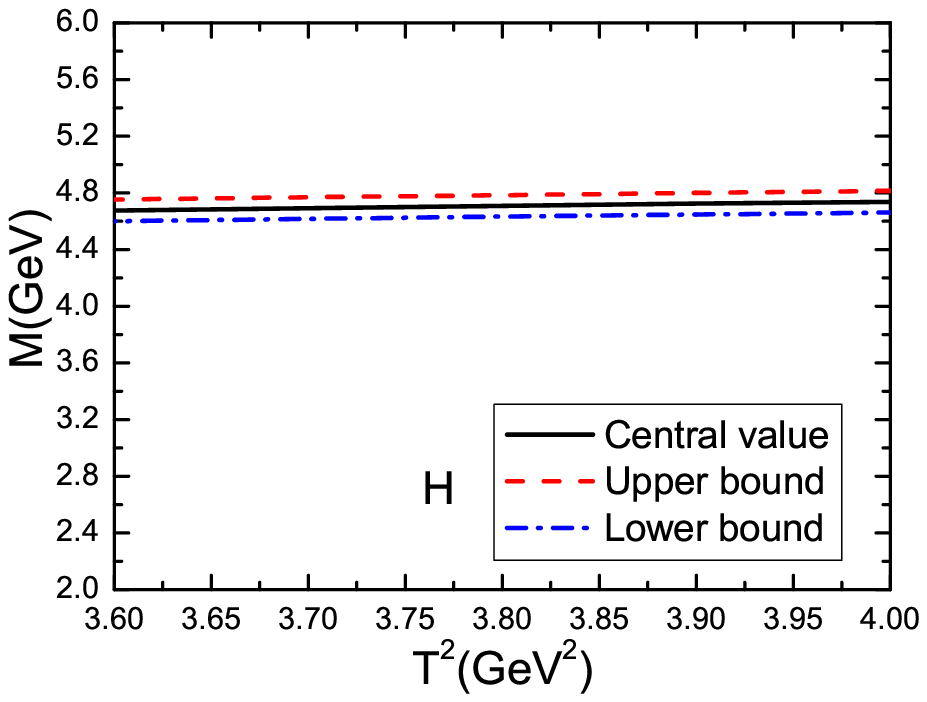}
 \caption{ The masses  of the pentaquark states  with variations of the Borel parameters $T^2$, where the $A$, $B$, $C$, $D$, $E$, $F$, $G$ and $H$ denote the pentaquark states $P_{uuu}^{11\frac{1}{2}}\left({\frac{1}{2}^-}\right)$,
$P_{uus}^{11\frac{1}{2}}\left({\frac{1}{2}^-}\right)$, $P_{uss}^{11\frac{1}{2}}\left({\frac{1}{2}^-}\right)$, $P_{sss}^{11\frac{1}{2}}\left({\frac{1}{2}^-}\right)$,
$P_{uuu}^{10\frac{1}{2}}\left({\frac{1}{2}^-}\right)$, $P_{uus}^{10\frac{1}{2}}\left({\frac{1}{2}^-}\right)$, $P_{uss}^{10\frac{1}{2}}\left({\frac{1}{2}^-}\right)$ and
$P_{sss}^{10\frac{1}{2}}\left({\frac{1}{2}^-}\right)$ with negative parity  respectively.  }
\end{figure}

\begin{figure}
 \centering
 \includegraphics[totalheight=5cm,width=6cm]{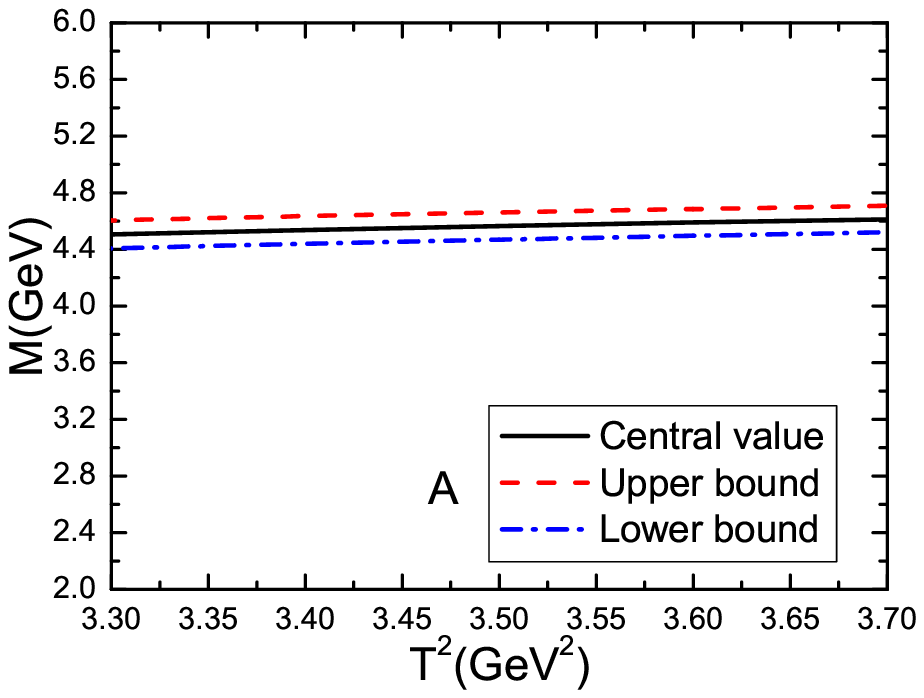}
 \includegraphics[totalheight=5cm,width=6cm]{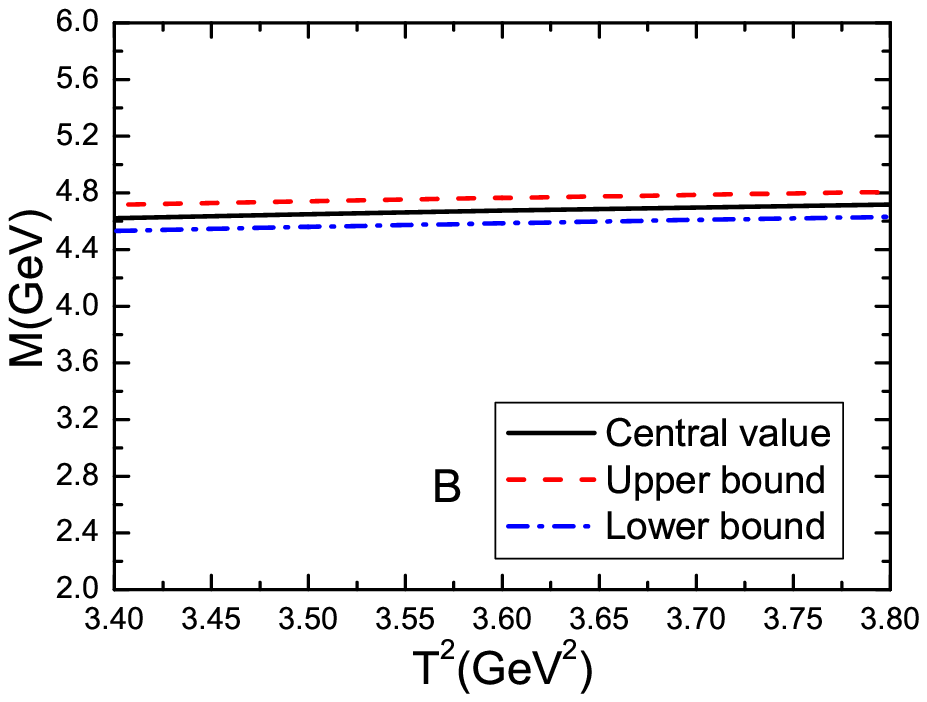}
 \includegraphics[totalheight=5cm,width=6cm]{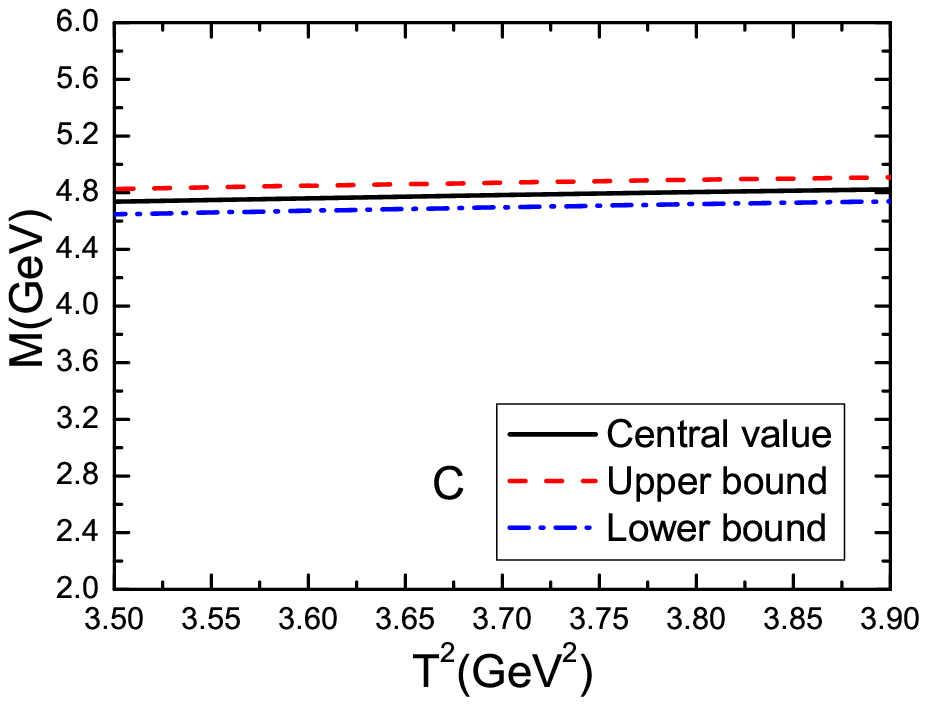}
 \includegraphics[totalheight=5cm,width=6cm]{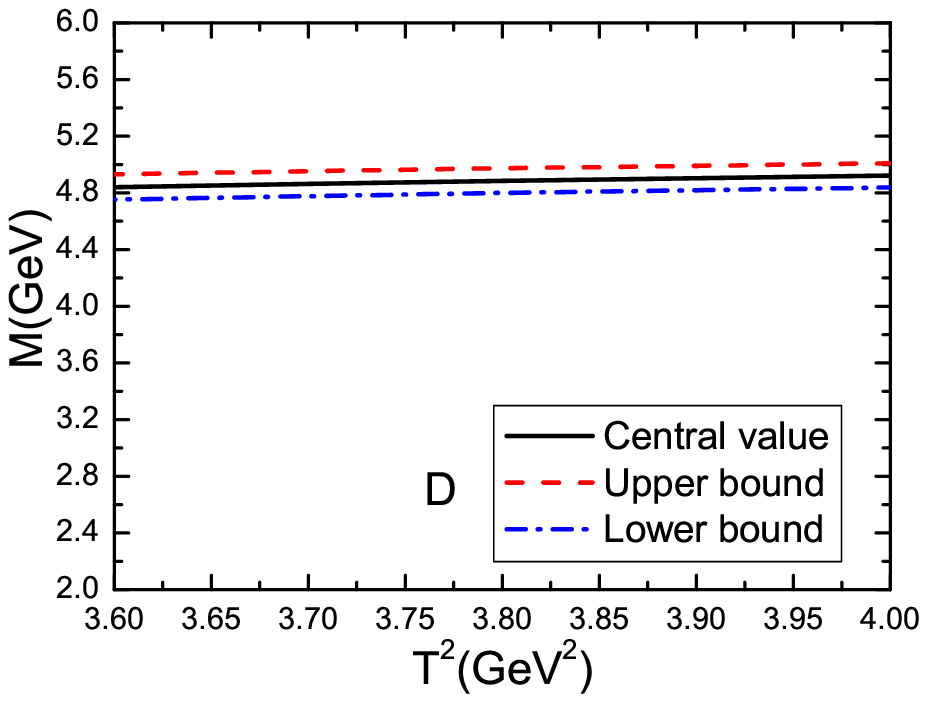}
 \includegraphics[totalheight=5cm,width=6cm]{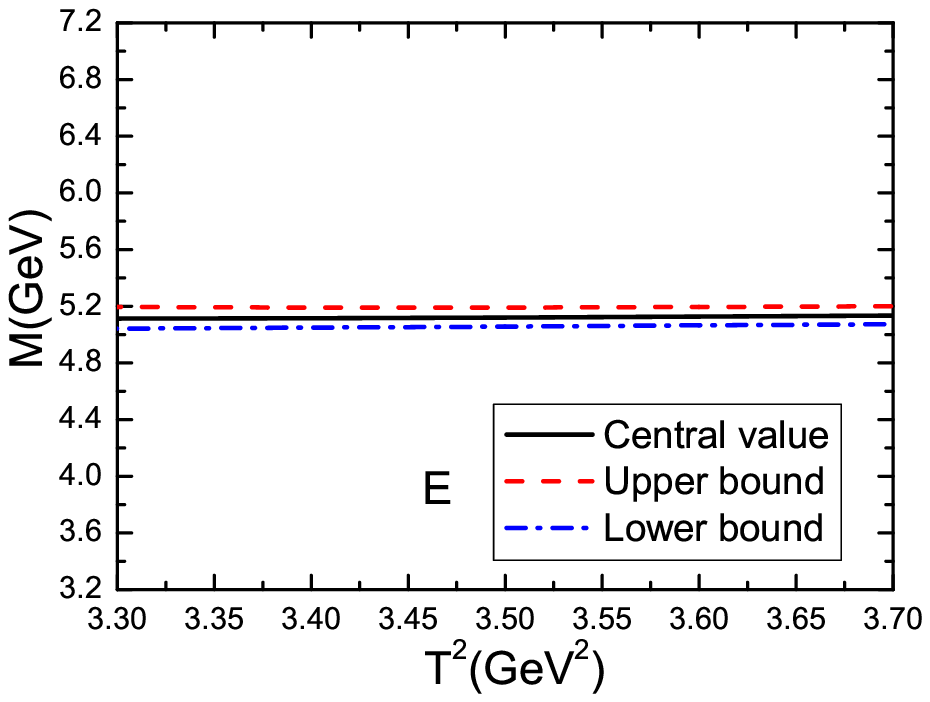}
 \includegraphics[totalheight=5cm,width=6cm]{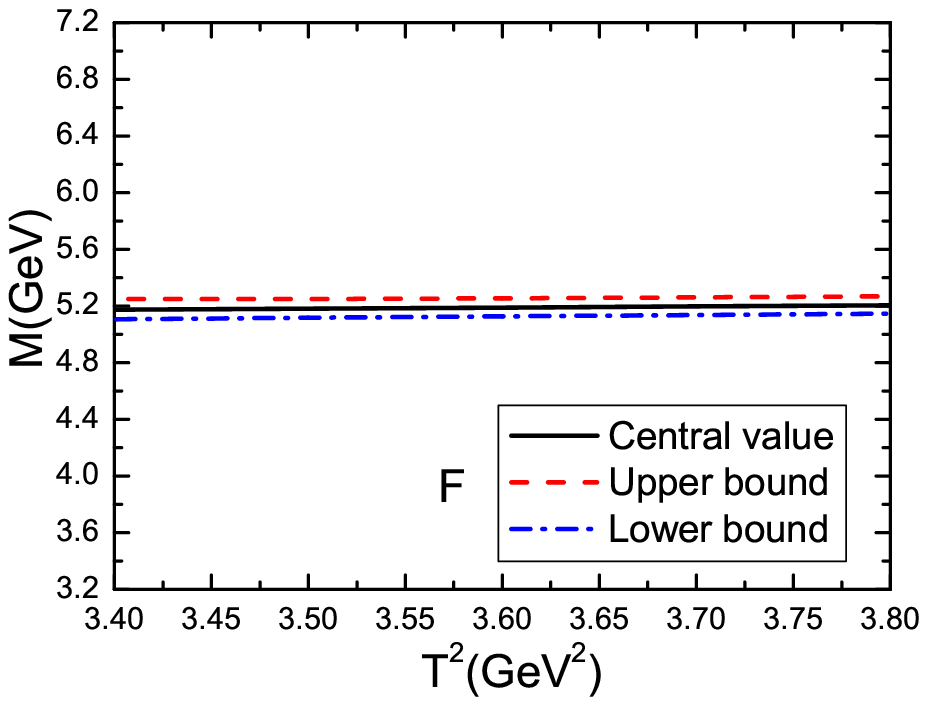}
 \includegraphics[totalheight=5cm,width=6cm]{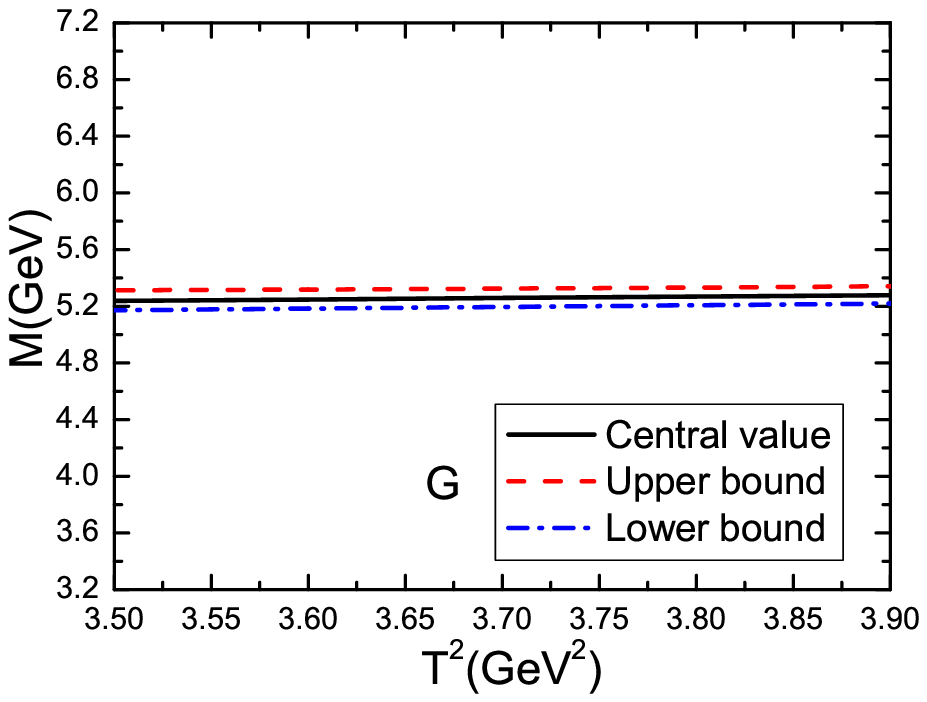}
 \includegraphics[totalheight=5cm,width=6cm]{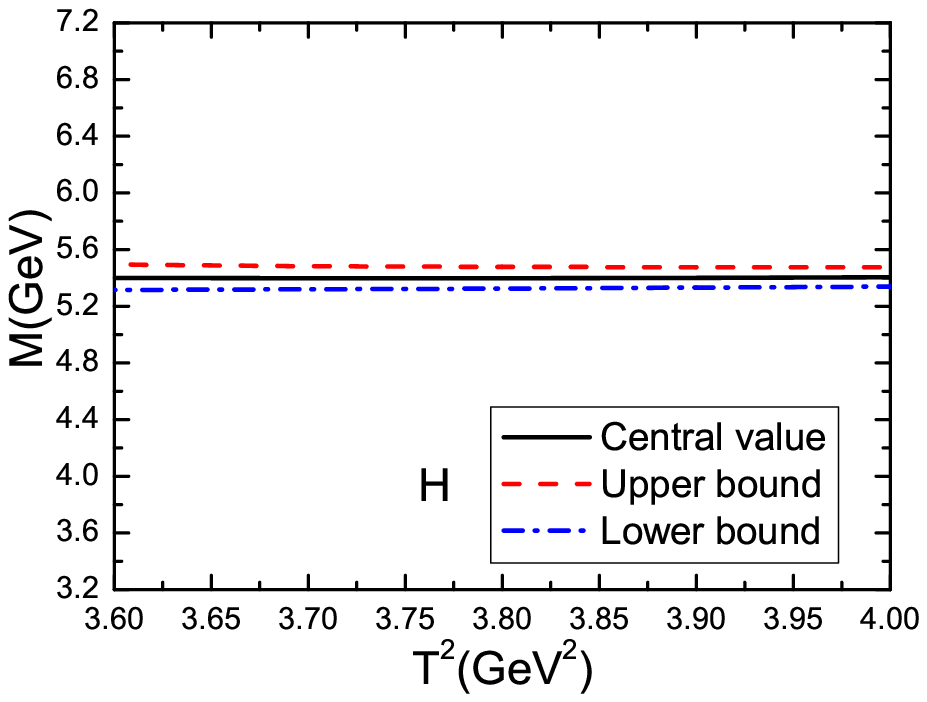}
        \caption{ The masses  of the pentaquark states  with variations of the Borel parameters $T^2$, where the $A$, $B$, $C$, $D$, $E$, $F$, $G$ and $H$ denote the pentaquark states $P_{uuu}^{11\frac{1}{2}}\left({\frac{1}{2}^+}\right)$,
$P_{uus}^{11\frac{1}{2}}\left({\frac{1}{2}^+}\right)$, $P_{uss}^{11\frac{1}{2}}\left({\frac{1}{2}^+}\right)$, $P_{sss}^{11\frac{1}{2}}\left({\frac{1}{2}^+}\right)$,
$P_{uuu}^{10\frac{1}{2}}\left({\frac{1}{2}^+}\right)$, $P_{uus}^{10\frac{1}{2}}\left({\frac{1}{2}^+}\right)$, $P_{uss}^{10\frac{1}{2}}\left({\frac{1}{2}^+}\right)$ and
$P_{sss}^{10\frac{1}{2}}\left({\frac{1}{2}^+}\right)$  with positive parity respectively.  }
\end{figure}

\section{Conclusion}
In this article, we choose the axialvector-diquark-axialvector-diquark-antiquark type and axialvector-diquark-scalar-diquark-antiquark type pentaquark configurations,  construct  both the axialvector-diquark-axialvector-diquark-antiquark type and axialvector-diquark-scalar-diquark-antiquark type interpolating currents, then calculate the contributions of the vacuum condensates up to dimension-10 in the operator product expansion, and  study the masses and pole residues of the  $J^P={\frac{1}{2}}^\pm$   hidden-charm pentaquark states    with the QCD sum rules in a systematic way. In calculations, we use the  formula $\mu=\sqrt{M^2_{P}-(2{\mathbb{M}}_c)^2}$  to determine  the energy scales of the QCD spectral densities, which works well in our previous work.  We take into account the $SU(3)$ breaking effects of the light  quarks, and obtain the masses of the hidden charm pentaquark states with the strangeness  $S=0,\,-1,\,-2,\,-3$, which can be confronted with the experimental data in the future. We can search for  those pentaquark  states in the decays of the bottom baryons to the $J/\psi$ plus octet (decuplet) baryons plus pseudoscalar (vector) mesons, or take  the  pole residues   as   basic input parameters to study relevant processes of the  pentaquark states with the three-point QCD sum rules.

\section*{Acknowledgements}
This  work is supported by National Natural Science Foundation,
Grant Number 11375063, and Natural Science Foundation of Hebei province, Grant Number A2014502017.

\section*{Appendix}
The QCD spectral densities $\rho^{11,1}_{sss}(s)$, $\widetilde{\rho}^{11,0}_{sss}(s)$, $\rho^{11,1}_{uss}(s)$, $\widetilde{\rho}^{11,0}_{uss}(s)$, $\rho^{11,1}_{uus}(s)$, $\widetilde{\rho}^{11,0}_{uus}(s)$, $\rho^{11,1}_{uuu}(s)$, $\widetilde{\rho}^{11,0}_{uuu}(s)$, $\rho^{10,1}_{sss}(s)$, $\widetilde{\rho}^{10,0}_{sss}(s)$, $\rho^{10,1}_{uss}(s)$, $\widetilde{\rho}^{10,0}_{uss}(s)$, $\rho^{10,1}_{uus}(s)$, $\widetilde{\rho}^{10,0}_{uus}(s)$, $\rho^{10,1}_{uuu}(s)$       and $\widetilde{\rho}^{10,0}_{uuu}(s)$ of the pentaquark states,

\begin{eqnarray}
\rho^{11,1}_{sss}(s)&=&\frac{1}{61440\pi^8}\int dydz \, yz(1-y-z)^4\left(s-\overline{m}_c^2\right)^4\left(8s-3\overline{m}_c^2 \right)  \nonumber\\
&&+\frac{m_s m_c}{12288\pi^8}\int dydz \, (y+z)(1-y-z)^3\left(s-\overline{m}_c^2\right)^4  \nonumber\\
&&- \frac{m_c\langle \bar{s}s\rangle}{768\pi^6}\int dydz \, (y+z)(1-y-z)^2\left(s-\overline{m}_c^2\right)^3  \nonumber \\
&&+ \frac{m_s\langle \bar{s}s\rangle}{128\pi^6}\int dydz \, yz(1-y-z)^2\left(s-\overline{m}_c^2\right)^2 \left(2s-\overline{m}_c^2 \right) \nonumber \\
&&+ \frac{3m_c\langle \bar{s}g_s\sigma Gs\rangle}{1024\pi^6}\int dydz  \, (y+z)(1-y-z) \left(s-\overline{m}_c^2 \right)^2 \nonumber\\
&&+ \frac{m_s\langle \bar{s}g_s\sigma Gs\rangle}{512\pi^6}\int dydz  \, yz(1-y-z) \left(s-\overline{m}_c^2 \right)\left(5s-3\overline{m}_c^2 \right) \nonumber\\
&&- \frac{m_s\langle \bar{s}g_s\sigma Gs\rangle}{1024\pi^6}\int dydz  \, (y+z)(1-y-z)^2 \left(s-\overline{m}_c^2 \right)\left(5s-3\overline{m}_c^2 \right) \nonumber\\
&&+\frac{\langle\bar{s}s\rangle^2}{48\pi^4}\int dydz \,  yz(1-y-z)\left(s-\overline{m}_c^2 \right)\left(5s-3\overline{m}_c^2 \right) \nonumber\\
&&+\frac{5 m_s m_c\langle\bar{s}s\rangle^2}{48\pi^4}\int dydz \,  (y+z)\left(s-\overline{m}_c^2 \right) \nonumber\\
&&-\frac{5\langle\bar{s}s\rangle\langle\bar{s}g_s\sigma Gs\rangle}{192\pi^4}\int dydz \,yz \left(4s-3\overline{m}_c^2\right)\nonumber\\
&&+\frac{\langle\bar{s}s\rangle\langle\bar{s}g_s\sigma Gs\rangle}{192\pi^4}\int dydz \,(y+z)(1-y-z) \left(4s-3\overline{m}_c^2\right)\nonumber\\
&&-\frac{121m_s m_c \langle\bar{s}s\rangle\langle\bar{s}g_s\sigma Gs\rangle}{2304\pi^4}\int dy   +\frac{m_s m_c\langle\bar{s}s\rangle\langle\bar{s}g_s\sigma Gs\rangle}{48\pi^4}\int dydz    \left( \frac{z}{y}+\frac{y}{z}\right)\nonumber\\
&&-\frac{m_c\langle\bar{s}s\rangle^3}{18\pi^2}\int dy   +\frac{m_s\langle\bar{s}s\rangle^3}{12\pi^2}\int dy \,y(1-y) \left[ 1+\frac{s}{3}\delta\left(s-\widetilde{m}_c^2 \right)\right]  \nonumber
\end{eqnarray}
\begin{eqnarray}
&&-\frac{3\langle\bar{s}g_s\sigma Gs\rangle^2}{512\pi^4}\int dydz \, (y+z)\left[ 1+\frac{s}{3}\delta\left(s-\overline{m}_c^2 \right)\right] \nonumber\\
&&+\frac{3\langle\bar{s}g_s\sigma Gs\rangle^2}{256\pi^4}\int dy \, y(1-y)\left[ 1+\frac{s}{3}\delta\left(s-\widetilde{m}_c^2 \right)\right] \nonumber\\
&&-\frac{m_s m_c\langle\bar{s}g_s\sigma Gs\rangle^2}{192\pi^4}\int dy \left( \frac{1-y}{y}+\frac{y}{1-y}\right)\delta\left(s-\widetilde{m}_c^2 \right) \nonumber\\
&&+\frac{m_s m_c\langle\bar{s}g_s\sigma Gs\rangle^2}{72\pi^4}\int dy \, y(1-y)\left( 1+\frac{s}{2M^2}\right)\delta\left(s-\widetilde{m}_c^2 \right) \nonumber\\
&&-\frac{m_s m_c\langle\bar{s}g_s\sigma Gs\rangle^2}{2304\pi^4}\int dy \,  \left( 1+\frac{s}{2M^2}\right)\delta\left(s-\widetilde{m}_c^2 \right)\, ,
\end{eqnarray}

\begin{eqnarray}
\widetilde{\rho}^{11,0}_{sss}(s)&=&\frac{1}{122880\pi^8}\int dydz \, (y+z)(1-y-z)^4\left(s-\overline{m}_c^2\right)^4\left(7s-2\overline{m}_c^2 \right)  \nonumber\\
&&+\frac{m_s m_c}{6144\pi^8}\int dydz \,(1-y-z)^3\left(s-\overline{m}_c^2\right)^4  \nonumber\\
&&- \frac{m_c\langle \bar{s}s\rangle}{384\pi^6}\int dydz \,  (1-y-z)^2\left(s-\overline{m}_c^2\right)^3  \nonumber \\
&&+ \frac{m_s\langle \bar{s}s\rangle}{768\pi^6}\int dydz \, (y+z)(1-y-z)^2\left(s-\overline{m}_c^2\right)^2 \left(5s-2\overline{m}_c^2 \right) \nonumber \\
&&+ \frac{3m_c\langle \bar{s}g_s\sigma Gs\rangle}{512\pi^6}\int dydz  \, (1-y-z) \left(s-\overline{m}_c^2 \right)^2 \nonumber\\
&&- \frac{m_s\langle \bar{s}g_s\sigma Gs\rangle}{256\pi^6}\int dydz  \, (1-y-z)^2 \left(s-\overline{m}_c^2 \right) \left(2s-\overline{m}_c^2 \right) \nonumber\\
&&+ \frac{m_s\langle \bar{s}g_s\sigma Gs\rangle}{512\pi^6}\int dydz  \, (y+z)(1-y-z) \left(s-\overline{m}_c^2 \right) \left(2s-\overline{m}_c^2 \right) \nonumber\\
&&+\frac{\langle\bar{s}s\rangle^2}{48\pi^4}\int dydz \,  (y+z)(1-y-z)\left(s-\overline{m}_c^2 \right)\left(2s-\overline{m}_c^2 \right) \nonumber\\
&&+\frac{5 m_s m_c\langle\bar{s}s\rangle^2}{24\pi^4}\int dydz \,   \left(s-\overline{m}_c^2 \right) \nonumber\\
&&-\frac{5\langle\bar{s}s\rangle\langle\bar{s}g_s\sigma Gs\rangle}{384\pi^4}\int dydz \,(y+z) \left(3s-2\overline{m}_c^2\right)\nonumber\\
&&+\frac{\langle\bar{s}s\rangle\langle\bar{s}g_s\sigma Gs\rangle}{96\pi^4}\int dydz \,(1-y-z) \left(3s-2\overline{m}_c^2\right)\nonumber\\
&&-\frac{121m_s m_c\langle\bar{s}s\rangle\langle\bar{s}g_s\sigma Gs\rangle}{1152\pi^4}\int dy    +\frac{m_s m_c\langle\bar{s}s\rangle\langle\bar{s}g_s\sigma Gs\rangle}{48\pi^4}\int dydz \left( \frac{1}{y}+\frac{1}{z}\right)     \nonumber\\
&&-\frac{m_c\langle\bar{s}s\rangle^3}{9\pi^2}\int dy   +\frac{m_s\langle\bar{s}s\rangle^3}{36\pi^2}\int dy \,  \left[ 1+\frac{s}{2}\delta\left(s-\widetilde{m}_c^2 \right)\right]  \nonumber\\
&&-\frac{\langle\bar{s}g_s\sigma Gs\rangle^2}{256\pi^4}\int dydz \,  \left[ 1+\frac{s}{2}\delta\left(s-\overline{m}_c^2 \right)\right] \nonumber\\
&&-\frac{m_s m_c\langle\bar{s}g_s\sigma Gs\rangle^2}{192\pi^4}\int dy \, \left( \frac{1}{y}+\frac{1}{1-y}\right)\delta\left(s-\widetilde{m}_c^2 \right) \nonumber\\
&&+\frac{31m_s m_c\langle\bar{s}g_s\sigma Gs\rangle^2}{2304\pi^4}\int dy \, \left( 1+\frac{s}{M^2}\right)\delta\left(s-\widetilde{m}_c^2 \right)  \, ,
\end{eqnarray}

\begin{eqnarray}
\rho^{11,1}_{uss}(s)&=&\frac{1}{61440\pi^8}\int dydz \, yz(1-y-z)^4\left(s-\overline{m}_c^2\right)^4\left(8s-3\overline{m}_c^2 \right)  \nonumber\\
&&+\frac{m_s m_c}{18432\pi^8}\int dydz \, (y+z)(1-y-z)^3\left(s-\overline{m}_c^2\right)^4  \nonumber\\
&&- \frac{m_c\left[\langle \bar{q}q\rangle+2\langle \bar{s}s\rangle\right]}{2304\pi^6}\int dydz \, (y+z)(1-y-z)^2\left(s-\overline{m}_c^2\right)^3  \nonumber \\
&&+ \frac{m_s\left[2\langle \bar{s}s\rangle-\langle \bar{q}q\rangle\right]}{192\pi^6}\int dydz \, yz(1-y-z)^2\left(s-\overline{m}_c^2\right)^2 \left(2s-\overline{m}_c^2 \right) \nonumber \\
&&+ \frac{m_c\left[\langle \bar{q}g_s\sigma Gq\rangle+2\langle \bar{s}g_s\sigma Gs\rangle\right]}{1024\pi^6}\int dydz  \, (y+z)(1-y-z) \left(s-\overline{m}_c^2 \right)^2 \nonumber\\
&&+ \frac{m_s\left[\langle \bar{q}g_s\sigma Gq\rangle-\langle \bar{s}g_s\sigma Gs\rangle\right]}{384\pi^6}\int dydz  \, yz(1-y-z) \left(s-\overline{m}_c^2 \right)\left(5s-3\overline{m}_c^2 \right) \nonumber\\
&&+ \frac{m_s\left[\langle \bar{q}g_s\sigma Gq\rangle+\langle \bar{s}g_s\sigma Gs\rangle\right]}{1536\pi^6}\int dydz  \, yz(1-y-z) \left(s-\overline{m}_c^2 \right)\left(5s-3\overline{m}_c^2 \right) \nonumber\\
&&- \frac{m_s\left[\langle \bar{q}g_s\sigma Gq\rangle+\langle \bar{s}g_s\sigma Gs\rangle\right]}{3072\pi^6}\int dydz  \, (y+z)(1-y-z)^2 \left(s-\overline{m}_c^2 \right)\left(5s-3\overline{m}_c^2 \right) \nonumber\\
&&+\frac{\langle\bar{s}s\rangle\left[2\langle\bar{q}q\rangle+\langle\bar{s}s\rangle \right]}{144\pi^4}\int dydz \,  yz(1-y-z)\left(s-\overline{m}_c^2 \right)\left(5s-3\overline{m}_c^2 \right) \nonumber\\
&&+\frac{ m_s m_c\langle\bar{s}s\rangle\left[11\langle\bar{q}q\rangle-\langle\bar{s}s\rangle\right]}{144\pi^4}\int dydz \,  (y+z)\left(s-\overline{m}_c^2 \right) \nonumber\\
&&-\frac{5\left[\langle\bar{q}q\rangle\langle\bar{s}g_s\sigma Gs\rangle+\langle\bar{s}s\rangle\langle\bar{q}g_s\sigma Gq\rangle+\langle\bar{s}s\rangle\langle\bar{s}g_s\sigma Gs\rangle\right]}{576\pi^4}\int dydz \,yz \left(4s-3\overline{m}_c^2\right)\nonumber\\
&&+\frac{\langle\bar{q}q\rangle\langle\bar{s}g_s\sigma Gs\rangle+\langle\bar{s}s\rangle\langle\bar{q}g_s\sigma Gq\rangle+\langle\bar{s}s\rangle\langle\bar{s}g_s\sigma Gs\rangle}{576\pi^4}\int dydz \,(y+z)(1-y-z) \left(4s-3\overline{m}_c^2\right)\nonumber\\
&&+\frac{m_s m_c\left[\langle\bar{q}q\rangle\langle\bar{s}g_s\sigma Gs\rangle+\langle\bar{s}s\rangle\langle\bar{q}g_s\sigma Gq\rangle\right]}{144\pi^4}\int dydz    \left( \frac{z}{y}+\frac{y}{z}\right)\nonumber\\
&&-\frac{m_s m_c \left[34\langle\bar{q}q\rangle\langle\bar{s}g_s\sigma Gs\rangle+33\langle\bar{s}s\rangle\langle\bar{q}g_s\sigma Gq\rangle-5\langle\bar{s}s\rangle\langle\bar{s}g_s\sigma Gs\rangle\right]}{1728\pi^4}\int dy \nonumber\\
&&+\frac{m_s m_c \langle\bar{s}s\rangle\left[\langle\bar{q}g_s\sigma Gq\rangle+\langle\bar{s}g_s\sigma Gs\rangle\right]}{2304\pi^4}\int dy \nonumber\\
&&-\frac{m_c\langle\bar{q}q\rangle\langle\bar{s}s\rangle^2}{18\pi^2}\int dy    +\frac{m_s\langle\bar{q}q\rangle\langle\bar{s}s\rangle^2}{18\pi^2}\int dy \,y(1-y) \left[ 1+\frac{s}{3}\delta\left(s-\widetilde{m}_c^2 \right)\right]  \nonumber\\
&&-\frac{\langle\bar{s}g_s\sigma Gs\rangle\left[2\langle\bar{q}g_s\sigma Gq\rangle+\langle\bar{s}g_s\sigma Gs\rangle \right]}{512\pi^4}\int dydz \, (y+z)\left[ 1+\frac{s}{3}\delta\left(s-\overline{m}_c^2 \right)\right] \nonumber\\
&&+\frac{\langle\bar{s}g_s\sigma Gs\rangle\left[2\langle\bar{q}g_s\sigma Gq\rangle+\langle\bar{s}g_s\sigma Gs\rangle \right]}{256\pi^4}\int dy \, y(1-y)\left[ 1+\frac{s}{3}\delta\left(s-\widetilde{m}_c^2 \right)\right] \nonumber\\
&&-\frac{m_s m_c\langle\bar{q}g_s\sigma Gq\rangle\langle\bar{s}g_s\sigma Gs\rangle}{288\pi^4}\int dy \left( \frac{1-y}{y}+\frac{y}{1-y}\right)\delta\left(s-\widetilde{m}_c^2 \right) \nonumber\\
&&+\frac{m_s m_c\langle\bar{s}g_s\sigma Gs\rangle\left[17\langle\bar{q}g_s\sigma Gq\rangle-\langle\bar{s}g_s\sigma Gs\rangle \right]}{1728\pi^4}\int dy \, \left( 1+\frac{s}{2M^2}\right)\delta\left(s-\widetilde{m}_c^2 \right) \nonumber\\
&&-\frac{m_s m_c\langle\bar{s}g_s\sigma Gs\rangle\left[ \langle\bar{q}g_s\sigma Gq\rangle+\langle\bar{s}g_s\sigma Gs\rangle\right]}{6912\pi^4}\int dy \,  \left( 1+\frac{s}{2M^2}\right)\delta\left(s-\widetilde{m}_c^2 \right)\, ,
\end{eqnarray}

\begin{eqnarray}
\widetilde{\rho}^{11,0}_{uss}(s)&=&\frac{1}{122880\pi^8}\int dydz \, (y+z)(1-y-z)^4\left(s-\overline{m}_c^2\right)^4\left(7s-2\overline{m}_c^2 \right)  \nonumber\\
&&+\frac{m_s m_c}{9216\pi^8}\int dydz \,(1-y-z)^3\left(s-\overline{m}_c^2\right)^4  \nonumber\\
&&- \frac{m_c\left[\langle \bar{q}q\rangle+2\langle \bar{s}s\rangle\right]}{1152\pi^6}\int dydz \,  (1-y-z)^2\left(s-\overline{m}_c^2\right)^3  \nonumber \\
&&+ \frac{m_s\left[2\langle \bar{s}s\rangle-\langle \bar{q}q\rangle\right]}{1152\pi^6}\int dydz \, (y+z)(1-y-z)^2\left(s-\overline{m}_c^2\right)^2 \left(5s-2\overline{m}_c^2 \right) \nonumber \\
&&+ \frac{m_c\left[\langle \bar{q}g_s\sigma Gq\rangle+2\langle \bar{s}g_s\sigma Gs\rangle\right]}{512\pi^6}\int dydz  \, (1-y-z) \left(s-\overline{m}_c^2 \right)^2 \nonumber\\
&&- \frac{m_s\left[\langle \bar{q}g_s\sigma Gq\rangle+\langle \bar{s}g_s\sigma Gs\rangle\right]}{768\pi^6}\int dydz  \, (1-y-z)^2 \left(s-\overline{m}_c^2 \right) \left(2s-\overline{m}_c^2 \right) \nonumber\\
&&+ \frac{m_s\left[\langle \bar{q}g_s\sigma Gq\rangle-\langle \bar{s}g_s\sigma Gs\rangle\right]}{384\pi^6}\int dydz  \,(y+z) (1-y-z)  \left(s-\overline{m}_c^2 \right) \left(2s-\overline{m}_c^2 \right) \nonumber\\
&&+ \frac{m_s\left[\langle \bar{q}g_s\sigma Gq\rangle+\langle \bar{s}g_s\sigma Gs\rangle\right]}{1536\pi^6}\int dydz  \, (y+z)(1-y-z) \left(s-\overline{m}_c^2 \right) \left(2s-\overline{m}_c^2 \right) \nonumber\\
&&+\frac{\langle\bar{s}s\rangle\left[ 2\langle\bar{q}q\rangle+\langle\bar{s}s\rangle\right]}{144\pi^4}\int dydz \,  (y+z)(1-y-z)\left(s-\overline{m}_c^2 \right)\left(2s-\overline{m}_c^2 \right) \nonumber\\
&&+\frac{ m_s m_c\langle\bar{s}s\rangle\left[11 \langle\bar{q}q\rangle-\langle\bar{s}s\rangle\right]}{72\pi^4}\int dydz \,   \left(s-\overline{m}_c^2 \right)
\nonumber\\
&&-\frac{5\left[\langle\bar{q}q\rangle\langle\bar{s}g_s\sigma Gs\rangle+\langle\bar{s}s\rangle\langle\bar{q}g_s\sigma Gq\rangle+\langle\bar{s}s\rangle\langle\bar{s}g_s\sigma Gs\rangle\right]}{1152\pi^4}\int dydz \,(y+z) \left(3s-2\overline{m}_c^2\right)\nonumber\\
&&+\frac{\left[\langle\bar{q}q\rangle\langle\bar{s}g_s\sigma Gs\rangle+\langle\bar{s}s\rangle\langle\bar{q}g_s\sigma Gq\rangle+\langle\bar{s}s\rangle\langle\bar{s}g_s\sigma Gs\rangle\right]}{288\pi^4}\int dydz \,(1-y-z) \left(3s-2\overline{m}_c^2\right)\nonumber\\
&&+\frac{m_s m_c\left[\langle\bar{q}q\rangle\langle\bar{s}g_s\sigma Gs\rangle+\langle\bar{s}s\rangle\langle\bar{q}g_s\sigma Gq\rangle\right]}{144\pi^4}\int dydz \left( \frac{1}{y}+\frac{1}{z}\right)     \nonumber\\
&&-\frac{m_s m_c\left[34\langle\bar{q}q\rangle\langle\bar{s}g_s\sigma Gs\rangle+33\langle\bar{s}s\rangle\langle\bar{q}g_s\sigma Gq\rangle-5\langle\bar{s}s\rangle\langle\bar{s}g_s\sigma Gs\rangle\right]}{864\pi^4}\int dy     \nonumber\\
&&+\frac{m_s m_c\langle\bar{s}s\rangle\left[\langle\bar{q}g_s\sigma Gq\rangle+\langle\bar{s}g_s\sigma Gs\rangle\right]}{1152\pi^4}\int dy     \nonumber\\
&&-\frac{m_c\langle\bar{q}q\rangle\langle\bar{s}s\rangle^2}{9\pi^2}\int dy    +\frac{m_s\langle\bar{q}q\rangle\langle\bar{s}s\rangle^2}{54\pi^2}\int dy \,  \left[ 1+\frac{s}{2}\delta\left(s-\widetilde{m}_c^2 \right)\right]  \nonumber\\
&&-\frac{\langle\bar{s}g_s\sigma Gs\rangle\left[2\langle\bar{q}g_s\sigma Gq\rangle+\langle\bar{s}g_s\sigma Gs\rangle \right]}{768\pi^4}\int dydz \,  \left[ 1+\frac{s}{2}\delta\left(s-\overline{m}_c^2 \right)\right] \nonumber\\
&&-\frac{m_s m_c\langle\bar{q}g_s\sigma Gq\rangle\langle\bar{s}g_s\sigma Gs\rangle}{288\pi^4}\int dy \, \left( \frac{1}{y}+\frac{1}{1-y}\right)\delta\left(s-\widetilde{m}_c^2 \right) \nonumber\\
&&+\frac{m_s m_c\langle\bar{s}g_s\sigma Gs\rangle\left[17\langle\bar{q}g_s\sigma Gq\rangle-\langle\bar{s}g_s\sigma Gs\rangle \right]}{1728\pi^4}\int dy \, \left( 1+\frac{s}{M^2}\right)\delta\left(s-\widetilde{m}_c^2 \right) \nonumber\\
&&-\frac{m_s m_c\langle\bar{s}g_s\sigma Gs\rangle\left[ \langle\bar{q}g_s\sigma Gq\rangle+\langle\bar{s}g_s\sigma Gs\rangle\right]}{6912\pi^4}\int dy \, \left( 1+\frac{s}{M^2}\right)\delta\left(s-\widetilde{m}_c^2 \right) \, ,
\end{eqnarray}

\begin{eqnarray}
\rho^{11,1}_{uus}(s)&=&\frac{1}{61440\pi^8}\int dydz \, yz(1-y-z)^4\left(s-\overline{m}_c^2\right)^4\left(8s-3\overline{m}_c^2 \right)  \nonumber\\
&&+\frac{m_s m_c}{36864\pi^8}\int dydz \, (y+z)(1-y-z)^3\left(s-\overline{m}_c^2\right)^4  \nonumber\\
&&- \frac{m_c\left[2\langle \bar{q}q\rangle+\langle \bar{s}s\rangle\right]}{2304\pi^6}\int dydz \, (y+z)(1-y-z)^2\left(s-\overline{m}_c^2\right)^3  \nonumber \\
&&+ \frac{m_s\left[3\langle \bar{s}s\rangle-2\langle \bar{q}q\rangle\right]}{384\pi^6}\int dydz \, yz(1-y-z)^2\left(s-\overline{m}_c^2\right)^2 \left(2s-\overline{m}_c^2 \right) \nonumber \\
&&+ \frac{m_c\left[2\langle \bar{q}g_s\sigma Gq\rangle+\langle \bar{s}g_s\sigma Gs\rangle\right]}{1024\pi^6}\int dydz  \, (y+z)(1-y-z) \left(s-\overline{m}_c^2 \right)^2 \nonumber\\
&&- \frac{m_s\left[3\langle \bar{q}g_s\sigma Gq\rangle+\langle \bar{s}g_s\sigma Gs\rangle\right]}{12288\pi^6}\int dydz  \, (y+z)(1-y-z)^2 \left(s-\overline{m}_c^2 \right)\left(5s-3\overline{m}_c^2 \right) \nonumber\\
&&+ \frac{m_s\left[\langle \bar{q}g_s\sigma Gq\rangle-\langle \bar{s}g_s\sigma Gs\rangle\right]}{384\pi^6}\int dydz  \, yz(1-y-z) \left(s-\overline{m}_c^2 \right)\left(5s-3\overline{m}_c^2 \right) \nonumber\\
&&+ \frac{m_s\langle \bar{q}g_s\sigma Gq\rangle}{1536\pi^6}\int dydz  \, yz(1-y-z) \left(s-\overline{m}_c^2 \right)\left(5s-3\overline{m}_c^2 \right) \nonumber\\
&&+\frac{\langle\bar{q}q\rangle \left[\langle\bar{q}q\rangle +2\langle\bar{s}s\rangle  \right]}{144\pi^4}\int dydz \,  yz(1-y-z)\left(s-\overline{m}_c^2 \right)\left(5s-3\overline{m}_c^2 \right) \nonumber\\
&&+\frac{  m_s m_c\langle\bar{q}q\rangle \left[6\langle\bar{q}q\rangle -\langle\bar{s}s\rangle  \right]}{144\pi^4}\int dydz \,  (y+z)\left(s-\overline{m}_c^2 \right) \nonumber\\
&&-\frac{5\left[\langle\bar{q}q\rangle\langle\bar{q}g_s\sigma Gq\rangle+\langle\bar{q}q\rangle\langle\bar{s}g_s\sigma Gs\rangle+\langle\bar{s}s\rangle\langle\bar{q}g_s\sigma Gq\rangle\right]}{576\pi^4}\int dydz \,yz \left(4s-3\overline{m}_c^2\right)\nonumber\\
&&+\frac{\langle\bar{q}q\rangle\langle\bar{q}g_s\sigma Gq\rangle+\langle\bar{q}q\rangle\langle\bar{s}g_s\sigma Gs\rangle+\langle\bar{s}s\rangle\langle\bar{q}g_s\sigma Gq\rangle}{576\pi^4}\int dydz \,(y+z)(1-y-z) \left(4s-3\overline{m}_c^2\right)\nonumber\\
 &&+\frac{m_s m_c\langle\bar{q}q\rangle\langle\bar{q}g_s\sigma Gq\rangle}{144\pi^4}\int dydz    \left( \frac{z}{y}+\frac{y}{z}\right)\nonumber\\
 &&-\frac{m_s m_c \left[36\langle\bar{q}q\rangle\langle\bar{q}g_s\sigma Gq\rangle-2\langle\bar{q}q\rangle\langle\bar{s}g_s\sigma Gs\rangle-3\langle\bar{s}s\rangle\langle\bar{q}g_s\sigma Gq\rangle\right]}{1728\pi^4}\int dy \nonumber\\
&&+\frac{m_s m_c \langle\bar{s}s\rangle\langle\bar{q}g_s\sigma Gq\rangle}{2304\pi^4}\int dy \nonumber\\
&&-\frac{m_c\langle\bar{q}q\rangle^2\langle\bar{s}s\rangle}{18\pi^2}\int dy   +\frac{m_s\langle\bar{q}q\rangle^2\langle\bar{s}s\rangle}{36\pi^2}\int dy \,y(1-y) \left[ 1+\frac{s}{3}\delta\left(s-\widetilde{m}_c^2 \right)\right]  \nonumber\\
&&-\frac{\langle\bar{q}g_s\sigma Gq\rangle \left[\langle\bar{q}g_s\sigma Gq\rangle +2\langle\bar{s}g_s\sigma Gs\rangle  \right]}{512\pi^4}\int dydz \, (y+z)\left[ 1+\frac{s}{3}\delta\left(s-\overline{m}_c^2 \right)\right] \nonumber\\
&&+\frac{\langle\bar{q}g_s\sigma Gq\rangle\left[\langle\bar{q}g_s\sigma Gq\rangle +2\langle\bar{s}g_s\sigma Gs\rangle\right] }{256\pi^4}\int dy \, y(1-y)\left[ 1+\frac{s}{3}\delta\left(s-\widetilde{m}_c^2 \right)\right] \nonumber\\
&&-\frac{m_s m_c\langle\bar{q}g_s\sigma Gq\rangle^2}{576\pi^4}\int dy \left( \frac{1-y}{y}+\frac{y}{1-y}\right)\delta\left(s-\widetilde{m}_c^2 \right) \nonumber\\
&&+\frac{m_s m_c\langle\bar{q}g_s\sigma Gq\rangle\left[ 9\langle\bar{q}g_s\sigma Gq\rangle-\langle\bar{s}g_s\sigma Gs\rangle\right]}{1728\pi^4}\int dy \, \left( 1+\frac{s}{2M^2}\right)\delta\left(s-\widetilde{m}_c^2 \right) \nonumber\\
&&-\frac{m_s m_c\langle\bar{q}g_s\sigma Gq\rangle \langle\bar{s}g_s\sigma Gs\rangle}{6912\pi^4}\int dy \,  \left( 1+\frac{s}{2M^2}\right)\delta\left(s-\widetilde{m}_c^2 \right)\, ,
\end{eqnarray}

\begin{eqnarray}
\widetilde{\rho}^{11,0}_{uus}(s)&=&\frac{1}{122880\pi^8}\int dydz \, (y+z)(1-y-z)^4\left(s-\overline{m}_c^2\right)^4\left(7s-2\overline{m}_c^2 \right)  \nonumber\\
&&+\frac{m_s m_c}{18432\pi^8}\int dydz \,(1-y-z)^3\left(s-\overline{m}_c^2\right)^4  \nonumber\\
&&- \frac{m_c\left[2\langle \bar{q}q\rangle+\langle \bar{s}s\rangle\right]}{1152\pi^6}\int dydz \,  (1-y-z)^2\left(s-\overline{m}_c^2\right)^3  \nonumber \\
&&+ \frac{m_s\left[3\langle \bar{s}s\rangle-2\langle \bar{q}q\rangle\right]}{2304\pi^6}\int dydz \, (y+z)(1-y-z)^2\left(s-\overline{m}_c^2\right)^2 \left(5s-2\overline{m}_c^2 \right) \nonumber \\
&&+ \frac{m_c\left[2\langle \bar{q}g_s\sigma Gq\rangle+\langle \bar{s}g_s\sigma Gs\rangle\right]}{512\pi^6}\int dydz  \, (1-y-z) \left(s-\overline{m}_c^2 \right)^2 \nonumber\\
&&- \frac{m_s\left[3\langle \bar{q}g_s\sigma Gq\rangle+\langle \bar{s}g_s\sigma Gs\rangle\right]}{3072\pi^6}\int dydz  \, (1-y-z)^2 \left(s-\overline{m}_c^2 \right) \left(2s-\overline{m}_c^2 \right) \nonumber\\
&&+ \frac{m_s\left[\langle \bar{q}g_s\sigma Gq\rangle-\langle \bar{s}g_s\sigma Gs\rangle\right]}{384\pi^6}\int dydz  \, (y+z)(1-y-z) \left(s-\overline{m}_c^2 \right) \left(2s-\overline{m}_c^2 \right) \nonumber\\
&&+ \frac{m_s\langle \bar{q}g_s\sigma Gq\rangle}{1536\pi^6}\int dydz  \, (y+z)(1-y-z) \left(s-\overline{m}_c^2 \right) \left(2s-\overline{m}_c^2 \right) \nonumber\\
&&+\frac{\langle\bar{q}q\rangle\left[\langle\bar{q}q\rangle+2\langle\bar{s}s\rangle \right]}{144\pi^4}\int dydz \,  (y+z)(1-y-z)\left(s-\overline{m}_c^2 \right)\left(2s-\overline{m}_c^2 \right) \nonumber\\
&&+\frac{ m_s m_c\langle\bar{q}q\rangle\left[6\langle\bar{q}q\rangle-\langle\bar{s}s\rangle \right]}{72\pi^4}\int dydz \,   \left(s-\overline{m}_c^2 \right)
\nonumber\\
&&-\frac{5\left[\langle\bar{q}q\rangle\langle\bar{q}g_s\sigma Gq\rangle+\langle\bar{q}q\rangle\langle\bar{s}g_s\sigma Gs\rangle+\langle\bar{s}s\rangle\langle\bar{q}g_s\sigma Gq\rangle\right]}{1152\pi^4}\int dydz \,(y+z) \left(3s-2\overline{m}_c^2\right)\nonumber\\
&&+\frac{\langle\bar{q}q\rangle\langle\bar{q}g_s\sigma Gq\rangle+\langle\bar{q}q\rangle\langle\bar{s}g_s\sigma Gs\rangle+\langle\bar{s}s\rangle\langle\bar{q}g_s\sigma Gq\rangle}{288\pi^4}\int dydz \,(1-y-z) \left(3s-2\overline{m}_c^2\right)\nonumber\\
&&+\frac{m_s m_c\langle\bar{q}q\rangle\langle\bar{q}g_s\sigma Gq\rangle}{144\pi^4}\int dydz \left( \frac{1}{y}+\frac{1}{z}\right)  +\frac{m_s m_c\langle\bar{s}s\rangle\langle\bar{q}g_s\sigma Gq\rangle}{1152\pi^4}\int dy    \nonumber\\
&&-\frac{m_s m_c\left[36\langle\bar{q}q\rangle\langle\bar{q}g_s\sigma Gq\rangle-2\langle\bar{q}q\rangle\langle\bar{s}g_s\sigma Gs\rangle-3\langle\bar{s}s\rangle\langle\bar{q}g_s\sigma Gq\rangle\right]}{864\pi^4}\int dy     \nonumber\\
&&-\frac{m_c\langle\bar{q}q\rangle^2\langle\bar{s}s\rangle}{9\pi^2}\int dy    +\frac{m_s\langle\bar{q}q\rangle^2\langle\bar{s}s\rangle}{108\pi^2}\int dy \,  \left[ 1+\frac{s}{2}\delta\left(s-\widetilde{m}_c^2 \right)\right]  \nonumber\\
&&-\frac{\langle\bar{q}g_s\sigma Gq\rangle \left[ \langle\bar{q}g_s\sigma Gq\rangle+2\langle\bar{s}g_s\sigma Gs\rangle\right]}{768\pi^4}\int dydz \,  \left[ 1+\frac{s}{2}\delta\left(s-\overline{m}_c^2 \right)\right] \nonumber\\
&&-\frac{m_s m_c\langle\bar{q}g_s\sigma Gq\rangle^2}{576\pi^4}\int dy \, \left( \frac{1}{y}+\frac{1}{1-y}\right)\delta\left(s-\widetilde{m}_c^2 \right) \nonumber\\
&&+\frac{m_s m_c\langle\bar{q}g_s\sigma Gq\rangle \left[9\langle\bar{q}g_s\sigma Gq\rangle-\langle\bar{s}g_s\sigma Gs\rangle \right]}{1728\pi^4}\int dy \, \left( 1+\frac{s}{M^2}\right)\delta\left(s-\widetilde{m}_c^2 \right) \nonumber\\
&&-\frac{m_s m_c\langle\bar{q}g_s\sigma Gq\rangle \langle\bar{s}g_s\sigma Gs\rangle}{6912\pi^4}\int dy \, \left( 1+\frac{s}{M^2}\right)\delta\left(s-\widetilde{m}_c^2 \right) \, ,
\end{eqnarray}

\begin{eqnarray}
\rho^{11,1}_{uuu}(s)&=&\rho^{11,1}_{sss}(s)\mid_{m_s \to 0,\,\,\langle\bar{s}s\rangle\to\langle\bar{q}q\rangle,\,\,\langle\bar{s}g_s\sigma Gs\rangle\to\langle\bar{q}g_s\sigma Gq\rangle} \, ,  \\
\widetilde{\rho}^{11,0}_{uuu}(s)&=&\widetilde{\rho}^{11,0}_{sss}(s)\mid_{m_s \to 0,\,\,\langle\bar{s}s\rangle\to\langle\bar{q}q\rangle,\,\,\langle\bar{s}g_s\sigma Gs\rangle\to\langle\bar{q}g_s\sigma Gq\rangle} \, ,
\end{eqnarray}

\begin{eqnarray}
\rho^{10,1}_{sss}(s)&=&\frac{1}{122880\pi^8}\int dydz \, yz(1-y-z)^4\left(s-\overline{m}_c^2\right)^4\left(8s-3\overline{m}_c^2 \right)  \nonumber\\
&&+\frac{1}{153600\pi^8}\int dydz \, yz(1-y-z)^5\left(s-\overline{m}_c^2\right)^3\left(18s^2-16s\overline{m}_c^2+3\overline{m}_c^4 \right)  \nonumber\\
&&+\frac{m_s m_c}{12288\pi^8}\int dydz \, (y+z)(1-y-z)^3\left(s-\overline{m}_c^2\right)^4  \nonumber\\
&&+\frac{m_s m_c}{49152\pi^8}\int dydz \, (y+z)(1-y-z)^4\left(s-\overline{m}_c^2\right)^3 \left(7s-3\overline{m}_c^2 \right) \nonumber\\
&&- \frac{m_c\langle \bar{s}s\rangle}{768\pi^6}\int dydz \, (y+z)(1-y-z)^2\left(s-\overline{m}_c^2\right)^3  \nonumber \\
&&- \frac{m_c\langle \bar{s}s\rangle}{768\pi^6}\int dydz \, (y+z)(1-y-z)^3\left(s-\overline{m}_c^2\right)^2\left(2s-\overline{m}_c^2\right)  \nonumber \\
&&- \frac{m_s\langle \bar{s}s\rangle}{256\pi^6}\int dydz \, yz(1-y-z)^2\left(s-\overline{m}_c^2\right)^2 \left(2s-\overline{m}_c^2 \right) \nonumber \\
&&+ \frac{m_s\langle \bar{s}s\rangle}{128\pi^6}\int dydz \, yz(1-y-z)^3\left(s-\overline{m}_c^2\right)
\left(7s^2-8s\overline{m}_c^2+2\overline{m}_c^4 \right) \nonumber\\
&&+ \frac{11m_c \langle \bar{s}g_s\sigma Gs\rangle }{4096\pi^6}\int dydz  \, (y+z)(1-y-z)  \left(s-\overline{m}_c^2 \right)^2 \nonumber\\
&&+ \frac{11m_c \langle \bar{s}g_s\sigma Gs\rangle }{8192\pi^6}\int dydz  \, \left(y+z\right)(1-y-z)^2 \left(s-\overline{m}_c^2 \right) \left(5s-3\overline{m}_c^2 \right) \nonumber\\
&&- \frac{7m_c \langle \bar{s}g_s\sigma Gs\rangle }{8192\pi^6}\int dydz  \, \left(\frac{z}{y}+\frac{y}{z}\right)(1-y-z)^2 \left(s-\overline{m}_c^2 \right)^2 \nonumber\\
&&- \frac{m_c \langle \bar{s}g_s\sigma Gs\rangle }{4096\pi^6}\int dydz  \, \left(\frac{z}{y}+\frac{y}{z}\right)(1-y-z)^3 \left(s-\overline{m}_c^2 \right) \left(5s-3\overline{m}_c^2 \right) \nonumber\\
&&+ \frac{11m_s\langle \bar{s}g_s\sigma Gs\rangle}{2048\pi^6}\int dydz  \, yz(1-y-z) \left(s-\overline{m}_c^2 \right)\left(5s-3\overline{m}_c^2 \right) \nonumber\\
&&- \frac{m_s\langle \bar{s}g_s\sigma Gs\rangle}{256\pi^6}\int dydz  \, yz(1-y-z)^2  \left(15s^2-20s\overline{m}_c^2+6\overline{m}_c^4 \right) \nonumber\\
&&+ \frac{3m_s\langle \bar{s}g_s\sigma Gs\rangle}{8192\pi^6}\int dydz  \, (y+z)(1-y-z)^2 \left(s-\overline{m}_c^2 \right)\left(5s-3\overline{m}_c^2 \right) \nonumber\\
&&+\frac{\langle\bar{s}s\rangle^2}{48\pi^4}\int dydz \,  yz(1-y-z)\left(s-\overline{m}_c^2 \right)\left(5s-3\overline{m}_c^2 \right) \nonumber\\
&&+\frac{ m_s m_c\langle\bar{s}s\rangle^2}{24\pi^4}\int dydz \,  (y+z)\left(s-\overline{m}_c^2 \right) \nonumber\\
&&-\frac{ m_s m_c\langle\bar{s}s\rangle^2}{48\pi^4}\int dydz \,  (y+z)(1-y-z)\left(4s-3\overline{m}_c^2 \right) \nonumber
\end{eqnarray}
\begin{eqnarray}
&&-\frac{19\langle\bar{s}s\rangle\langle\bar{s}g_s\sigma Gs\rangle}{768\pi^4}\int dydz \,yz \left(4s-3\overline{m}_c^2\right)\nonumber\\
&&-\frac{\langle\bar{s}s\rangle\langle\bar{s}g_s\sigma Gs\rangle}{512\pi^4}\int dydz \,(y+z)(1-y-z) \left(4s-3\overline{m}_c^2\right)\nonumber\\
&&+\frac{89m_s m_c \langle\bar{s}s\rangle\langle\bar{s}g_s\sigma Gs\rangle}{3072\pi^4}\int dydz \,(y+z)\left[ 1+\frac{s}{3}\delta\left(s-\overline{m}_c^2 \right)\right] \nonumber\\
&&-\frac{181m_s m_c \langle\bar{s}s\rangle\langle\bar{s}g_s\sigma Gs\rangle}{9216\pi^4}\int dy  +\frac{3m_s m_c\langle\bar{s}s\rangle\langle\bar{s}g_s\sigma Gs\rangle}{512\pi^4}\int dydz    \left( \frac{z}{y}+\frac{y}{z}\right)\nonumber\\
&&-\frac{3m_s m_c\langle\bar{s}s\rangle\langle\bar{s}g_s\sigma Gs\rangle}{256\pi^4}\int dydz    \left( \frac{z}{y}+\frac{y}{z}\right)(1-y-z)\left[ 1+\frac{s}{3}\delta\left(s-\overline{m}_c^2 \right)\right]\nonumber\\
&&-\frac{m_c\langle\bar{s}s\rangle^3}{36\pi^2}\int dy    +\frac{m_s\langle\bar{s}s\rangle^3}{12\pi^2}\int dy \,y(1-y) \left[ 1+\frac{s}{3}\delta\left(s-\widetilde{m}_c^2 \right)\right]  \nonumber\\
&&+\frac{43\langle\bar{s}g_s\sigma Gs\rangle^2}{18432\pi^4}\int dydz \, (y+z)\left[ 1+\frac{s}{3}\delta\left(s-\overline{m}_c^2 \right)\right] \nonumber\\
&&+\frac{11\langle\bar{s}g_s\sigma Gs\rangle^2}{1024\pi^4}\int dy \, y(1-y)\left[ 1+\frac{s}{3}\delta\left(s-\widetilde{m}_c^2 \right)\right] \nonumber\\
&&+\frac{m_s m_c\langle\bar{s}g_s\sigma Gs\rangle^2}{768\pi^4}\int dydz    \left( \frac{z}{y}+\frac{y}{z}\right) \left( 1+\frac{s}{2M^2}\right)\delta\left(s-\overline{m}_c^2 \right)\nonumber\\
&&-\frac{17m_s m_c\langle\bar{s}g_s\sigma Gs\rangle^2}{9216\pi^4}\int dy \,  \left( \frac{1-y}{y}+\frac{y}{1-y}\right)\delta\left(s-\widetilde{m}_c^2 \right) \nonumber\\
&&+\frac{17m_s m_c\langle\bar{s}g_s\sigma Gs\rangle^2}{4608\pi^4}\int dy   \left( 1+\frac{s}{2M^2}\right)\delta\left(s-\widetilde{m}_c^2 \right) \, ,
\end{eqnarray}

\begin{eqnarray}
\widetilde{\rho}^{10,0}_{sss}(s)&=&\frac{1}{122880\pi^8}\int dydz \, (y+z)(1-y-z)^4\left(s-\overline{m}_c^2\right)^4\left(7s-2\overline{m}_c^2 \right)  \nonumber\\
&&-\frac{1}{614400\pi^8}\int dydz \, (y+z)(1-y-z)^5\left(s-\overline{m}_c^2\right)^3\left(28s^2-21s\overline{m}_c^2+3\overline{m}_c^4 \right)  \nonumber\\
&&+\frac{m_s m_c}{3072\pi^8}\int dydz \, (1-y-z)^3\left(s-\overline{m}_c^2\right)^4  \nonumber\\
&&-\frac{m_s m_c}{12288\pi^8}\int dydz \, (1-y-z)^4\left(s-\overline{m}_c^2\right)^3 \left(3s-\overline{m}_c^2 \right) \nonumber\\
&&- \frac{m_c\langle \bar{s}s\rangle}{192\pi^6}\int dydz \, (1-y-z)^2\left(s-\overline{m}_c^2\right)^3  \nonumber \\
&&+ \frac{m_c\langle \bar{s}s\rangle}{1152\pi^6}\int dydz \, (1-y-z)^3\left(s-\overline{m}_c^2\right)^2\left(5s-2\overline{m}_c^2\right)  \nonumber \\
&&- \frac{m_s\langle \bar{s}s\rangle}{768\pi^6}\int dydz \, (y+z)(1-y-z)^2\left(s-\overline{m}_c^2\right)^2 \left(5s-2\overline{m}_c^2 \right) \nonumber \\
&&- \frac{m_s\langle \bar{s}s\rangle}{256\pi^6}\int dydz \, (y+z)(1-y-z)^3\left(s-\overline{m}_c^2\right)
\left(5s^2-5s\overline{m}_c^2+\overline{m}_c^4 \right) \nonumber
\end{eqnarray}
\begin{eqnarray}
&&+ \frac{11m_c\langle \bar{s}g_s\sigma Gs\rangle}{1024\pi^6}\int dydz  \, (1-y-z) \left(s-\overline{m}_c^2 \right)^2 \nonumber\\
&&- \frac{11m_c\langle \bar{s}g_s\sigma Gs\rangle}{2048\pi^6}\int dydz  \, (1-y-z)^2 \left(s-\overline{m}_c^2 \right)\left(2s-\overline{m}_c^2 \right)  \nonumber\\
&&- \frac{7m_c\langle \bar{s}g_s\sigma Gs\rangle}{4096\pi^6}\int dydz  \, \left(\frac{1}{y}+\frac{1}{z}\right)(1-y-z)^2 \left(s-\overline{m}_c^2 \right)^2 \nonumber\\
&&+ \frac{m_c\langle \bar{s}g_s\sigma Gs\rangle}{2048\pi^6}\int dydz  \, \left(\frac{1}{y}+\frac{1}{z}\right)(1-y-z)^3 \left(s-\overline{m}_c^2 \right)\left(2s-\overline{m}_c^2 \right)  \nonumber\\
&&+ \frac{m_s\langle \bar{s}g_s\sigma Gs\rangle}{512\pi^6}\int dydz  \, (y+z)(1-y-z)^2  \left(10s^2-12s\overline{m}_c^2+3\overline{m}_c^4 \right) \nonumber\\
&&+ \frac{11m_s\langle \bar{s}g_s\sigma Gs\rangle}{1024\pi^6}\int dydz  \, (y+z)(1-y-z) \left(s-\overline{m}_c^2 \right)\left(2s-\overline{m}_c^2 \right) \nonumber\\
&&+\frac{\langle\bar{s}s\rangle^2}{24\pi^4}\int dydz \,  (y+z)(1-y-z)\left(s-\overline{m}_c^2 \right)\left(2s-\overline{m}_c^2 \right) \nonumber\\
&&+\frac{ m_s m_c\langle\bar{s}s\rangle^2}{6\pi^4}\int dydz \,  \left(s-\overline{m}_c^2 \right) \nonumber\\
&&+\frac{ m_s m_c\langle\bar{s}s\rangle^2}{24\pi^4}\int dydz \,  (1-y-z)\left(3s-2\overline{m}_c^2 \right)
\nonumber\\
&&-\frac{19\langle\bar{s}s\rangle\langle\bar{s}g_s\sigma Gs\rangle}{768\pi^4}\int dydz \,(y+z) \left(3s-2\overline{m}_c^2\right)\nonumber\\
&&-\frac{89m_s m_c \langle\bar{s}s\rangle\langle\bar{s}g_s\sigma Gs\rangle}{2304\pi^4}\int dydz  \left[ 1+\frac{s}{2}\delta\left(s-\overline{m}_c^2 \right)\right] -\frac{199m_s m_c \langle\bar{s}s\rangle\langle\bar{s}g_s\sigma Gs\rangle}{2304\pi^4}\int dy  \nonumber\\
&&+\frac{3m_s m_c\langle\bar{s}s\rangle\langle\bar{s}g_s\sigma Gs\rangle}{256\pi^4}\int dydz    \left( \frac{1}{y}+\frac{1}{z}\right)\nonumber\\
&&+\frac{m_s m_c\langle\bar{s}s\rangle\langle\bar{s}g_s\sigma Gs\rangle}{128\pi^4}\int dydz    \left( \frac{1}{y}+\frac{1}{z}\right)(1-y-z)\left[1+\frac{s}{2}\delta\left(s-\overline{m}_c^2 \right) \right]\nonumber\\
&&-\frac{m_c\langle\bar{s}s\rangle^3}{9\pi^2}\int dy   +\frac{m_s\langle\bar{s}s\rangle^3}{18\pi^2}\int dy  \left[ 1+\frac{s}{2}\delta\left(s-\widetilde{m}_c^2 \right)\right]  \nonumber\\
&&+\frac{11\langle\bar{s}g_s\sigma Gs\rangle^2}{1536\pi^4}\int dy  \left[ 1+\frac{s}{2}\delta\left(s-\widetilde{m}_c^2 \right)\right] \nonumber\\
&&-\frac{m_s m_c\langle\bar{s}g_s\sigma Gs\rangle^2}{1536\pi^4}\int dydz    \left( \frac{1}{y}+\frac{1}{z}\right) \left(1+\frac{s}{M^2} \right)\delta\left(s-\overline{m}_c^2 \right)\nonumber\\
&&-\frac{17m_s m_c\langle\bar{s}g_s\sigma Gs\rangle^2}{4608\pi^4}\int dy \left( \frac{1}{y}+\frac{1}{1-y}\right)  \delta\left(s-\widetilde{m}_c^2 \right) \nonumber\\
&&+\frac{125m_s m_c\langle\bar{s}g_s\sigma Gs\rangle^2}{9216\pi^4}\int dy   \left( 1+\frac{s}{M^2}\right)\delta\left(s-\widetilde{m}_c^2 \right) \, ,
\end{eqnarray}

\begin{eqnarray}
\rho^{10,1}_{uss}(s)&=&\frac{1}{122880\pi^8}\int dydz \, yz(1-y-z)^4\left(s-\overline{m}_c^2\right)^4\left(8s-3\overline{m}_c^2 \right)  \nonumber\\
&&+\frac{1}{153600\pi^8}\int dydz \, yz(1-y-z)^5\left(s-\overline{m}_c^2\right)^3\left(18s^2-16s\overline{m}_c^2+3\overline{m}_c^4 \right)  \nonumber\\
&&+\frac{m_s m_c}{18432\pi^8}\int dydz \, (y+z)(1-y-z)^3\left(s-\overline{m}_c^2\right)^4  \nonumber\\
&&+\frac{m_s m_c}{73728\pi^8}\int dydz \, (y+z)(1-y-z)^4\left(s-\overline{m}_c^2\right)^3 \left(7s-3\overline{m}_c^2 \right) \nonumber\\
&&- \frac{m_c\left[\langle \bar{q}q\rangle+2\langle \bar{s}s\rangle\right]}{2304\pi^6}\int dydz \, (y+z)(1-y-z)^2\left(s-\overline{m}_c^2\right)^3  \nonumber \\
&&- \frac{m_c\left[\langle \bar{q}q\rangle+2\langle \bar{s}s\rangle\right]}{2304\pi^6}\int dydz \, (y+z)(1-y-z)^3\left(s-\overline{m}_c^2\right)^2\left(2s-\overline{m}_c^2\right)  \nonumber \\
&&- \frac{m_s\left[2\langle \bar{q}q\rangle-\langle \bar{s}s\rangle\right]}{384\pi^6}\int dydz \, yz(1-y-z)^2\left(s-\overline{m}_c^2\right)^2 \left(2s-\overline{m}_c^2 \right) \nonumber \\
&&+ \frac{m_s\langle \bar{s}s\rangle}{192\pi^6}\int dydz \, yz(1-y-z)^3\left(s-\overline{m}_c^2\right)
\left(7s^2-8s\overline{m}_c^2+2\overline{m}_c^4 \right) \nonumber\\
&&+ \frac{11m_c\left[\langle \bar{q}g_s\sigma Gq\rangle+2\langle \bar{s}g_s\sigma Gs\rangle\right]}{12288\pi^6}\int dydz  \, (y+z)(1-y-z) \left(s-\overline{m}_c^2 \right)^2 \nonumber\\
&&+ \frac{11m_c \left[\langle \bar{q}g_s\sigma Gq\rangle+2\langle \bar{s}g_s\sigma Gs\rangle\right] }{24576\pi^6}\int dydz  \, \left(y+z\right)(1-y-z)^2 \left(s-\overline{m}_c^2 \right) \left(5s-3\overline{m}_c^2 \right) \nonumber\\
&&-\frac{7m_c \left[\langle \bar{q}g_s\sigma Gq\rangle+2\langle \bar{s}g_s\sigma Gs\rangle\right] }{24576\pi^6}\int dydz  \, \left(\frac{z}{y}+\frac{y}{z}\right)(1-y-z)^2 \left(s-\overline{m}_c^2 \right)^2 \nonumber\\
&&- \frac{m_c \left[\langle \bar{q}g_s\sigma Gq\rangle+2\langle \bar{s}g_s\sigma Gs\rangle\right] }{12288\pi^6}\int dydz  \, \left(\frac{z}{y}+\frac{y}{z}\right)(1-y-z)^3 \left(s-\overline{m}_c^2 \right) \left(5s-3\overline{m}_c^2 \right) \nonumber\\
&&- \frac{m_s\langle \bar{s}g_s\sigma Gs\rangle}{384\pi^6}\int dydz  \, yz(1-y-z)^2  \left(15s^2-20s\overline{m}_c^2+6\overline{m}_c^4 \right) \nonumber\\
&&+ \frac{m_s\left[\langle \bar{q}g_s\sigma Gq\rangle+\langle \bar{s}g_s\sigma Gs\rangle\right]}{8192\pi^6}\int dydz  \, (y+z)(1-y-z)^2 \left(s-\overline{m}_c^2 \right)\left(5s-3\overline{m}_c^2 \right) \nonumber\\
&&+ \frac{m_s\langle \bar{q}g_s\sigma Gq\rangle}{384\pi^6}\int dydz  \, yz(1-y-z) \left(s-\overline{m}_c^2 \right)\left(5s-3\overline{m}_c^2 \right) \nonumber\\
&&+ \frac{m_s\left[\langle \bar{q}g_s\sigma Gq\rangle+\langle \bar{s}g_s\sigma Gs\rangle\right]}{2048\pi^6}\int dydz  \, yz(1-y-z) \left(s-\overline{m}_c^2 \right)\left(5s-3\overline{m}_c^2 \right) \nonumber\\
&&+\frac{\langle\bar{s}s\rangle \left[2\langle\bar{q}q\rangle+\langle\bar{s}s\rangle \right]}{144\pi^4}\int dydz \,  yz(1-y-z)\left(s-\overline{m}_c^2 \right)\left(5s-3\overline{m}_c^2 \right) \nonumber\\
&&-\frac{ m_s m_c\langle\bar{s}s\rangle\left[5\langle\bar{q}q\rangle-\langle\bar{s}s\rangle \right]}{144\pi^4}\int dydz \,  (y+z)\left(s-\overline{m}_c^2 \right) \nonumber\\
&&-\frac{ m_s m_c\langle\bar{s}s\rangle \left[\langle\bar{q}q\rangle+\langle\bar{s}s\rangle \right]}{144\pi^4}\int dydz \,  (y+z)(1-y-z)\left(4s-3\overline{m}_c^2 \right) \nonumber
\end{eqnarray}
\begin{eqnarray}
&&-\frac{19\left[\langle\bar{q}q\rangle\langle\bar{s}g_s\sigma Gs\rangle+\langle\bar{s}s\rangle\langle\bar{q}g_s\sigma Gq\rangle+\langle\bar{s}s\rangle\langle\bar{s}g_s\sigma Gs\rangle\right]}{2304\pi^4}\int dydz \,yz \left(4s-3\overline{m}_c^2\right)\nonumber\\
&&-\frac{\langle\bar{q}q\rangle\langle\bar{s}g_s\sigma Gs\rangle+\langle\bar{s}s\rangle\langle\bar{q}g_s\sigma Gq\rangle+\langle\bar{s}s\rangle\langle\bar{s}g_s\sigma Gs\rangle}{1536\pi^4}\int dydz \,(y+z)(1-y-z) \left(4s-3\overline{m}_c^2\right)\nonumber\\
&&+\frac{m_s m_c \left[2\langle\bar{q}q\rangle\langle\bar{s}g_s\sigma Gs\rangle+3\langle\bar{s}s\rangle\langle\bar{q}g_s\sigma Gq\rangle+5\langle\bar{s}s\rangle\langle\bar{s}g_s\sigma Gs\rangle\right]}{576\pi^4}\int dydz \,(y+z)\left[ 1+\frac{s}{3}\delta\left(s-\overline{m}_c^2 \right)\right] \nonumber\\
&&+\frac{m_s m_c \langle\bar{s}s\rangle\left[\langle\bar{q}g_s\sigma Gq\rangle+\langle\bar{s}g_s\sigma Gs\rangle\right]}{1024\pi^4}\int dydz \left( y+z\right)  \left[ 1+\frac{s}{3}\delta\left(s-\overline{m}_c^2 \right)\right]\nonumber\\
&&+\frac{m_s m_c \langle\bar{s}s\rangle\left[\langle\bar{q}g_s\sigma Gq\rangle+\langle\bar{s}g_s\sigma Gs\rangle\right]}{3072\pi^4}\int dy    \nonumber\\
&&-\frac{m_s m_c \left[16\langle\bar{q}q\rangle\langle\bar{s}g_s\sigma Gs\rangle+15\langle\bar{s}s\rangle\langle\bar{q}g_s\sigma Gq\rangle-5\langle\bar{s}s\rangle\langle\bar{s}g_s\sigma Gs\rangle\right]}{1728\pi^4}\int dy  \nonumber\\
&&+\frac{m_s m_c\left[2\langle\bar{q}q\rangle\langle\bar{s}g_s\sigma Gs\rangle+\langle\bar{s}s\rangle\langle\bar{q}g_s\sigma Gq\rangle-\langle\bar{s}s\rangle\langle\bar{s}g_s\sigma Gs\rangle\right]}{576\pi^4}\int dydz    \left( \frac{z}{y}+\frac{y}{z}\right)\nonumber\\
&&+\frac{m_s m_c \langle\bar{s}s\rangle\left[\langle\bar{q}g_s\sigma Gq\rangle+\langle\bar{s}g_s\sigma Gs\rangle\right]}{4608\pi^4}\int dydz \left( \frac{z}{y}+\frac{y}{z}\right) \nonumber\\
&&-\frac{m_s m_c\langle\bar{s}s\rangle\left[\langle\bar{q}g_s\sigma Gq\rangle+\langle\bar{s}g_s\sigma Gs\rangle\right]}{256\pi^4}\int dydz    \left( \frac{z}{y}+\frac{y}{z}\right)(1-y-z)\left[ 1+\frac{s}{3}\delta\left(s-\overline{m}_c^2 \right)\right]\nonumber\\
&&-\frac{m_c\langle\bar{q}q\rangle\langle\bar{s}s\rangle^2}{36\pi^2}\int dy   +\frac{m_s\langle\bar{q}q\rangle\langle\bar{s}s\rangle^2}{18\pi^2}\int dy \,y(1-y) \left[ 1+\frac{s}{3}\delta\left(s-\widetilde{m}_c^2 \right)\right]  \nonumber\\
&&+\frac{13\langle\bar{s}g_s\sigma Gs\rangle\left[2\langle\bar{q}g_s\sigma Gq\rangle+\langle\bar{s}g_s\sigma Gs\rangle \right]}{18432\pi^4}\int dydz \, (y+z)\left[ 1+\frac{s}{3}\delta\left(s-\overline{m}_c^2 \right)\right] \nonumber\\
&&+\frac{11\langle\bar{s}g_s\sigma Gs\rangle\left[2\langle\bar{q}g_s\sigma Gq\rangle+\langle\bar{s}g_s\sigma Gs\rangle \right]}{3072\pi^4}\int dy \, y(1-y)\left[ 1+\frac{s}{3}\delta\left(s-\widetilde{m}_c^2 \right)\right] \nonumber\\
&&+\frac{m_s m_c\langle\bar{s}g_s\sigma Gs\rangle\left[\langle\bar{q}g_s\sigma Gq\rangle+\langle\bar{s}g_s\sigma Gs\rangle \right]}{2304\pi^4}\int dydz    \left( \frac{z}{y}+\frac{y}{z}\right) \left( 1+\frac{s}{2M^2}\right)\delta\left(s-\overline{m}_c^2 \right)\nonumber\\
&&-\frac{m_s m_c\langle\bar{s}g_s\sigma Gs\rangle \left[ 5\langle\bar{q}g_s\sigma Gq\rangle-\langle\bar{s}g_s\sigma Gs\rangle\right]}{3456\pi^4}\int dy    \left( \frac{1-y}{y}+\frac{y}{1-y}\right)  \delta\left(s-\widetilde{m}_c^2 \right)\nonumber\\
&&-\frac{m_s m_c\langle\bar{s}g_s\sigma Gs\rangle\left[\langle\bar{q}g_s\sigma Gq\rangle+\langle\bar{s}g_s\sigma Gs\rangle \right]}{27648\pi^4}\int dy \,  \left( \frac{1-y}{y}+\frac{y}{1-y}\right)\delta\left(s-\widetilde{m}_c^2 \right) \nonumber\\
&&+\frac{m_s m_c\langle\bar{s}g_s\sigma Gs\rangle\left[7\langle\bar{q}g_s\sigma Gq\rangle-2\langle\bar{s}g_s\sigma Gs\rangle \right]}{1728\pi^4}\int dy   \left( 1+\frac{s}{2M^2}\right)\delta\left(s-\widetilde{m}_c^2 \right) \nonumber\\
&&-\frac{m_s m_c\langle\bar{s}g_s\sigma Gs\rangle\left[\langle\bar{q}g_s\sigma Gq\rangle+\langle\bar{s}g_s\sigma Gs\rangle \right]}{4608\pi^4}\int dy \,  \left( 1+\frac{s}{2M^2}\right)\delta\left(s-\widetilde{m}_c^2 \right)\, ,
\end{eqnarray}

\begin{eqnarray}
\widetilde{\rho}^{10,0}_{uss}(s)&=&\frac{1}{122880\pi^8}\int dydz \, (y+z)(1-y-z)^4\left(s-\overline{m}_c^2\right)^4\left(7s-2\overline{m}_c^2 \right)  \nonumber\\
&&-\frac{1}{614400\pi^8}\int dydz \, (y+z)(1-y-z)^5\left(s-\overline{m}_c^2\right)^3\left(28s^2-21s\overline{m}_c^2+3\overline{m}_c^4 \right)  \nonumber\\
&&+\frac{m_s m_c}{4608\pi^8}\int dydz \, (1-y-z)^3\left(s-\overline{m}_c^2\right)^4  \nonumber\\
&&-\frac{m_s m_c}{18432\pi^8}\int dydz \, (1-y-z)^4\left(s-\overline{m}_c^2\right)^3 \left(3s-\overline{m}_c^2 \right) \nonumber\\
&&- \frac{m_c\left[\langle \bar{q}q\rangle+2\langle \bar{s}s\rangle\right]}{576\pi^6}\int dydz \, (1-y-z)^2\left(s-\overline{m}_c^2\right)^3  \nonumber \\
&&+ \frac{m_c\left[\langle \bar{q}q\rangle+2\langle \bar{s}s\rangle\right]}{3456\pi^6}\int dydz \, (1-y-z)^3\left(s-\overline{m}_c^2\right)^2\left(5s-2\overline{m}_c^2\right)  \nonumber \\
&&- \frac{m_s\left[2\langle \bar{q}q\rangle-\langle \bar{s}s\rangle\right]}{1152\pi^6}\int dydz \, (y+z)(1-y-z)^2\left(s-\overline{m}_c^2\right)^2 \left(5s-2\overline{m}_c^2 \right) \nonumber \\
&&- \frac{m_s\langle \bar{s}s\rangle}{384\pi^6}\int dydz \, (y+z)(1-y-z)^3\left(s-\overline{m}_c^2\right)
 \left(5s^2-5s\overline{m}_c^2+\overline{m}_c^4 \right) \nonumber\\
&&+ \frac{11m_c\left[\langle \bar{q}g_s\sigma Gq\rangle+2\langle \bar{s}g_s\sigma Gs\rangle\right]}{3072\pi^6}\int dydz  \, (1-y-z) \left(s-\overline{m}_c^2 \right)^2 \nonumber\\
&&- \frac{11m_c\left[\langle \bar{q}g_s\sigma Gq\rangle+2\langle \bar{s}g_s\sigma Gs\rangle\right]}{6144\pi^6}\int dydz  \, (1-y-z)^2 \left(s-\overline{m}_c^2 \right)\left(2s-\overline{m}_c^2 \right)  \nonumber\\
&&- \frac{7m_c\left[\langle \bar{q}g_s\sigma Gq\rangle+2\langle \bar{s}g_s\sigma Gs\rangle\right]}{12288\pi^6}\int dydz  \, \left(\frac{1}{y}+\frac{1}{z}\right)(1-y-z)^2 \left(s-\overline{m}_c^2 \right)^2 \nonumber\\
&&+ \frac{m_c\left[\langle \bar{q}g_s\sigma Gq\rangle+2\langle \bar{s}g_s\sigma Gs\rangle\right]}{6144\pi^6}\int dydz  \, \left(\frac{1}{y}+\frac{1}{z}\right)(1-y-z)^3 \left(s-\overline{m}_c^2 \right)\left(2s-\overline{m}_c^2 \right)  \nonumber\\
&&+ \frac{m_s\langle \bar{s}g_s\sigma Gs\rangle}{768\pi^6}\int dydz  \, (y+z)(1-y-z)^2  \left(10s^2-12s\overline{m}_c^2+3\overline{m}_c^4 \right) \nonumber\\
&&+ \frac{m_s\langle \bar{q}g_s\sigma Gq\rangle}{192\pi^6}\int dydz  \, (y+z)(1-y-z) \left(s-\overline{m}_c^2 \right)\left(2s-\overline{m}_c^2 \right) \nonumber\\
&&+ \frac{m_s\left[\langle \bar{q}g_s\sigma Gq\rangle+\langle \bar{s}g_s\sigma Gs\rangle\right]}{1024\pi^6}\int dydz  \, (y+z)(1-y-z) \left(s-\overline{m}_c^2 \right)\left(2s-\overline{m}_c^2 \right) \nonumber\\
&&+\frac{\langle\bar{s}s\rangle\left[2\langle\bar{q}q\rangle+\langle\bar{s}s\rangle \right]}{72\pi^4}\int dydz \,  (y+z)(1-y-z)\left(s-\overline{m}_c^2 \right)\left(2s-\overline{m}_c^2 \right) \nonumber\\
&&+\frac{ m_s m_c\langle\bar{s}s\rangle\left[5\langle\bar{q}q\rangle-\langle\bar{s}s\rangle \right]}{36\pi^4}\int dydz \,  \left(s-\overline{m}_c^2 \right) \nonumber\\
&&+\frac{ m_s m_c\langle\bar{s}s\rangle\left[\langle\bar{q}q\rangle+\langle\bar{s}s\rangle \right]}{72\pi^4}\int dydz \,  (1-y-z)\left(3s-2\overline{m}_c^2 \right) \nonumber
\end{eqnarray}
\begin{eqnarray}
&&-\frac{19\left[\langle\bar{q}q\rangle\langle\bar{s}g_s\sigma Gs\rangle+\langle\bar{s}s\rangle\langle\bar{q}g_s\sigma Gq\rangle+\langle\bar{s}s\rangle\langle\bar{s}g_s\sigma Gs\rangle\right]}{2304\pi^4}\int dydz \,(y+z) \left(3s-2\overline{m}_c^2\right)\nonumber\\
&&-\frac{m_s m_c \left[2\langle\bar{q}q\rangle\langle\bar{s}g_s\sigma Gs\rangle+3\langle\bar{s}s\rangle\langle\bar{q}g_s\sigma Gq\rangle+5\langle\bar{s}s\rangle\langle\bar{s}g_s\sigma Gs\rangle\right]}{432\pi^4}\int dydz  \left[ 1+\frac{s}{2}\delta\left(s-\overline{m}_c^2 \right)\right] \nonumber\\
&&-\frac{m_s m_c\langle\bar{s}s\rangle\left[\langle\bar{q}g_s\sigma Gq\rangle+\langle\bar{s}g_s\sigma Gs\rangle\right]}{768\pi^4}\int dydz     \left[1+\frac{s}{2}\delta\left(s-\overline{m}_c^2 \right) \right]\nonumber\\
&&+\frac{m_s m_c\langle\bar{s}s\rangle\left[\langle\bar{q}g_s\sigma Gq\rangle+\langle\bar{s}g_s\sigma Gs\rangle\right]}{768\pi^4}\int dy  \nonumber\\
&&-\frac{m_s m_c \left[16\langle\bar{q}q\rangle\langle\bar{s}g_s\sigma Gs\rangle+15\langle\bar{s}s\rangle\langle\bar{q}g_s\sigma Gq\rangle-5\langle\bar{s}s\rangle\langle\bar{s}g_s\sigma Gs\rangle\right]}{432\pi^4}\int dy  \nonumber\\
&&+\frac{m_s m_c\langle\bar{s}s\rangle\left[\langle\bar{q}g_s\sigma Gq\rangle+\langle\bar{s}g_s\sigma Gs\rangle\right]}{2304\pi^4}\int dydz    \left( \frac{1}{y}+\frac{1}{z}\right)\nonumber\\
&&+\frac{m_s m_c\left[2\langle\bar{q}q\rangle\langle\bar{s}g_s\sigma Gs\rangle+\langle\bar{s}s\rangle\langle\bar{q}g_s\sigma Gq\rangle-\langle\bar{s}s\rangle\langle\bar{s}g_s\sigma Gs\rangle\right]}{288\pi^4}\int dydz    \left( \frac{1}{y}+\frac{1}{z}\right)\nonumber\\
&&+\frac{m_s m_c\langle\bar{s}s\rangle\left[\langle\bar{q}g_s\sigma Gq\rangle+\langle\bar{s}g_s\sigma Gs\rangle\right]}{384\pi^4}\int dydz    \left( \frac{1}{y}+\frac{1}{z}\right)(1-y-z)\left[1+\frac{s}{2}\delta\left(s-\overline{m}_c^2 \right) \right]\nonumber\\
&&-\frac{m_c\langle\bar{q}q\rangle\langle\bar{s}s\rangle^2}{9\pi^2}\int dy    +\frac{m_s\langle\bar{q}q\rangle\langle\bar{s}s\rangle^2}{27\pi^2}\int dy  \left[ 1+\frac{s}{2}\delta\left(s-\widetilde{m}_c^2 \right)\right]  \nonumber\\
&&+\frac{11\langle\bar{s}g_s\sigma Gs\rangle\left[2\langle\bar{q}g_s\sigma Gq\rangle+\langle\bar{s}g_s\sigma Gs\rangle \right]}{4608\pi^4}\int dy  \left[ 1+\frac{s}{2}\delta\left(s-\widetilde{m}_c^2 \right)\right] \nonumber\\
&&-\frac{m_s m_c\langle\bar{s}g_s\sigma Gs\rangle\left[\langle\bar{q}g_s\sigma Gq\rangle+\langle\bar{s}g_s\sigma Gs\rangle \right]}{4608\pi^4}\int dydz    \left( \frac{1}{y}+\frac{1}{z}\right) \left(1+\frac{s}{M^2} \right)\delta\left(s-\overline{m}_c^2 \right)\nonumber\\
&&-\frac{m_s m_c\langle\bar{s}g_s\sigma Gs\rangle\left[5\langle\bar{q}g_s\sigma Gq\rangle-\langle\bar{s}g_s\sigma Gs\rangle \right]}{1728\pi^4}\int dy     \left( \frac{1}{y}+\frac{1}{1-y}\right) \delta\left(s-\widetilde{m}_c^2 \right)\nonumber\\
&&-\frac{m_s m_c\langle\bar{s}g_s\sigma Gs\rangle\left[\langle\bar{q}g_s\sigma Gq\rangle+\langle\bar{s}g_s\sigma Gs\rangle\right]}{13824\pi^4}\int dy \left( \frac{1}{y}+\frac{1}{1-y}\right)  \delta\left(s-\widetilde{m}_c^2 \right) \nonumber\\
&&+\frac{m_s m_c\langle\bar{s}g_s\sigma Gs\rangle\left[15\langle\bar{q}g_s\sigma Gq\rangle-3\langle\bar{s}g_s\sigma Gs\rangle \right]}{1728\pi^4}\int dy   \left( 1+\frac{s}{M^2}\right)\delta\left(s-\widetilde{m}_c^2 \right) \nonumber\\
&&-\frac{m_s m_c\langle\bar{s}g_s\sigma Gs\rangle\left[\langle\bar{q}g_s\sigma Gq\rangle+\langle\bar{s}g_s\sigma Gs\rangle \right]}{9216\pi^4}\int dy \,  \left( 1+\frac{s}{M^2}\right)\delta\left(s-\widetilde{m}_c^2 \right)\, ,
\end{eqnarray}

\begin{eqnarray}
\rho^{10,1}_{uus}(s)&=&\frac{1}{122880\pi^8}\int dydz \, yz(1-y-z)^4\left(s-\overline{m}_c^2\right)^4\left(8s-3\overline{m}_c^2 \right)  \nonumber\\
&&+\frac{1}{153600\pi^8}\int dydz \, yz(1-y-z)^5\left(s-\overline{m}_c^2\right)^3\left(18s^2-16s\overline{m}_c^2+3\overline{m}_c^4 \right)  \nonumber\\
&&+\frac{m_s m_c}{36864\pi^8}\int dydz \, (y+z)(1-y-z)^3\left(s-\overline{m}_c^2\right)^4  \nonumber\\
&&+\frac{m_s m_c}{147456\pi^8}\int dydz \, (y+z)(1-y-z)^4\left(s-\overline{m}_c^2\right)^3 \left(7s-3\overline{m}_c^2 \right) \nonumber\\
&&- \frac{m_c\left[2\langle \bar{q}q\rangle+\langle \bar{s}s\rangle\right]}{2304\pi^6}\int dydz \, (y+z)(1-y-z)^2\left(s-\overline{m}_c^2\right)^3  \nonumber \\
&&- \frac{m_c\left[2\langle \bar{q}q\rangle+\langle \bar{s}s\rangle\right]}{2304\pi^6}\int dydz \, (y+z)(1-y-z)^3\left(s-\overline{m}_c^2\right)^2\left(2s-\overline{m}_c^2\right)  \nonumber \\
&&- \frac{m_s\left[4\langle \bar{q}q\rangle-3\langle \bar{s}s\rangle\right]}{768\pi^6}\int dydz \, yz(1-y-z)^2\left(s-\overline{m}_c^2\right)^2 \left(2s-\overline{m}_c^2 \right) \nonumber \\
&&+ \frac{m_s\langle \bar{s}s\rangle}{384\pi^6}\int dydz \, yz(1-y-z)^3\left(s-\overline{m}_c^2\right)
 \left(7s^2-8s\overline{m}_c^2+2\overline{m}_c^4 \right) \nonumber
\end{eqnarray}
\begin{eqnarray}
&&+ \frac{11m_c\left[2\langle \bar{q}g_s\sigma Gq\rangle+\langle \bar{s}g_s\sigma Gs\rangle\right]}{12288\pi^6}\int dydz  \, (y+z)(1-y-z) \left(s-\overline{m}_c^2 \right)^2 \nonumber\\
&&+ \frac{11m_c \left[2\langle \bar{q}g_s\sigma Gq\rangle+\langle \bar{s}g_s\sigma Gs\rangle\right] }{24576\pi^6}\int dydz  \, \left(y+z\right)(1-y-z)^2 \left(s-\overline{m}_c^2 \right) \left(5s-3\overline{m}_c^2 \right) \nonumber\\
&&- \frac{7m_c \left[2\langle \bar{q}g_s\sigma Gq\rangle+\langle \bar{s}g_s\sigma Gs\rangle\right] }{24576\pi^6}\int dydz  \, \left(\frac{z}{y}+\frac{y}{z}\right)(1-y-z)^2 \left(s-\overline{m}_c^2 \right)^2 \nonumber\\
&&- \frac{m_c \left[2\langle \bar{q}g_s\sigma Gq\rangle+\langle \bar{s}g_s\sigma Gs\rangle\right] }{12288\pi^6}\int dydz  \, \left(\frac{z}{y}+\frac{y}{z}\right)(1-y-z)^3 \left(s-\overline{m}_c^2 \right) \left(5s-3\overline{m}_c^2 \right) \nonumber\\
&&- \frac{m_s\langle \bar{s}g_s\sigma Gs\rangle}{768\pi^6}\int dydz  \, yz(1-y-z)^2  \left(15s^2-20s\overline{m}_c^2+6\overline{m}_c^4 \right) \nonumber\\
&&+ \frac{m_s\langle \bar{q}g_s\sigma Gq\rangle}{8192\pi^6}\int dydz  \, (y+z)(1-y-z)^2 \left(s-\overline{m}_c^2 \right)\left(5s-3\overline{m}_c^2 \right) \nonumber\\
&&+ \frac{m_s\left[2\langle \bar{q}g_s\sigma Gq\rangle-\langle \bar{s}g_s\sigma Gs\rangle\right]}{768\pi^6}\int dydz  \, yz(1-y-z) \left(s-\overline{m}_c^2 \right)\left(5s-3\overline{m}_c^2 \right) \nonumber\\
&&+ \frac{m_s\langle \bar{q}g_s\sigma Gq\rangle}{2048\pi^6}\int dydz  \, yz(1-y-z) \left(s-\overline{m}_c^2 \right)\left(5s-3\overline{m}_c^2 \right) \nonumber\\
&&+\frac{\langle\bar{q}q\rangle\left[\langle\bar{q}q\rangle+2\langle\bar{s}s\rangle\right]}{144\pi^4}\int dydz \,  yz(1-y-z)\left(s-\overline{m}_c^2 \right)\left(5s-3\overline{m}_c^2 \right) \nonumber\\
&&+\frac{ m_s m_c\langle\bar{q}q\rangle\left[3\langle\bar{q}q\rangle-\langle\bar{s}s\rangle\right]}{144\pi^4}\int dydz \,  (y+z)\left(s-\overline{m}_c^2 \right) \nonumber\\
&&-\frac{ m_s m_c\langle\bar{q}q\rangle\langle\bar{s}s\rangle}{144\pi^4}\int dydz \,  (y+z)(1-y-z)\left(4s-3\overline{m}_c^2 \right)
\nonumber\\
&&-\frac{19\left[\langle\bar{q}q\rangle\langle\bar{q}g_s\sigma Gq\rangle+\langle\bar{q}q\rangle\langle\bar{s}g_s\sigma Gs\rangle+\langle\bar{s}s\rangle\langle\bar{q}g_s\sigma Gq\rangle\right]}{2304\pi^4}\int dydz \,yz \left(4s-3\overline{m}_c^2\right)\nonumber\\
&&-\frac{\langle\bar{q}q\rangle\langle\bar{q}g_s\sigma Gq\rangle+\langle\bar{q}q\rangle\langle\bar{s}g_s\sigma Gs\rangle+\langle\bar{s}s\rangle\langle\bar{q}g_s\sigma Gq\rangle}{1536\pi^4}\int dydz \,(y+z)(1-y-z) \left(4s-3\overline{m}_c^2\right)\nonumber\\
&&+\frac{m_s m_c \left[2\langle\bar{q}q\rangle\langle\bar{s}g_s\sigma Gs\rangle+3\langle\bar{s}s\rangle\langle\bar{q}g_s\sigma Gq\rangle\right]}{576\pi^4}\int dydz \,(y+z)\left[ 1+\frac{s}{3}\delta\left(s-\overline{m}_c^2 \right)\right] \nonumber\\
&&+\frac{m_s m_c \langle\bar{s}s\rangle\langle\bar{q}g_s\sigma Gq\rangle}{1024\pi^4}\int dydz \left( y+z\right)  \left[ 1+\frac{s}{3}\delta\left(s-\overline{m}_c^2 \right)\right]\nonumber\\
&&+\frac{m_s m_c \langle\bar{s}s\rangle\langle\bar{q}g_s\sigma Gq\rangle}{3072\pi^4}\int dy    \nonumber\\
&&-\frac{m_s m_c \left[18\langle\bar{q}q\rangle\langle\bar{q}g_s\sigma Gq\rangle-2\langle\bar{q}q\rangle\langle\bar{s}g_s\sigma Gs\rangle-3\langle\bar{s}s\rangle\langle\bar{q}g_s\sigma Gq\rangle\right]}{1728\pi^4}\int dy  \nonumber\\
&&+\frac{m_s m_c\left[2 \langle\bar{q}q\rangle-\langle\bar{s}s\rangle\right]\langle\bar{q}g_s\sigma Gq\rangle}{576\pi^4}\int dydz    \left( \frac{z}{y}+\frac{y}{z}\right)\nonumber\\
&&+\frac{m_s m_c \langle\bar{s}s\rangle\langle\bar{q}g_s\sigma Gq\rangle}{4608\pi^4}\int dydz \left( \frac{z}{y}+\frac{y}{z}\right) \nonumber\\
&&-\frac{m_s m_c\langle\bar{s}s\rangle\langle\bar{q}g_s\sigma Gq\rangle}{256\pi^4}\int dydz    \left( \frac{z}{y}+\frac{y}{z}\right)(1-y-z)\left[ 1+\frac{s}{3}\delta\left(s-\overline{m}_c^2 \right)\right]\nonumber\\
&&-\frac{m_c\langle\bar{q}q\rangle^2\langle\bar{s}s\rangle}{36\pi^2}\int dy   +\frac{m_s\langle\bar{q}q\rangle^2\langle\bar{s}s\rangle}{36\pi^2}\int dy \,y(1-y) \left[ 1+\frac{s}{3}\delta\left(s-\widetilde{m}_c^2 \right)\right]  \nonumber
\end{eqnarray}
\begin{eqnarray}
&&+\frac{13\langle\bar{q}g_s\sigma Gq\rangle \left[\langle\bar{q}g_s\sigma Gq\rangle+2\langle\bar{s}g_s\sigma Gs\rangle \right]}{18432\pi^4}\int dydz \, (y+z)\left[ 1+\frac{s}{3}\delta\left(s-\overline{m}_c^2 \right)\right] \nonumber\\
&&+\frac{11\langle\bar{q}g_s\sigma Gq\rangle \left[\langle\bar{q}g_s\sigma Gq\rangle+2\langle\bar{s}g_s\sigma Gs\rangle \right]}{3072\pi^4}\int dy \, y(1-y)\left[ 1+\frac{s}{3}\delta\left(s-\widetilde{m}_c^2 \right)\right] \nonumber\\
&&+\frac{m_s m_c\langle\bar{q}g_s\sigma Gq\rangle\langle\bar{s}g_s\sigma Gs\rangle }{2304\pi^4}\int dydz    \left( \frac{z}{y}+\frac{y}{z}\right) \left( 1+\frac{s}{2M^2}\right)\delta\left(s-\overline{m}_c^2 \right)\nonumber\\
&&-\frac{m_s m_c\langle\bar{q}g_s\sigma Gq\rangle\left[3\langle\bar{q}g_s\sigma Gq\rangle -\langle\bar{s}g_s\sigma Gs\rangle\right] }{3456\pi^4}\int dy    \left( \frac{1-y}{y}+\frac{y}{1-y}\right)  \delta\left(s-\widetilde{m}_c^2 \right)\nonumber\\
&&-\frac{m_s m_c\langle\bar{q}g_s\sigma Gq\rangle \langle\bar{s}g_s\sigma Gs\rangle}{27648\pi^4}\int dy \,  \left( \frac{1-y}{y}+\frac{y}{1-y}\right)\delta\left(s-\widetilde{m}_c^2 \right) \nonumber\\
&&+\frac{m_s m_c\langle\bar{q}g_s\sigma Gq\rangle \left[9\langle\bar{q}g_s\sigma Gq\rangle-4\langle\bar{s}g_s\sigma Gs\rangle \right]}{3456\pi^4}\int dy   \left( 1+\frac{s}{2M^2}\right)\delta\left(s-\widetilde{m}_c^2 \right) \nonumber\\
&&-\frac{m_s m_c\langle\bar{q}g_s\sigma Gq\rangle\langle\bar{s}g_s\sigma Gs\rangle }{4608\pi^4}\int dy \,  \left( 1+\frac{s}{2M^2}\right)\delta\left(s-\widetilde{m}_c^2 \right)\, ,
\end{eqnarray}

\begin{eqnarray}
\widetilde{\rho}^{10,0}_{uus}(s)&=&\frac{1}{122880\pi^8}\int dydz \, (y+z)(1-y-z)^4\left(s-\overline{m}_c^2\right)^4\left(7s-2\overline{m}_c^2 \right)  \nonumber\\
&&-\frac{1}{614400\pi^8}\int dydz \, (y+z)(1-y-z)^5\left(s-\overline{m}_c^2\right)^3\left(28s^2-21s\overline{m}_c^2+3\overline{m}_c^4 \right)  \nonumber\\
&&+\frac{m_s m_c}{9216\pi^8}\int dydz \, (1-y-z)^3\left(s-\overline{m}_c^2\right)^4  \nonumber\\
&&-\frac{m_s m_c}{36864\pi^8}\int dydz \, (1-y-z)^4\left(s-\overline{m}_c^2\right)^3 \left(3s-\overline{m}_c^2 \right) \nonumber\\
&&- \frac{m_c\left[2\langle \bar{q}q\rangle+\langle \bar{s}s\rangle\right]}{576\pi^6}\int dydz \, (1-y-z)^2\left(s-\overline{m}_c^2\right)^3  \nonumber \\
&&+ \frac{m_c\left[2\langle \bar{q}q\rangle+\langle \bar{s}s\rangle\right]}{3456\pi^6}\int dydz \, (1-y-z)^3\left(s-\overline{m}_c^2\right)^2\left(5s-2\overline{m}_c^2\right)  \nonumber \\
&&- \frac{m_s\left[4\langle \bar{q}q\rangle-3\langle \bar{s}s\rangle\right]}{2304\pi^6}\int dydz \, (y+z)(1-y-z)^2\left(s-\overline{m}_c^2\right)^2 \left(5s-2\overline{m}_c^2 \right) \nonumber \\
&&- \frac{m_s\langle \bar{s}s\rangle}{768\pi^6}\int dydz \, (y+z)(1-y-z)^3\left(s-\overline{m}_c^2\right)
\left(5s^2-5s\overline{m}_c^2+\overline{m}_c^4 \right) \nonumber
\end{eqnarray}
\begin{eqnarray}
&&+ \frac{11m_c\left[2\langle \bar{q}g_s\sigma Gq\rangle+\langle \bar{s}g_s\sigma Gs\rangle\right]}{3072\pi^6}\int dydz  \, (1-y-z) \left(s-\overline{m}_c^2 \right)^2 \nonumber\\
&&- \frac{11m_c\left[2\langle \bar{q}g_s\sigma Gq\rangle+\langle \bar{s}g_s\sigma Gs\rangle\right]}{6144\pi^6}\int dydz  \, (1-y-z)^2 \left(s-\overline{m}_c^2 \right)\left(2s-\overline{m}_c^2 \right)  \nonumber\\
&&- \frac{7m_c\left[2\langle \bar{q}g_s\sigma Gq\rangle+\langle \bar{s}g_s\sigma Gs\rangle\right]}{12288\pi^6}\int dydz  \, \left(\frac{1}{y}+\frac{1}{z}\right)(1-y-z)^2 \left(s-\overline{m}_c^2 \right)^2 \nonumber\\
&&+ \frac{m_c\left[2\langle \bar{q}g_s\sigma Gq\rangle+\langle \bar{s}g_s\sigma Gs\rangle\right]}{6144\pi^6}\int dydz  \, \left(\frac{1}{y}+\frac{1}{z}\right)(1-y-z)^3 \left(s-\overline{m}_c^2 \right)\left(2s-\overline{m}_c^2 \right)  \nonumber\\
&&+ \frac{m_s\langle \bar{s}g_s\sigma Gs\rangle}{1536\pi^6}\int dydz  \, (y+z)(1-y-z)^2  \left(10s^2-12s\overline{m}_c^2+3\overline{m}_c^4 \right) \nonumber\\
&&+ \frac{m_s\left[2\langle \bar{q}g_s\sigma Gq\rangle-\langle \bar{s}g_s\sigma Gs\rangle\right]}{384\pi^6}\int dydz  \, (y+z)(1-y-z) \left(s-\overline{m}_c^2 \right)\left(2s-\overline{m}_c^2 \right) \nonumber\\
&&+ \frac{m_s\langle \bar{q}g_s\sigma Gq\rangle}{1024\pi^6}\int dydz  \, (y+z)(1-y-z) \left(s-\overline{m}_c^2 \right)\left(2s-\overline{m}_c^2 \right) \nonumber\\
&&+\frac{\langle\bar{q}q\rangle\left[ \langle\bar{q}q\rangle+2\langle\bar{s}s\rangle\right]}{72\pi^4}\int dydz \,  (y+z)(1-y-z)\left(s-\overline{m}_c^2 \right)\left(2s-\overline{m}_c^2 \right) \nonumber\\
&&+\frac{ m_s m_c\langle\bar{q}q\rangle\left[ 3\langle\bar{q}q\rangle-\langle\bar{s}s\rangle\right]}{36\pi^4}\int dydz \,  \left(s-\overline{m}_c^2 \right) \nonumber\\
&&+\frac{ m_s m_c\langle\bar{q}q\rangle\langle\bar{s}s\rangle }{72\pi^4}\int dydz \,  (1-y-z)\left(3s-2\overline{m}_c^2 \right) \nonumber\\
&&-\frac{19\left[\langle\bar{q}q\rangle\langle\bar{q}g_s\sigma Gq\rangle+\langle\bar{q}q\rangle\langle\bar{s}g_s\sigma Gs\rangle+\langle\bar{s}s\rangle\langle\bar{q}g_s\sigma Gq\rangle\right]}{2304\pi^4}\int dydz \,(y+z) \left(3s-2\overline{m}_c^2\right)\nonumber\\
&&-\frac{m_s m_c \left[2\langle\bar{q}q\rangle\langle\bar{s}g_s\sigma Gs\rangle+3\langle\bar{s}s\rangle\langle\bar{q}g_s\sigma Gq\rangle\right]}{432\pi^4}\int dydz  \left[ 1+\frac{s}{2}\delta\left(s-\overline{m}_c^2 \right)\right] \nonumber\\
&&-\frac{m_s m_c\langle\bar{s}s\rangle\langle\bar{q}g_s\sigma Gq\rangle}{768\pi^4}\int dydz     \left[1+\frac{s}{2}\delta\left(s-\overline{m}_c^2 \right) \right]\nonumber\\
&&+\frac{m_s m_c\langle\bar{s}s\rangle\langle\bar{q}g_s\sigma Gq\rangle}{768\pi^4}\int dy  \nonumber\\
&&-\frac{m_s m_c \left[18\langle\bar{q}q\rangle\langle\bar{q}g_s\sigma Gq\rangle-2\langle\bar{q}q\rangle\langle\bar{s}g_s\sigma Gs\rangle-3\langle\bar{s}s\rangle\langle\bar{q}g_s\sigma Gq\rangle\right]}{432\pi^4}\int dy  \nonumber\\
&&+\frac{m_s m_c\left[2\langle\bar{q}q\rangle-\langle\bar{s}s\rangle\right]\langle\bar{q}g_s\sigma Gq\rangle}{288\pi^4}\int dydz    \left( \frac{1}{y}+\frac{1}{z}\right)\nonumber\\
&&+\frac{m_s m_c\langle\bar{s}s\rangle\langle\bar{q}g_s\sigma Gq\rangle}{2304\pi^4}\int dydz    \left( \frac{1}{y}+\frac{1}{z}\right)\nonumber\\
&&+\frac{m_s m_c\langle\bar{s}s\rangle\langle\bar{q}g_s\sigma Gq\rangle}{384\pi^4}\int dydz    \left( \frac{1}{y}+\frac{1}{z}\right)(1-y-z)\left[1+\frac{s}{2}\delta\left(s-\overline{m}_c^2 \right) \right]\nonumber\\
&&-\frac{m_c\langle\bar{q}q\rangle^2\langle\bar{s}s\rangle}{9\pi^2}\int dy    +\frac{m_s\langle\bar{q}q\rangle^2\langle\bar{s}s\rangle}{54\pi^2}\int dy  \left[ 1+\frac{s}{2}\delta\left(s-\widetilde{m}_c^2 \right)\right]  \nonumber
\end{eqnarray}
\begin{eqnarray}
&&+\frac{11\langle\bar{q}g_s\sigma Gq\rangle \left[\langle\bar{q}g_s\sigma Gq\rangle+2\langle\bar{s}g_s\sigma Gs\rangle \right]}{4608\pi^4}\int dy  \left[ 1+\frac{s}{2}\delta\left(s-\widetilde{m}_c^2 \right)\right] \nonumber\\
&&-\frac{m_s m_c\langle\bar{q}g_s\sigma Gq\rangle \langle\bar{s}g_s\sigma Gs\rangle}{4608\pi^4}\int dydz    \left( \frac{1}{y}+\frac{1}{z}\right) \left(1+\frac{s}{M^2} \right)\delta\left(s-\overline{m}_c^2 \right)\nonumber\\
&&-\frac{m_s m_c\langle\bar{q}g_s\sigma Gq\rangle \left[3\langle\bar{q}g_s\sigma Gq\rangle -\langle\bar{s}g_s\sigma Gs\rangle\right]}{1728\pi^4}\int dy     \left( \frac{1}{y}+\frac{1}{1-y}\right) \delta\left(s-\widetilde{m}_c^2 \right)\nonumber\\
&&-\frac{m_s m_c\langle\bar{q}g_s\sigma Gq\rangle \langle\bar{s}g_s\sigma Gs\rangle}{13824\pi^4}\int dy \left( \frac{1}{y}+\frac{1}{1-y}\right)  \delta\left(s-\widetilde{m}_c^2 \right) \nonumber\\
&&+\frac{m_s m_c\langle\bar{q}g_s\sigma Gq\rangle \left[ 9\langle\bar{q}g_s\sigma Gq\rangle-\langle\bar{s}g_s\sigma Gs\rangle\right]}{1728\pi^4}\int dy   \left( 1+\frac{s}{M^2}\right)\delta\left(s-\widetilde{m}_c^2 \right) \nonumber\\
&&-\frac{m_s m_c\langle\bar{q}g_s\sigma Gq\rangle \langle\bar{s}g_s\sigma Gs\rangle}{9216\pi^4}\int dy \,  \left( 1+\frac{s}{M^2}\right)\delta\left(s-\widetilde{m}_c^2 \right)\, ,
\end{eqnarray}

\begin{eqnarray}
\rho^{10,1}_{uuu}(s)&=&\rho^{10,1}_{sss}(s)\mid_{m_s \to 0,\,\,\langle\bar{s}s\rangle\to\langle\bar{q}q\rangle,\,\,\langle\bar{s}g_s\sigma Gs\rangle\to\langle\bar{q}g_s\sigma Gq\rangle} \, ,  \\
\widetilde{\rho}^{10,0}_{uuu}(s)&=&\widetilde{\rho}^{10,0}_{sss}(s)\mid_{m_s \to 0,\,\,\langle\bar{s}s\rangle\to\langle\bar{q}q\rangle,\,\,\langle\bar{s}g_s\sigma Gs\rangle\to\langle\bar{q}g_s\sigma Gq\rangle} \, ,
\end{eqnarray}
where $\int dydz=\int_{y_i}^{y_f}dy \int_{z_i}^{1-y}dz$, $\int dy=\int_{y_i}^{y_f}dy$, $y_{f}=\frac{1+\sqrt{1-4m_c^2/s}}{2}$,
$y_{i}=\frac{1-\sqrt{1-4m_c^2/s}}{2}$, $z_{i}=\frac{y
m_c^2}{y s -m_c^2}$, $\overline{m}_c^2=\frac{(y+z)m_c^2}{yz}$,
$ \widetilde{m}_c^2=\frac{m_c^2}{y(1-y)}$, $\int_{y_i}^{y_f}dy \to \int_{0}^{1}dy$, $\int_{z_i}^{1-y}dz \to \int_{0}^{1-y}dz$ when the $\delta$ functions $\delta\left(s-\overline{m}_c^2\right)$ and $\delta\left(s-\widetilde{m}_c^2\right)$ appear.

\end{document}